\documentclass{aa}  

\usepackage{graphicx}
\usepackage[labelfont=bf,font=small]{caption}
\usepackage{subcaption}
\usepackage{amsmath}
\usepackage{xcolor}
\usepackage{float}
\usepackage[utf8]{inputenc}
\usepackage[T1]{fontenc}
\usepackage{txfonts}
\usepackage{fix2col}

\usepackage{color}
\usepackage{colortbl} 
\definecolor{pinegreen}{RGB}{1, 121, 111}
\definecolor{salmon}{RGB}{255,160,122}
\definecolor{azul}{RGB}{0,45,234}
\definecolor{rojo}{RGB}{238,90,76}
\usepackage[breaklinks=true,colorlinks,citecolor=azul,urlcolor=rojo]{hyperref}         
\usepackage[hyphenbreaks]{breakurl}
\usepackage{natbib}
\usepackage{booktabs}    
\usepackage{morefloats}

\newcommand\kms{\ensuremath{\text{km~s}^{-1}}}

\newcommand\gaia{\textit{Gaia}}
\newcommand\gdr[1]{\gaia~DR#1}
\newcommand\hip{\textsc{Hipparcos}}

\newcommand\secref[1]{Sect.~\ref{#1}}

\newcommand\figref[1]{Fig.~\ref{#1}}

\newcommand\figrefalt[1]{Figure~\ref{#1}}

\newcommand\equref[1]{Eq.~\eqref{#1}}
\newcommand\equrefalt[1]{Equation~\eqref{#1}}
\newcommand\tabref[1]{Table~\ref{#1}}

\graphicspath{ {./images/} }

\begin{document} 

     \title{Galactic tide and local stellar perturbations on the Oort cloud: creation of interstellar comets}


    \author{Santiago Torres\inst{1}
          \and
          Maxwell X.~Cai\inst{1}
          \and
          A.~G.~A.~Brown\inst{1}
          \and
          S.~Portegies Zwart\inst{1}
          }
        \institute{Leiden Observatory, Leiden University, P.O.~Box 9513, 2300 RA, Leiden, The Netherlands\\
              \email{storres@strw.leidenuniv.nl}}

   \date{Received 12 March 2019 / Accepted 24 July 2019}

 
  \abstract
   {Comets in the Oort cloud evolve under the influence of internal and external perturbations, such
     as giant planets, stellar passages, and the Galactic gravitational tidal field. We aim to study
     the dynamical evolution of the comets in the Oort cloud, accounting for the perturbation of the
     Galactic tidal field and passing stars. We base our study on three main approaches; analytic,
     observational, and numerical.  We first construct an analytical  model of stellar
       encounters. We find that individual perturbations do not modify the dynamics of the comets in
       the cloud unless very close ($<$ 0.5~pc) encounters occur. Using proper motions,
       parallaxes, and radial velocities from \gdr{2} and combining them with the radial velocities
       from other surveys, we then construct an astrometric catalogue of the $14\,659$ stars that are
       within $50$~pc of the Sun. For all these stars we calculate the time and distance of
     closest approach to the Sun. We find that the cumulative effect of relatively distant ($\leq1$~pc) passing
   stars can perturb the comets in the Oort cloud. Finally, we study the dynamical evolution of the
   comets in the Oort cloud under the influence of multiple stellar encounters from stars that pass
   within $2.5$~pc of the Sun and the Galactic tidal field over $\pm10$~Myr. We use the
   Astrophysical Multipurpose Software Environment (\texttt{AMUSE}), and the GPU-accelerated direct
   N-body code \texttt{ABIE}.  We considered two models for the Oort cloud, compact  
   ($a \leq0.25$~pc) and extended ($a \leq0.5$~pc). We find that the cumulative effect of stellar encounters 
   is the major perturber of the Oort cloud for a compact configuration while for the extended
   configuration the Galactic tidal field is the major perturber. In both cases the cumulative effect of distant stellar encounters
   together with the Galactic tidal field raises the semi-major axis of $\sim1.1$\%  of the comets
   at the edge of the Oort cloud up to interstellar regions ($a >0.5$~pc) over the $20$~Myr period
   considered. This leads to the creation of transitional interstellar comets (TICs), which might 
   become interstellar objects due to external perturbations.  This raises the question
   of the formation, evolution, and current status of the Oort cloud as well as the existence of
   a ``cloud'' of objects in the interstellar space that might overlap with our Oort cloud, when
   considering that other planetary systems should undergo similar processes leading to the ejection of comets.}
   

   \titlerunning{Dynamical Evolution of the Oort cloud}
   \authorrunning{Torres et~al.}

   \maketitle
%
\section{Introduction}

The outer region of the solar system is populated by a large number of planetesimals. Further away,
more than $1\,000$ AU from the Sun, and almost extending to the nearest stars, is the Oort cloud. Its
existence was proposed in the late 1950s by the Dutch astronomer Jan Hendrik Oort, who realised that
long-term comets (with orbital semi-major axes $a>40$~AU) bound to the Sun must come from an area
well beyond Neptune. \citet{J.H.Oort1950} pointed out that a spike in the distribution of $1/a$ of
the long-period comets with $a>10^4$ AU, and isotropic inclinations in $\cos i$, $\omega$, and
$\Omega$, would argue for the existence of a reservoir of objects in quasi-spherical symmetry
surrounding the solar system. The Oort cloud has remained unobserved to date.

There have been numerous studies aimed at trying to explain the formation, evolution, and structure of the
Oort cloud, mostly through numerical simulation (e.g. \citealt{Hills1981, Heisler1986a,
DuncanM.J.QuinnT.1987, Weissman1996, Wiegert1999, Garcia-Sanchez1999,Dybczynski2002, Levison2004,
Dones2004a, Morbidelli2008, Duncan2008, Brasser2006, Fouchard2006, Kaib2008, Brasser2013,
Shannon2014, Dones2015}). There is general agreement on some properties of the Oort cloud, in
particular that it is composed of the residual planetesimals after the planet formation epoch. The
Oort cloud is divided into two regions: the {\it inner} Oort cloud is usually reserved for comets
with semi-major axes $a<20\,000$ AU and is invisible unless there is a
comet shower. The {\it outer} Oort cloud refers to comets with semi-major axes $a>20\,000$ AU (e.g.
\citealt{Dones2015}). Its shape is thought to be nearly spherical and limited at $0.5$ pc mainly by the influence
of the Galactic tidal field and stellar flybys (e.g. \citealt{Heisler1986a}). The Oort cloud is
thought to contain around $10^{12}$ objects with a total mass of $\sim3\times 10^{25}$~kg (e.g.
\citealt{Morbidelli2008}). However, these estimations are highly uncertain. The above-mentioned
studies also concluded that in order for long-period comets to still exist today they need to be
replenished. Otherwise they would have been depleted on a timescale much shorter than the lifetime
of the solar system.    

The orbits of the comets in the Oort cloud form a frozen record of the evolution of the solar system
and preserve the memory of its birth environment (\citealt{PortegiesZwart2015a,
Martinez-Barbosa2016,Fouchard2011,Fouchard2018}). External perturbations such as Galactic tides,
stellar flybys, and molecular clouds play an important role in the understanding of the formation and
evolution of the Oort cloud and Oort cloud-like structures in other planetary systems (see e.g.  \citealt{Veras2013,Veras2014}).  
Passing stars can perturb the comets, changing their perihelion
distances much more than they change the overall size of the orbit, changing the cometary
trajectories and injecting the comets into the inner solar system (\citealt{Morbidelli2008} and
\citealt{Duncan2008}). The outer Oort cloud has been affected quite substantially by external
influences. Not only by passing stars in the parental cluster of the Sun but also by occasional
relatively close encounters that have occurred after the Sun has left its birth cluster
(\citealt{Jilkova2016a}). \citet{Jilkova2015} pointed out that the planetesimals
\textit{Sedna} and \textit{2008PV113} belong to the inner Oort cloud and that they may have been
captured during an encounter with another star in the birth cluster of the Sun. This star is conjectured
to have passed the solar system within about 340 AU and would have deposited approximately $1\,400$ other
planetesimals together with the two currently known objects in this family. The orbital
characteristics of these objects share similar properties which can be used to reconstruct the
encounter. 

Close encounters with the solar system have been studied by a number of authors (e.g.
\citealt{Rickman1976, Matthews1994, Weissman1996, Dehnen1998, Garcia-Sanchez1999,
Levison2004,Jimenez-Torres2011a, Bailer-Jones2015, Dybczynski2015, Higuchi2015, Feng2015,
Berski2016}). Most of them calculated the closest encounters with the solar system within
$\pm10$~Myr using the astrometric data of the stars in the solar neighbourhood ($<50$ pc)
provided by {\hip} mission (\citealt{Perryman1997}). They find that the closest approach
($\sim0.3$ pc) in the future ($\sim1.3$ Myr from the present) will be with the star HIP 8982
(GJ~710), which will cause minor changes in the perihelion distance of the comets. The most recent
close stellar encounter was with the so-called Scholz's star ($M_\star \simeq 0.15$ M$_\odot$\ at
a distance of $0.25^{+0.11}_{-0.07}$ pc, \citealt{Scholz2014}, \citealt{Mamajek2015}). All of the
studies cited above were limited by the observational data due to the incompleteness of the {\hip}
survey. 

The first data release (\gdr{1}) of the European Space Agency's {\gaia} mission
(\citealt{Gaiacollaboration2017, GC_Prusti2016}) opened a new window for understanding the Milky Way. 
In the particular case of the solar system, {\gaia} detected nearly all of the local star systems within $50$~pc of the Sun
(compared to the 20\% detected by \hip). Using \gdr{1}, several authors \citep{Berski2016,
Bobylev2017,Torres2018,Bailer-Jones_1} re-computed the orbit of the closest stars to the Sun. They
found new stars and new parameters for some of the very well known encounters, such as GJ~710, which
gets closer ($0.064$ pc) based on the \gdr{1} data. The recent second {\gaia} data release -- \gdr{2}
(\citealt{Gaiacollaboration2018}) --, provided $7.2$ million radial velocities. This provided an
opportunity to find new and more accurately characterised stellar encounters. Using \gdr{2,}
\citet{Bailer-Jones2018} found 693 new stars with closest-encounter distances within $5$ pc and $15$
Myrs from now; accounting for the incompleteness they also re-calculate the present rate of
encounters, which within $\sim 1$ pc of the Sun is estimated to be $20\pm2$ Myr$^{-1}$. From {\hip}
data \citet{Garcia-Sanchez1999} derived $11.7\pm1.3$ Myr$^{-1}$ within $\sim 1$ pc and
\citet{Martinez-Barbosa2017} employed simulations to derive rates of 21, 39, and 63 Myr$^{-1}$
within $\sim 2$ pc for three different scenarios (orbital migration from the Milky Way inner disk,
migration from the outer disk, and no migration, respectively).

We aim to obtain a conservative estimate of the combined effects of stellar encounters and the
Galactic tidal field on the Oort cloud, by only considering the encounters from stars listed in
\gdr{2} within $\pm10$ Myr from the present.  The latter sample is incomplete and thus provides a
lower limit on the effects of passing stars.  In \secref{sec2} we present a simple analytical
model for stellar encounters and discuss the cumulative effect of passing stars on the Oort cloud,
using  the impulse approximation. In \secref{sec3} we present a catalogue of nearby stars and
we calculate the effect of individuals encounters with stars within $2.5$ pc of the Sun. In
\secref{sec4} we present a numerical model for multiple stellar encounters and study the
dynamical evolution of a simulated Oort cloud after the interaction with the nearby stars and the
Galactic tidal field. Finally, in \secref{sec5} we present our summary and conclusions. 


\section{Model for stellar encounters}
\label{sec2}

The estimated extent of the Oort cloud is $\sim$ 0.5pc (\citealt{J.H.Oort1950,Dones2015}), which
means that the orbital velocity of bodies in the Oort cloud is limited to $0.13$ {\kms}. 
This implies that comets at the edge of the Oort cloud are barely bound to the
Sun and thus the condition for a comet ejection due to an external perturbation, $\Delta
v_{\bot}>v_\mathrm{esc}$ (where $v_\mathrm{esc}$ is the escape velocity) is easily met. The Galactic
tidal field is the most important perturbation to the outer Oort cloud at large distances
(\citealt{Heisler1986a}). However close encounters with stars also play an important role in the
evolution of the Oort cloud.

 \subsection{Analytic model}
 \label{an_model}

A simple analytical model of stellar encounters can help us to better understand the effect of passing stars on the Oort cloud. To construct such a model, we followed the works of
\citet{Rickman1976,Garcia-Sanchez2001,Rickman2004, Rickman2008, Martinez-Barbosa2017}. We first
compiled data for the mass, velocity dispersion, and the space density of the stars in the solar
neighbourhood for 13 spectral types, as in Table 8 in \citet{Garcia-Sanchez2001}. The mass of the
stars corresponding to the spectral types B0V to M5V was taken from the data compiled by
\citet{Mamajeck18}\footnote{\url{http://www.pas.rochester.edu/~emamajek/EEM_dwarf_UBVIJHK_colors_Teff.txt}}
(see also \citealt{Pecaut2013}). While the mean value for white dwarfs (WD) was taken from
\citet{Jimenez-Esteban2018}.  The peculiar velocity of the Sun ($v_{\odot}$) and the velocity
dispersion of the stars ($v_{*}$) were taken from \citet{Rickman2008}. The space density of spectral
types A to K and Giants was obtained from \citet{Bovy2017}. For  B and M type stars the values
were obtained from \citet{Rickman2008}, and for the WD  from \citet{Jimenez-Esteban2018}. The
compiled data are shown in \tabref{catalogue_st}.

\begin{table*}
\centering
 \caption{ Stellar parameters. Columns represent the spectral type of the stars followed by
 their mass, velocity dispersion, and the peculiar velocity of the Sun with respect to each spectral type.
 The relative velocity of the encounter within the Sun--comet system and the star is shown in column
 5. The number density of stars in the solar neighbourhood is shown in column 6.} 
 \label{catalogue_st}
%
\begin{tabular}{cccccc}
 \hline
 \hline
S.T  & M$_{*}$[M$_{\odot}$]  & $v_{*}$[\kms] & $v_{\odot}$[\kms] & $v_{enc}$[\kms] & $\rho_{*}[$10$^{-3}$ pc$^{-3}$] \\
  \hline
 B0V & 15 & 14.7 & 18.6 & 24.6 & 0.06 \\
A0V & 2.3 & 19.7 & 17.1 & 27.5 & 0.26  \\
A5V & 1.85 & 23.7 & 13.7 & 29.3 & 0.34  \\
F0V & 1.59 & 29.1 & 17.1 & 36.5 & 0.61   \\
F5V & 1.33 & 36.2 & 17.1 & 43.6 & 1.51   \\
G0V & 1.08 & 37.4 & 26.4 & 49.8 &  1.61  \\
G5V & 0.98 & 39.2 & 23.9 & 49.6 &1.73  \\
K0V & 0.87 & 34.1 & 19.8 & 42.6 & 4.21  \\
K5V & 0.68 & 43.4 & 25.0 & 54.3 & 5.26  \\
M0V & 0.55 & 42.7 & 17.3 & 50.0 & 8.72  \\
M5V & 0.16 & 41.8 & 23.3 & 51.8 & 41.55  \\
WD & 0.6 & 63.4 & 38.3 & 80.2 & 4.9  \\
Giants & 2.2 & 41.0 & 21.0 & 49.7 & 3.9   \\
 \hline
  \hline
 \end{tabular}
\end{table*}

 We consider the effect of the stars with different masses ( $M_{*}$) and
spectral types in the solar neighbourhood on the comets in the Oort cloud.  We assume that the stars
move on a straight line trajectory, and with a constant velocity relative to the Sun ($v_{*}$). For
high stellar velocities, we can assume  that the comet is at rest during the stellar passage. Using
the impulse approximation \citep{J.H.Oort1950, Rickman1976}, we then calculate the
change of the velocity  ($\Delta V_{\bot}$) imparted to a comet in the Oort cloud due to a random
stellar encounter by integrating the perpendicular force generated by each passing star:

\begin{equation}
  \Delta V_{\bot} \approx \frac{2GM_{*}}{v_{*}} \left[ \frac{\bf r_{c}}{r_{c}^{2}} - \frac{\bf r_{\odot}}{r_{\odot}^{2}} \right] \,,
  \label{impulse_aprox-comet}
\end{equation}

\noindent  where  $r_{c}$ and $r_{\odot}$ correspond to the vectors from the comet and the Sun to
the point of closest approach of the star (assuming that the comet has not been deflected by the
gravity of the star). If we consider $r$ the heliocentric distance of the comet and we assume that the
distance of the encounter is large enough compared to the distance Sun--comet, we can
approximate \equrefalt{impulse_aprox-comet} with:   

\begin{equation}
\Delta V_{\bot}  \propto \frac{M_{*}r}{v_\mathrm{*}  {r_{\odot}^{2}}} \,.
  \label{sun-comet_impulse}
\end{equation}

 For the case of a very close encounter with the comet, \equrefalt{sun-comet_impulse} can be approximated as 
\begin{equation}
\Delta V_{\bot}  \propto \frac{M_{*}}{v_\mathrm{*}  {r_{c}}} \,.
  \label{star-comet_impulse}
\end{equation}

 It is important to stress that the impulse approximation is based on a number of simplifying
assumptions, and therefore it should be used for statistical analysis only. For our propose, it gives
us a general idea of the effect of the different stars in the solar neighbourhood on the comets in
the Oort cloud. 

Following  \citet{Rickman1976} we can calculate the frequency of the stellar encounters using 

\begin{equation}
  f= \pi r_{*}^{2}v_\mathrm{enc}\rho_{\ast}\,,
\label{frec}
\end{equation}

where $r_{*}$ is the distance of the encounter, $v_\mathrm{enc}$=$\sqrt{v^{2}_{\odot} +
v^{2}_{\ast}}$ is the relative velocity of the Sun and a random passing star ($v_{\odot}$ represents
the peculiar velocity of the Sun, and $v_{*}$ the velocity dispersion of the parent population of
the passing star), and $\rho_{\ast}$ is the number density of stars of a given spectral type in the
solar neighbourhood. \equrefalt{frec} can be used to determine the number of stars passing by
within a sphere of radius $r_\mathrm{s}$ centred on the Sun or a random comet 
\citep[assuming that stars of the solar neighbourhood are uniformly distributed at any time and the 
stellar velocities relative to the Sun are constant,][]{Rickman1976}:

\begin{equation}
N_{\ast}= r_\mathrm{s}^{2} \textit{f} t \,.
\label{numb}
\end{equation}

 \subsection{Perturbations on the Oort cloud}
 \label{pert_pc_an}

Using \equref{sun-comet_impulse} and the values in \tabref{catalogue_st}, we calculated the frequency of the stars passing within a distance $r_{*}$ from the
Oort cloud. We find that the total frequency of stars passing within 1~pc is around $12.5$ Myr$^{-1}$ (see
\tabref{stellar_model}). Following the same method, \citet{Garcia-Sanchez1999} found a lower value
($11.7$ Myr$^{-1}$). The main difference with our result is due to the updated values for the mass and
density of the stars used in this work. The most probable perturber of the Oort cloud is the low
mass, high-relative-velocity stars.

\begin{figure}[h]       
        \includegraphics[width=\columnwidth]{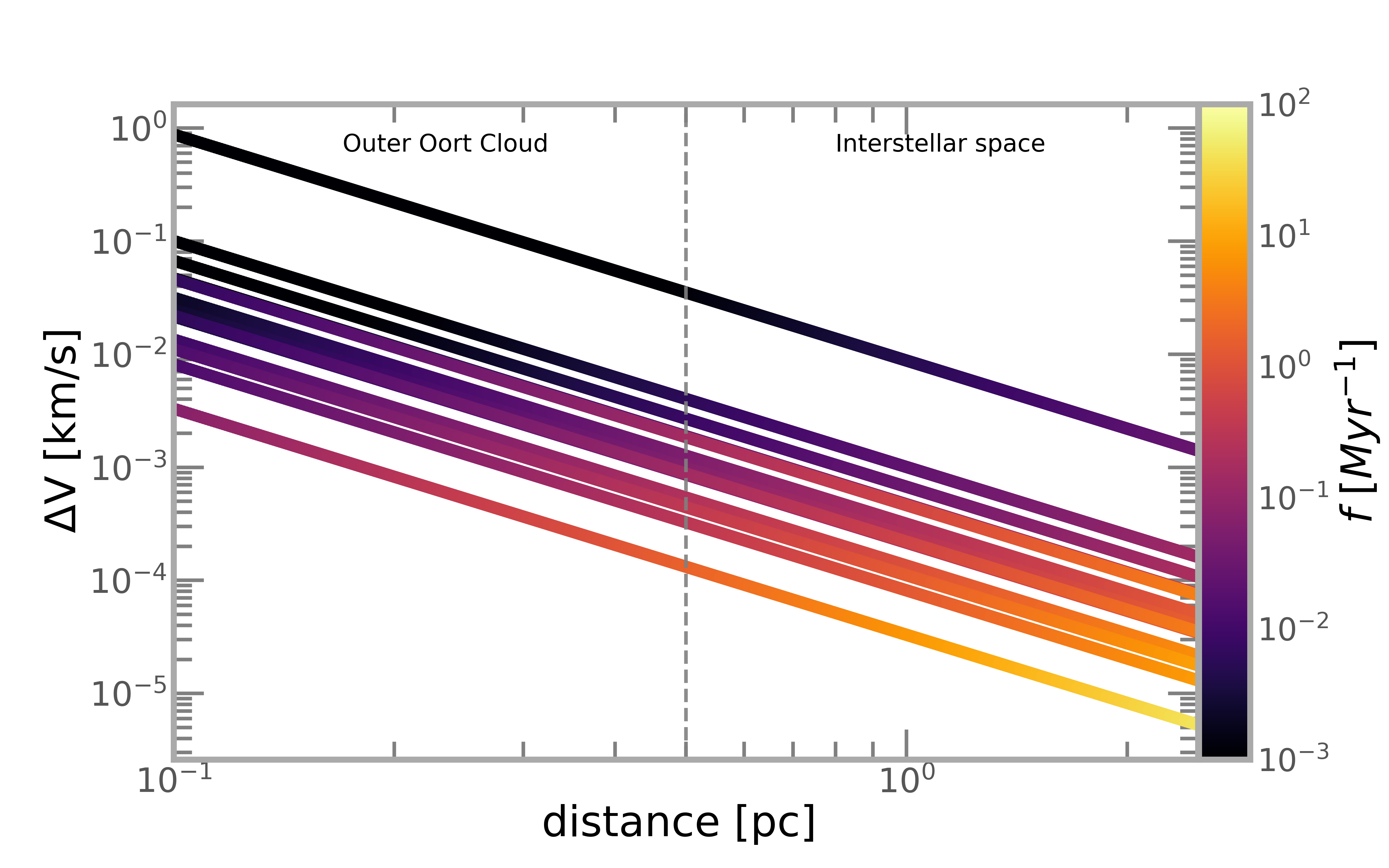}
    \caption{Change of the velocity of a comet due to random stellar encounters for the 13 stellar spectral
    types listed in \tabref{catalogue_st} as a function of the encounter
  distance. The colour coding of the lines represents the frequency of stellar encounters as a function of the
distance of the encounter for the corresponding types of stars. The lines represent the
different spectral types in the order listed in \tabref{catalogue_st}. The mass of the stars
decreases from the top to the bottom lines.}
\label{dv_s}
\end{figure}

Using \equref{sun-comet_impulse},  in \figref{dv_s} we show the change of the velocity of a comet due
to an encounter with a star for different spectral types and as a function of the distance of the
encounter for an interval of $0.1$--$2.5$ pc. The lower distance corresponds to the inner Oort
cloud, while the larger distance corresponds to the limit where a passing star can start perturbing
a comet at the edge of the cloud. In the rest of this work we refer to the latter distance as the
\textit{critical radius}.

\begin{table}
\centering
 \caption{ Analytical model for stellar encounters for an encounter distance of 1~pc. The spectral type is shown in column 1, the frequency of the stellar encounters  is shown in column 2,
  the change in the velocity of a comet  due to an interaction with a star is shown in column 3, and the total number of stars entering a sphere of radius 1~pc around the Sun--comet system over a time 
  interval of 1\ $Gyr$ is shown in column 4.} 
\label{stellar_model}
\begin{tabular}{cccc}
\hline
\hline
S.T  & $f^{1\mathrm{pc}}_{*}$ [Myr$^{-1}$] & $\Delta V_{\bot,*}^{1\mathrm{pc}}$ [\kms] &
$N^{1\mathrm{pc}}_{*}$ \\
 \hline
 B0V &  0.005 &   8.77e-03 & 4.742 \\
A0V & 0.023 &    1.005e-03 & 22.973 \\
A5V &  0.032 & 6.716e-04 & 32.007 \\
F0V &  0.072 & 4.701e-04 & 71.536 \\
F5V &  0.212 & 3.161e-04 & 211.527 \\
G0V & 0.258 & 2.485e-04 & 257.608 \\
G5V &  0.276 & 2.151e-04 &275.696 \\
K0V &  0.576 & 2.195e-04 & 576.229 \\
K5V &  0.918 & 1.348e-04 & 917.675 \\
M0V &  1.401 & 1.108e-04 & 1400.844 \\
M5V &  6.915 & 3.293e-05  & 6915.189 \\
WD & 1.263 & 8.143e-05 & 1262.623 \\
Giants & 0.623 & 4.617e-04 & 622.765 \\
 \hline
\hline
\end{tabular}
\end{table}

As shown in \figref{dv_s} the change induced by a single encounter is relatively small.  Massive
stars are effective in exciting the object in the Oort cloud, but they are rare. Low-mass stars are
very common, but their effect on the orbits of a comet is small. However the number of stars
encountering the solar system increases over time.

 The model presented in this section is based on a number of simplifying assumptions.
  Specifically, the impulse approximation provides a quick but inaccurate estimate of the effect of a
  random passing star on a comet in the
  Oort cloud. As we show in \tabref{stellar_model} the effect of individual stars is relative
  small. However, considering the frequency and the number of stars approaching the Sun to within
  1~pc over 1~Gyr,
  their cumulative effect might change the structure and dynamics of the Oort cloud. In order to
  have a better understanding of the evolution of the Oort cloud it is necessary to employ a
  detailed numerical model which accounts for the effects of the Galactic tidal field and stellar
  distribution of stars around the Sun. In \secref{sec3} we present a list of nearby stars within
  50 pc of the Sun for which \gdr{2} astrometry and radial velocities (including from other surveys)
  are available. This provides us with accurate kinematic information on nearby stars that could
influence the Oort cloud in the recent past or near future. In \secref{sec4} we use numerical
simulations to analyse their
effect on the comets in the Oort cloud and estimate their cumulative effect over $\pm10$ Myr,
including the effect of the Galactic tidal field.


\section{Close encounters with the solar system}
\label{sec3}

Our knowledge of close stellar encounters in the recent past or near
future has been limited by the availability of precise and accurate astrometry and radial velocities
for the nearby stars.  The Gaia mission has  considerably increased the availability of astrometric
and radial velocity data for the closest stars, even if about 20$\%$ of the stars with high proper 
motions are not listed in \gdr{2} and those tend to be close to the Sun (\citealt{Bailer-Jones2018}).

\begin{table}[ht]
\centering
 \caption{Overview of the catalogue of stars within 50 pc of the Sun for which encounter
 parameters were calculated. The full catalogue can be download from: \url{https://home.strw.leidenuniv.nl/~storres/\#Research}}
 \label{catalogue}
 \begin{tabular}{lc}
 \hline
 \hline
 Input catalogue(s) & \multicolumn{1}{c}{No.~Stars} \\
 & $\leq50$ pc  \\
 \hline
\gdr{2} & 10,744   \\
\gdr{2} + RAVE DR5 & $2356$   \\
\gdr{2} + GALAH DR2 & $11$ \\
\gdr{2} + LAMOST DR3 & $307$  \\
\gdr{2} + APOGEE DR14 & $1092$ \\
\gdr{2} + XHIP &  $149$ \\
\hline
Total & $14\,659$  \\
 \hline
 \hline
 \end{tabular}
\end{table}

\subsection{Observational model}
\label{obs_model}

To construct the list of stars within 50 pc of the Sun for which the encounter parameters (closest
approach distance, velocity, and time) can be calculated we used the data from the \gdr{2}
catalogue. To increase the number of stars for which radial velocity information is available we
cross-matched \gdr{2} with the following catalogues: {RAVE-DR5} (\citealt{Kunder2016}), {GALAH DR2}
(\citealt{Buder2018}), {LAMOST DR3} (\citealt{Zhao2012}), {APOGEE DR14} (\citealt{Abolfathi2017}),
and {XHIP} (\citealt{Anderson2012}). We selected only stars with relative uncertainty on the
parallax ($\varpi$) smaller than 20\%, such that $1/\varpi$ is a good estimator of the distance to the
stars. Following \citet{Lindegren2018a} we further filtered the list of stars according to

\begin{equation}
  u^{2}<1.44 \times \text{max}[1,\exp(-0.4(G-19.5))] \label{c1}
,\end{equation}
and
\begin{equation}
  1.0 +0.015(G_\mathrm{BP}-G_\mathrm{RP})^{2}< E < 1.3 + 0.06(G_\mathrm{BP}-G_\mathrm{RP})^{2}\,
  \label{c2}
,\end{equation}

where $G$, $G_\mathrm{BP}$, and $G_\mathrm{RP}$ correspond to the photometric measurements,
covering a wavelength from the near-ultraviolet to the near-infrared for the $G$ passband, 330 to
680nm, and 630 to 1050~nm for $G_\mathrm{BP}$, and $G_\mathrm{RP}$, respectively.  The
$u=(\chi^{2}/\nu)^{1/2}$ corresponds to the unit weight error,  and $E$ is the flux excess factor.
This filter selects sources with high-quality astrometry and weeds out stars which appear to be
nearby because of spuriously high values of the parallax \citep[see appendix C in
][]{Lindegren2018a}. The resulting catalogue contains $14\,659$ stars within 50~pc of the Sun
(\tabref{catalogue}).

For the selected stars we estimated the distance, time, and velocity of closest approach using the
linear approximation method of \citet{Matthews1994} in the formulation presented in \cite{Bailer-Jones2015}:
\begin{align}
v_\mathrm{tot} & = \sqrt{v_\mathrm{T}^{2}+v_\mathrm{rad}^{2}}
\label{vtotal}\\
t_\mathrm{ph} & = - \frac{c v_\mathrm{rad}}{\varpi v_\mathrm{tot}^{2}}
\label{time} \\
d_\mathrm{ph} & = \frac{10^{3}}{\varpi}\frac{v_\mathrm{T}}{v_\mathrm{tot}}\,,
\label{dist}
\end{align}
where $v_\mathrm{T}=4.74 \left[({\mu_{\alpha*}}^{2} + {{\mu_{\delta}}}^{2})^{0.5} / \varpi \right]$
is the transverse velocity, $v_\mathrm{rad}$ is the radial velocity of the star, $\varpi$ is the
parallax, $c=10^{3}$~pc~km$^{-1}$~yr$^{-1}$, and the subscript `ph' stands for perihelion. We
estimated the mass of the stars using the effective temperature provided in \gdr{2}
\citep{Andrae2018} and linearly interpolating in the Tables in \citet{Mamajeck18} and
\citet{Pecaut2013}.

\begin{figure}[ht]      
\includegraphics[width=\columnwidth]{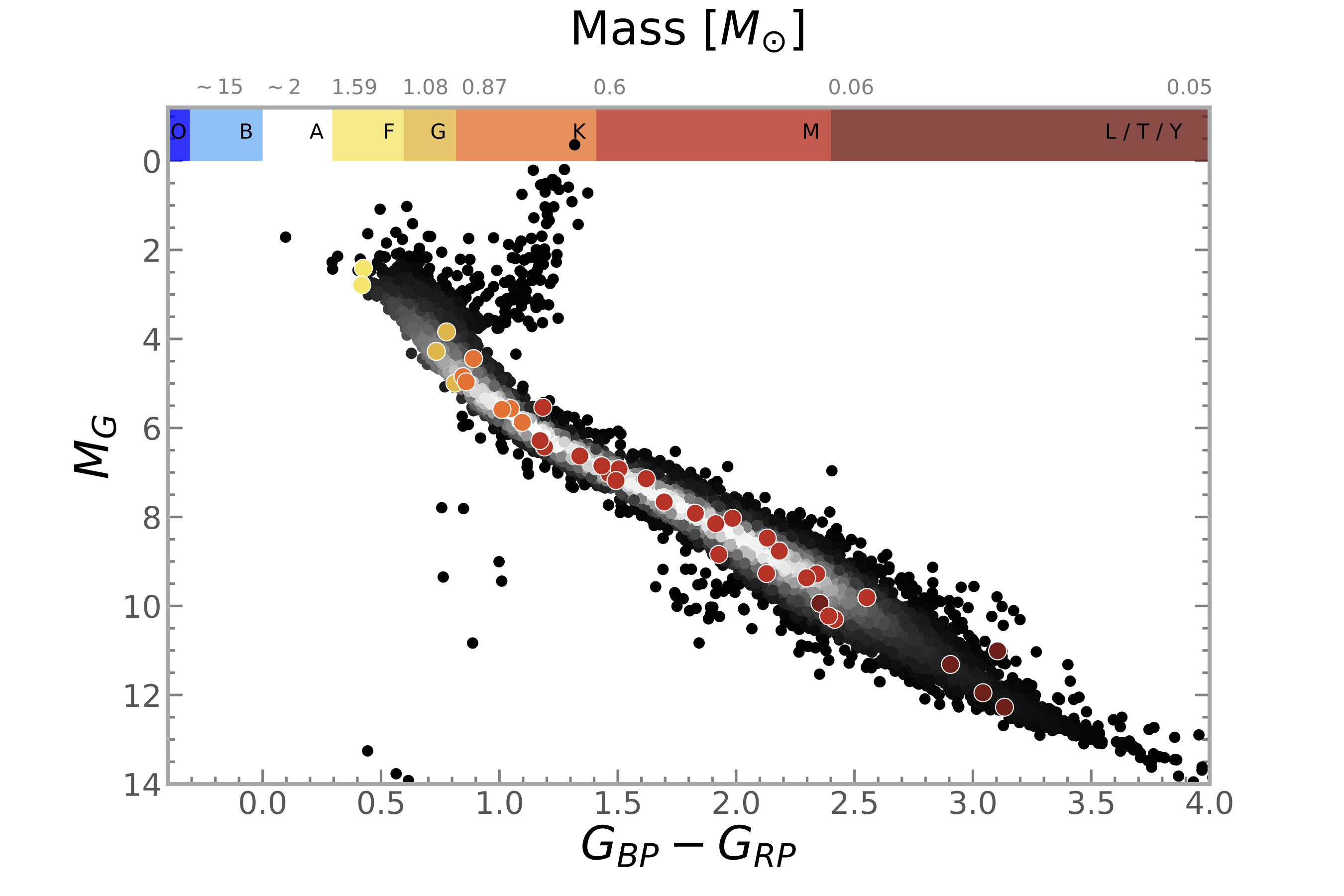}
\caption{Observational HR diagram of the nearby stars contained in the catalogue summarised in
\tabref{catalogue}. The big dots represent the stars within $2.5$~pc of the Sun colour coded
according to their spectral type. The density map shows all the stars in our sample.}
\label{hr}
 \end{figure}

\begin{figure*}[ht]
\centering
\includegraphics[width=0.8\textwidth]{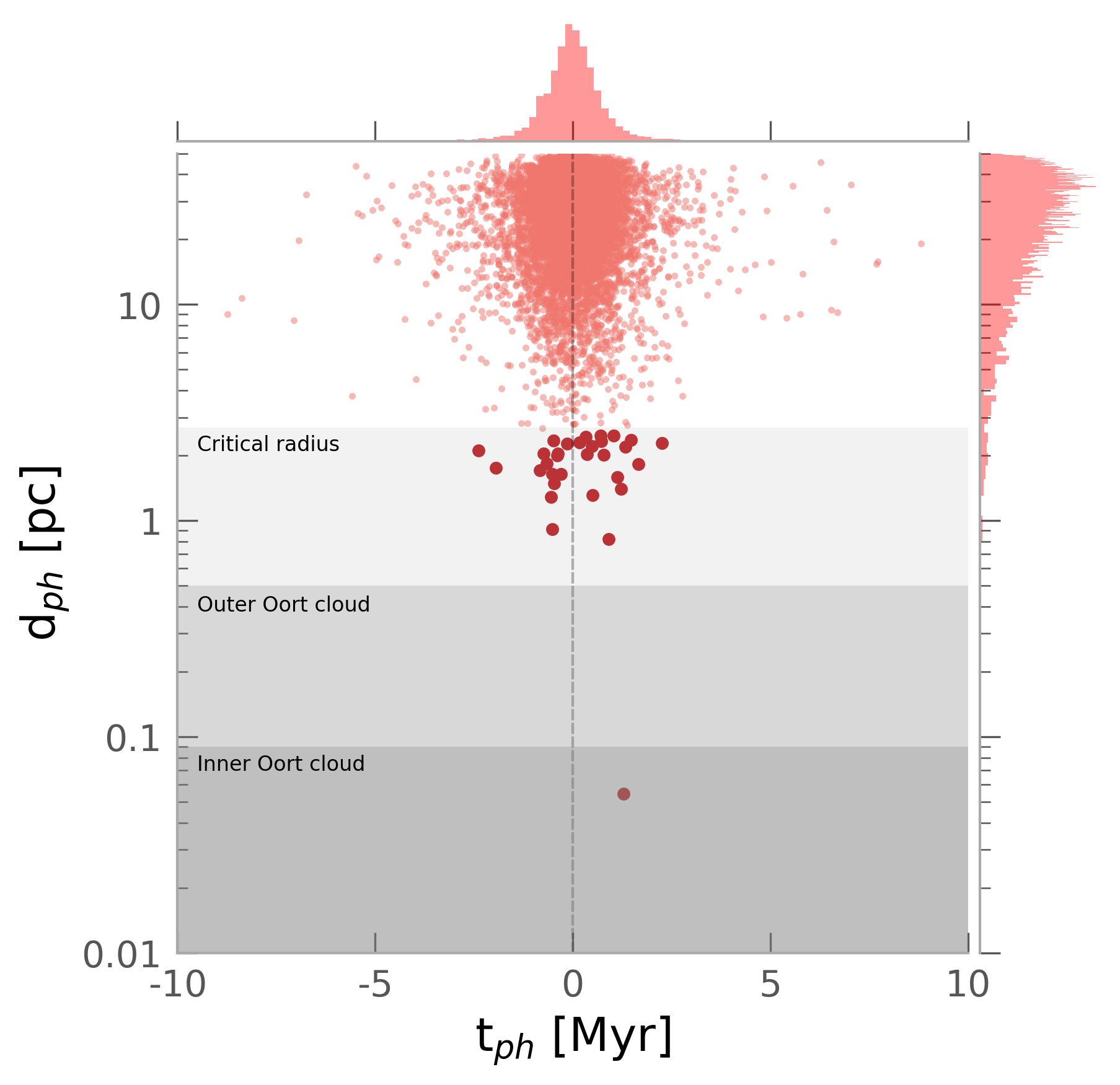}
\caption{Closest approach distance vs.\, closest approach time. The shaded areas represent the
critical radius within which stars can affect an object in the Oort cloud, and the boundaries of the
outer and inner Oort cloud, respectively. The big dots correspond to those stars that pass within $2.5$~pc of the Sun.}
\label{stars50pc}
\end{figure*}

Of the $14\,659$ stars within $50$~pc there are 31 that pass within $2.5$~pc of the Sun (\figref{hr},
big dots) over a
period of $20$~Myr centred on the present (i.e.\ $10$~Myr in the past and $10$~Myr in the future).
\figrefalt{hr} shows the observational Hertzsprung-Russell diagram of our sample.  In    
 \figref{stars50pc} we show the closest approach distance and time of the stars in our sample with respect to 
the Sun. The limited distance range of the stars under study only allows us to find very close encounters
 within $\pm3$~Myr.

In \figref{stars50pc} the large dots show the distribution of the stars passing within $2.5$~pc and
those tend to be the major perturber of the Oort cloud (referred to as `\gaia\ stars' below). The closest encounter with the solar system
is GJ~710 which will penetrate deep inside the inner Oort cloud.
As shown in \figref{hr} most of the closest encounters involve M dwarfs, with a considerable
fraction of solar type stars. This implies that the effect of a single encounter with the Oort cloud
will be minimal, mainly due to the low mass of the perturber and its high velocity with respect to the Sun.

\citet{Bailer-Jones2018} find 62 new stellar encounters, which partially overlap with our list.
Their list of encounters is larger than ours mostly because they did not apply the strict filtering
on the astrometric quality of the \gdr{2} data employed in this work. For stars appearing in both
studies we find similar results.

 We stress that the sample of the closest stars presented here is incomplete. The
  observational incompleteness is evident in the decrease in encounter frequency as one moves away
  from the present epoch in time. A complete census of stellar encounters requires all the stars
  within a certain distance to be identified. The main limitations in using the \gaia\ survey for
  finding the closest encounters are the survey magnitude limit, which prevents the identification of
  encounters with faint low-mass stars, and the lack of radial velocities.  The \gdr{2} radial
  velocity survey is limited to effective temperatures in the range $\sim3550$--$6900$~K and to
stars brighter than $G=14$ mag (see e.g. \citealt{Bailer-Jones2018}). An additional limitation is
that some of the brightest stars in the sky are missing from the \gdr{2} catalogue \citep{Gaiacollaboration2018}

 A detailed study correcting for incompleteness in \gdr{2} was carried out by
  \citet{Bailer-Jones2018}. These latter authors constructed a completeness map (Fig. 12,
  \citealt{Bailer-Jones2018}), interpreted as the probability of detecting a given close encounter
  in the \gdr{2} sample. They found that only $15$\% of the encounters within $5$~pc in a period of
  $5$~Myr have been identified.  Using this result, the authors used a simulated Milky Way galaxy to infer
  the encounter rate averaged over $5$~Myr, in the past and future.  They found that the encounter
  rate of stellar encounters within  $1$~pc is $20\pm2$~Myr$^{-1}$. 

\subsection{Stellar encounters with the Solar System}
\label{pert_oc}

In Section \ref{obs_model} we employed a simple method to estimate the perihelion distances and times for
stars approaching the Sun by assuming the stars follow a uniform motion along straight lines with
respect to the Sun \citep[see also][]{Bailer-Jones2015}. We now seek a better estimation of the
perihelion distance through the joint integration of the orbits of the Sun and the stars that are
predicted to approach to within $2.5$~pc (\tabref{se_2pc}) backwards and forwards in time for
$10$~Myr. We first transformed the astrometric and radial velocity data into galactocentric
Cartesian frame using \texttt{Astropy} \citep{TheAstropyCollaboration2018}. We adopted the
position of the Sun and the local circular velocity parameters from \cite{Reid2014}: $Z_{\odot}=27$~pc,
$R_{\odot}=8.34$~kpc, and $V_{c,\odot}=240$ \kms; while the peculiar velocity of the Sun was adopted from
\cite{Schonrich2010}: $(U_{\odot}, V_{\odot}, W_{\odot}) = (11.1,12.24,7.25)$ \kms. We used the
\texttt{Gala} (\citealt{Gala2017}) package to perform the orbital integration. The Milky Way
potential used is described by an analytic axisymmetric model which contains a spherical nucleus and
bulge (\citealt{Hernquist1990}), a Miyamoto-Nagai disk (\citealt{Miyamoto-Nagai1975,Bovy2015}), and
a spherical Navarro-Frenk-White (NFW)  dark matter halo (\citealt{NFW1995}).

\begin{figure}[ht]
  \includegraphics[width=\columnwidth]{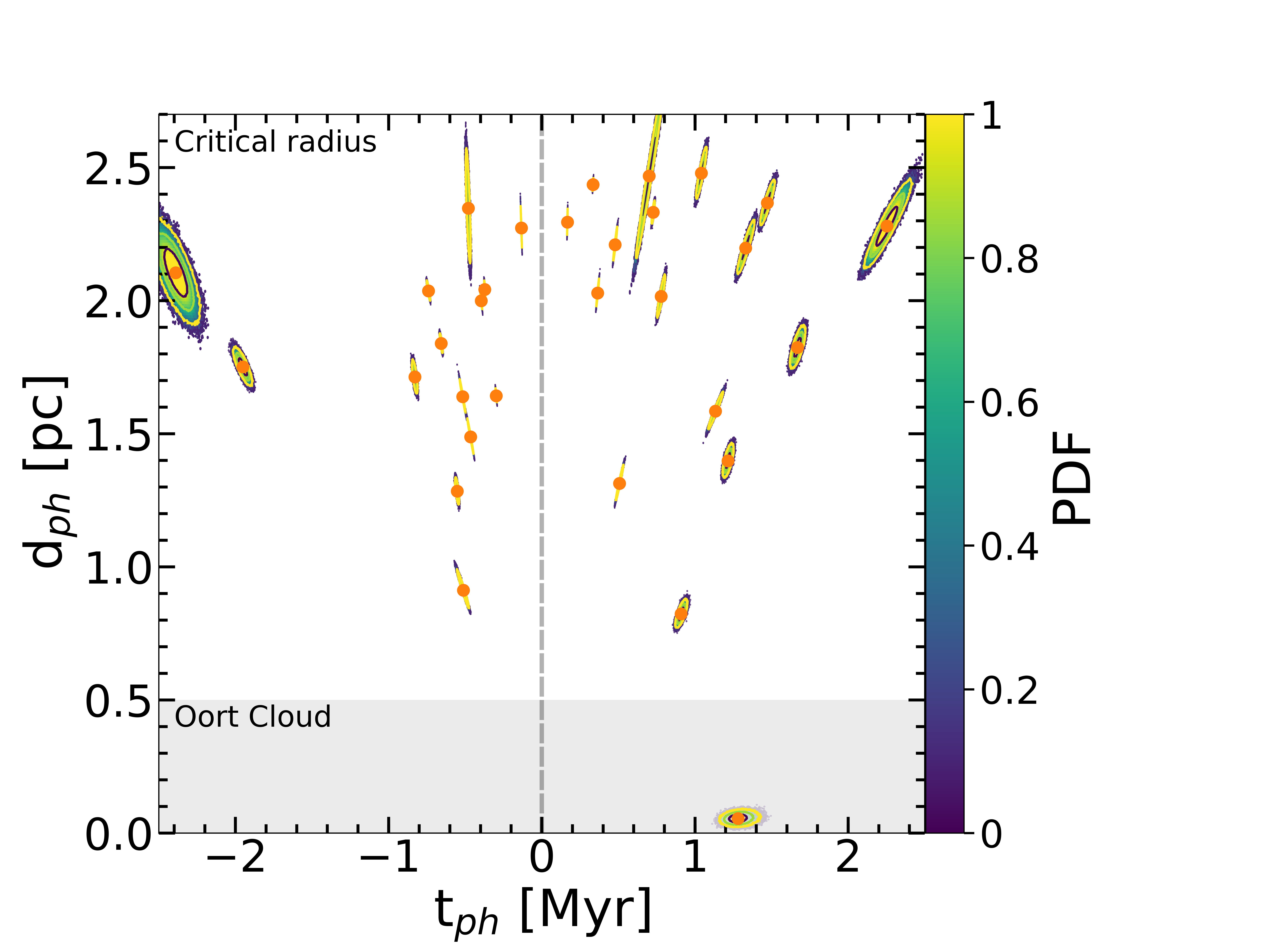}
  \caption{Joint probability density of the time and distance of closest approach for those stars
    that are predicted to pass within $2.5$~pc of the Sun (listed in \tabref{catalogue}). The
    contour levels indicate regions enclosing 0.6, 0.9, and 0.99 \% cumulative probability (colour bar). 
    The shape of each PDF is affected by the relative errors in the observational data of each star, particularly the
  errors on parallax and radial velocity.} 
  \label{td_2pc}
\end{figure}

\begin{figure}[ht]
  \includegraphics[width=\columnwidth]{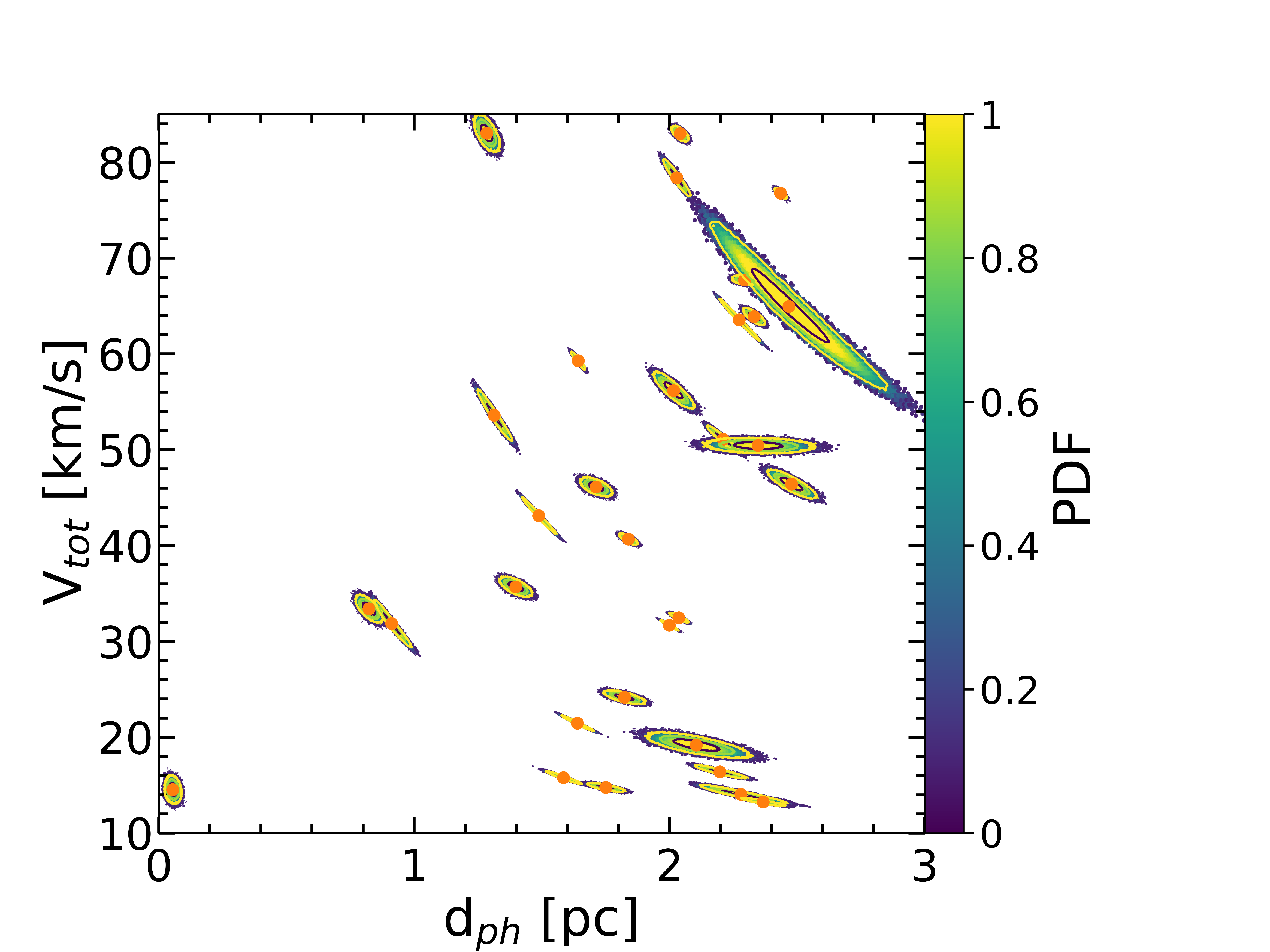}
  \caption{Joint probability density of the time and relative velocity of closest approach for the
  stars in \tabref{catalogue}.} 
  \label{dv_2pc}
\end{figure}

\begin{figure}[ht]
  \includegraphics[width=\columnwidth]{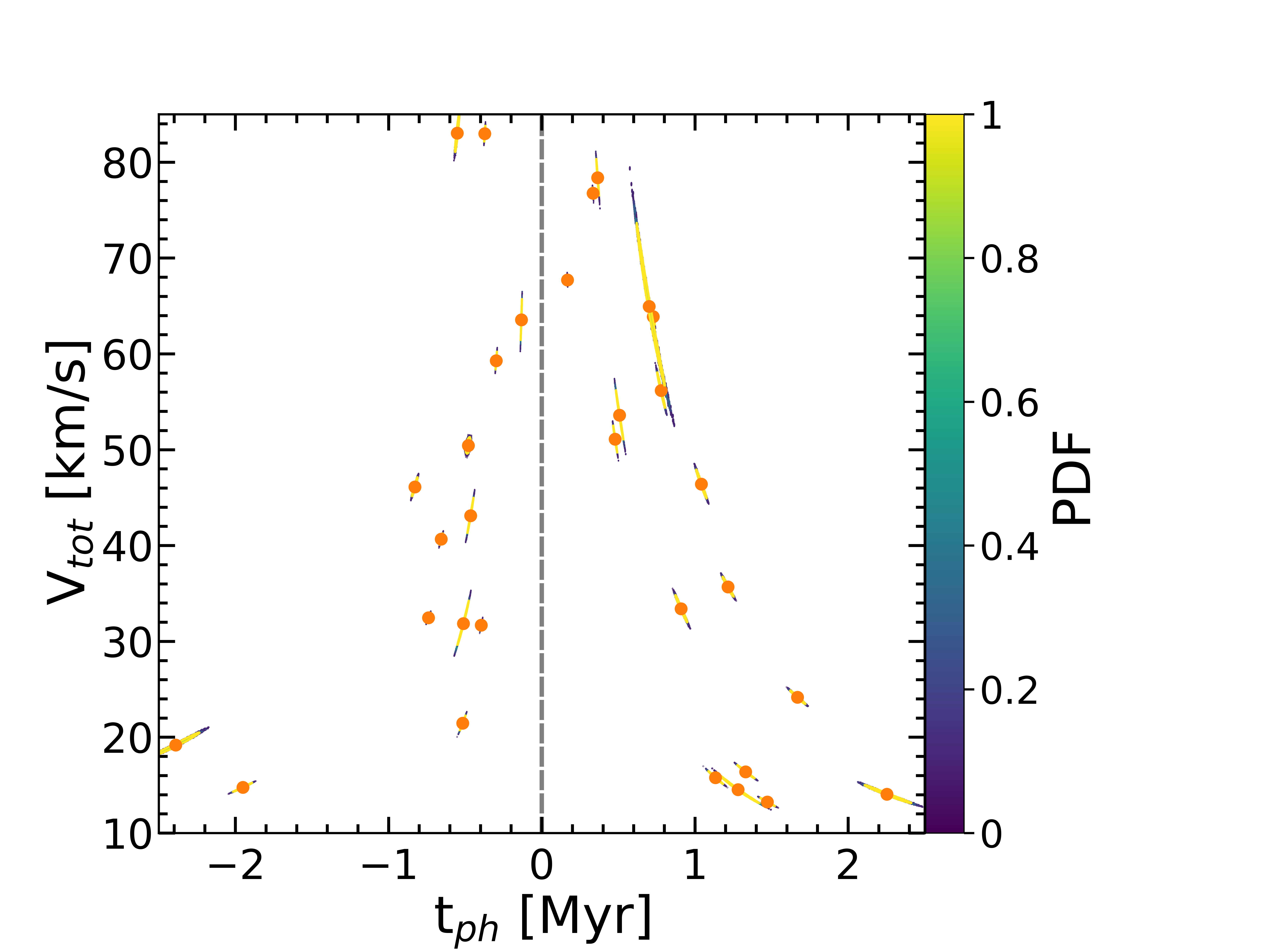}
  \caption{Joint probability density of the distance and relative velocity of closest approach for
  the stars in \tabref{catalogue}.} 
  \label{tv_2pc}
\end{figure}

 To account for the observational uncertainties we sample the astrometric and
radial velocity observables for each star, taking the full covariance matrix into account. For each star, $10^{6}$
samples of the astrometry and radial velocity are drawn and for each of these the above described
orbit integration is carried out. The end result is a sampling of the distribution of possible
perihelion distances and times. This distribution obtained through Monte Carlo sampling is then
treated as the probability density function (PDF) of the encounter parameters. The shape of the confidence 
regions is mainly affected by the relative errors on parallax and radial velocity. The relative
error in the proper motion likewise affects the shape of the confidence regions around the mean.
Figures \ref{td_2pc}, \ref{dv_2pc}, and \ref{tv_2pc} show the resulting PDFs. 

\begin{table*}
\caption{Stars predicted to approach the Sun within $2.5$~pc over the $\pm$10~Myr from today. The
  columns represent the \gdr{2} ID, the time, distance and velocity at the perihelion with its
  respective confidence interval. These are followed by the effective temperature listed in \gdr{2}, the estimated mass of
  the star, and the change in the velocity due to the encounter with the Sun for each star.}
\label{se_2pc}
{\tiny
\hskip-0.8cm\begin{tabular}{crcccccccc}
\hline
\hline
\gdr{2} ID & $t_\mathrm{ph}$ [Myr] & $t_\mathrm{sample}$ & $d_\mathrm{ph}$[pc] &
  $d_\mathrm{sample}$ & $v_\mathrm{tot}$ [\kms] & $v_\mathrm{sample}$ & $T_\mathrm{eff}$ [K] & Mass
  [M$_{\odot}$] & $\Delta v_{\bot}$ [\kms] \\
\hline
4270814637616488064 &  $1.282$ & [ $1.123$ , $1.488$ ] &    0.054 & [ 0.006 , 0.107 ] &   14.525 & [ 12.515 , 16.564 ] &       4116 &    0.654 & 1.305e-01 \\
553219967007245312  &  $1.670$ & [ $1.595$ , $1.742$ ] &    1.824 & [ 1.714 , 1.952 ] &   24.163 & [ 23.187 , 25.284 ] &       5175 &    0.851 & 9.106e-05 \\
258179971749627776  &  $0.365$ & [ $0.352$ , $0.378$ ] &    2.028 & [ 1.955 , 2.107 ] &   78.377 & [ 75.746 , 81.470 ] &       4507 &    0.717 & 1.915e-05 \\
4575928186606190336 &  $0.479$ & [ $0.462$ , $0.500$ ] &    2.210 & [ 2.126 , 2.307 ] &   51.102 & [ 48.983 , 52.973 ] &       3795 &    0.559 & 1.925e-05 \\
3240424426786618624 & $-0.552$ & [ $-0.572$ , $-0.534$ ] &    1.284 & [ 1.214 , 1.368 ] &   83.029 & [ 80.325 , 85.666 ] &       3836 &    0.580 & 3.647e-05 \\
4795598309045006208 & $-0.739$ & [ $-0.757$ , $-0.723$ ] &    2.037 & [ 1.985 , 2.091 ] &   32.470 & [ 31.698 , 33.194 ] &       5343 &    0.901 & 5.755e-05 \\
3274130814728561792 & $-2.390$ & [ $-2.645$ , $-2.158$ ] &    2.105 & [ 1.834 , 2.447 ] &   19.171 & [ 17.362 , 21.165 ] &       4471 &    0.715 & 7.240e-05 \\
981375326780564608  &  $0.508$ & [ $0.476$ , $0.548$ ] &    1.314 & [ 1.229 , 1.422 ] &   53.595 & [ 49.613 , 57.211 ] &       3875 &    0.601 & 5.589e-05 \\
6684504722300935680 & $-0.465$ & [ $-0.495$ , $-0.436$ ] &    1.488 & [ 1.400 , 1.587 ] &   43.104 & [ 40.449 , 45.973 ] &       3619 &    0.494 & 4.451e-05 \\
4430238051199001216 &  $0.167$ & [ $0.165$ , $0.171$ ] &    2.295 & [ 2.225 , 2.379 ] &   67.699 & [ 66.901 , 68.396 ] &       6017 &    1.128 & 2.724e-05 \\
2417069815934357248 &  $2.253$ & [ $2.049$ , $2.486$ ] &    2.280 & [ 2.068 , 2.531 ] &   14.047 & [ 12.746 , 15.408 ] &       4613 &    0.741 & 8.735e-05 \\
3089711447388931584 & $-0.133$ & [ $-0.139$ , $-0.127$ ] &    2.273 & [ 2.164 , 2.389 ] &   63.544 & [ 60.457 , 66.818 ] &       3820 &    0.571 & 1.497e-05 \\
3339921875389105152 & $-0.516$ & [ $-0.544$ , $-0.485$ ] &    1.639 & [ 1.541 , 1.732 ] &   21.455 & [ 20.317 , 22.843 ] &       4105 &    0.653 & 9.746e-05 \\
1134618591670426112 &  $0.728$ & [ $0.713$ , $0.742$ ] &    2.331 & [ 2.273 , 2.394 ] &   63.885 & [ 62.639 , 65.223 ] &       4887 &    0.787 & 1.950e-05 \\
5861048509766415616 & $-0.297$ & [ $-0.304$ , $-0.290$ ] &    1.643 & [ 1.601 , 1.682 ] &   59.275 & [ 57.962 , 60.702 ] &       3795 &    0.559 & 3.005e-05 \\
681999884156922368  &  $1.134$ & [ $1.066$ , $1.209$ ] &    1.585 & [ 1.483 , 1.697 ] &   15.785 & [ 14.795 , 16.794 ] &       3956 &    0.615 & 1.336e-04 \\
3260079227925564160 &  $0.910$ & [ $0.851$ , $0.968$ ] &    0.823 & [ 0.754 , 0.905 ] &   33.395 & [ 31.465 , 35.704 ] &       3998 &    0.629 & 2.395e-04 \\
2648914040357320576 &  $1.473$ & [ $1.412$ , $1.544$ ] &    2.367 & [ 2.257 , 2.502 ] &   13.235 & [ 12.616 , 13.823 ] &       5630 &    0.976 & 1.133e-04 \\
3972130276695660288 & $-0.511$ & [ $-0.572$ , $-0.458$ ] &    0.912 & [ 0.818 , 1.025 ] &   31.845 & [ 28.459 , 35.551 ] &       3980 &    0.623 & 2.025e-04 \\
2118161219075485824 &  $0.779$ & [ $0.743$ , $0.816$ ] &    2.016 & [ 1.908 , 2.126 ] &   56.180 & [ 53.650 , 58.906 ] &       4122 &    0.654 & 2.466e-05 \\
1392610405193517952 &  $0.702$ & [ $0.586$ , $0.882$ ] &    2.468 & [ 2.067 , 3.095 ] &   64.956 & [ 51.343 , 77.613 ] &       5057 &    0.823 & 1.790e-05 \\
2089889682751105536 &  $1.041$ & [ $0.997$ , $1.090$ ] &    2.478 & [ 2.346 , 2.633 ] &   46.407 & [ 44.351 , 48.454 ] &       3965 &    0.618 & 1.866e-05 \\
2272191085754928768 &  $0.335$ & [ $0.331$ , $0.338$ ] &    2.436 & [ 2.401 , 2.469 ] &   76.760 & [ 75.892 , 77.614 ] &       5859 &    1.060 & 2.003e-05 \\
4758877919212831104 & $-0.395$ & [ $-0.405$ , $-0.385$ ] &    2.000 & [ 1.952 , 2.050 ] &   31.680 & [ 30.850 , 32.497 ] &       4893 &    0.788 & 5.352e-05 \\
4839132097557586560 & $-0.828$ & [ $-0.856$ , $-0.803$ ] &    1.713 & [ 1.624 , 1.815 ] &   46.116 & [ 44.687 , 47.522 ] &       3865 &    0.597 & 3.798e-05 \\
5076269164798852864 & $-0.479$ & [ $-0.503$ , $-0.456$ ] &    2.347 & [ 2.063 , 2.673 ] &   50.434 & [ 49.277 , 51.577 ] &       4837 &    0.781 & 2.418e-05 \\
4546557031272743680 &  $1.216$ & [ $1.163$ , $1.272$ ] &    1.398 & [ 1.311 , 1.490 ] &   35.688 & [ 34.137 , 37.245 ] &       4305 &    0.695 & 8.573e-05 \\
875071278432954240  &  $1.331$ & [ $1.250$ , $1.413$ ] &    2.198 & [ 2.057 , 2.351 ] &   16.389 & [ 15.429 , 17.455 ] &       5714 &    0.998 & 1.085e-04 \\
2924339469735490560 & $-1.951$ & [ $-2.039$ , $-1.862$ ] &    1.751 & [ 1.646 , 1.864 ] &   14.766 & [ 14.128 , 15.488 ] &       5743 &    1.009 & 1.918e-04 \\
3371908043029299840 & $-0.372$ & [ $-0.379$ , $-0.366$ ] &    2.041 & [ 1.994 , 2.085 ] &   82.969 & [ 81.685 , 84.203 ] &       4330 &    0.700 & 1.742e-05 \\
3369088315397965056 & $-0.656$ & [ $-0.670$ , $-0.643$ ] &    1.839 & [ 1.788 , 1.894 ] &   40.659 & [ 39.870 , 41.541 ] &       6020 &    1.130 & 7.066e-05 \\
\hline
\hline
 \end{tabular}
 }
\end{table*}


In \figref{td_2pc} we show the distribution of time and distance of closest approach for the time
interval $\pm3$~Myr from the present.The closest encounter found is, as expected, the very well
known case of GJ~710.In \figref{dv_2pc} we show the distribution of the total relative velocity and distance of closest
approach. Most stars in our sample have high velocities ($20$ to $80$~\kms) meaning that their effect
on the Oort cloud is small. \figrefalt{tv_2pc} shows the distribution of the time and relative
velocity of closest approach, showing a triangular shape with a peak toward high velocities and the
present time. This is a selection effect caused by our limitation of the total studied sample to
stars that are currently within 50~pc of the Sun (this means that very fast-moving stars that
would approach the Sun far in the past or the future are currently not in the 50~pc volume).

 We calculate the effect on a comet of a passing star that approaches to within $2.5$~pc (\tabref{se_2pc}), using the impulse approximation (\equref{sun-comet_impulse}). We find
  that the change in the velocity of a comet is relatively small (in the order of $10^{-3}$--$10^{-4}$~\kms, \tabref{se_2pc}). The 
  exception is for the passage of GJ~710, which causes a velocity change of $\sim$ $0.13$~\kms, creating  an important  
  perturbation in the inner  Oort cloud. Overall if only individual encounters are considered  the Oort cloud comets barely 
  feel the effect of passing stars. The impulse approximation is based on a number of simplifying
  assumptions, but this approach gives us a general panorama of the individual effect of the
  nearby stars on a comet in the Oort cloud. In order to quantify the global effect of passing stars, 
  it is necessary to integrate their orbits backwards and forward in time (see \secref{obs_model}). Such a scenario 
  is shown in the third row of \figrefalt{ef_af_qf}. The cumulative effect of nearby stars is strong enough to lift the
perihelion of $\sim$ 0.38\% of the objects in the Oort cloud (\figref{ef_af_qf}, third row). Particularly the effect of GJ~710 is
strong (\figref{perilif_gj710}), but encounters within  $\sim$1~pc also have an important contribution.

\subsection{The case of GJ~710/HIP 89825}

\begin{figure*}
  \includegraphics[width=0.31\textwidth,trim=0 0 5cm 0,clip]{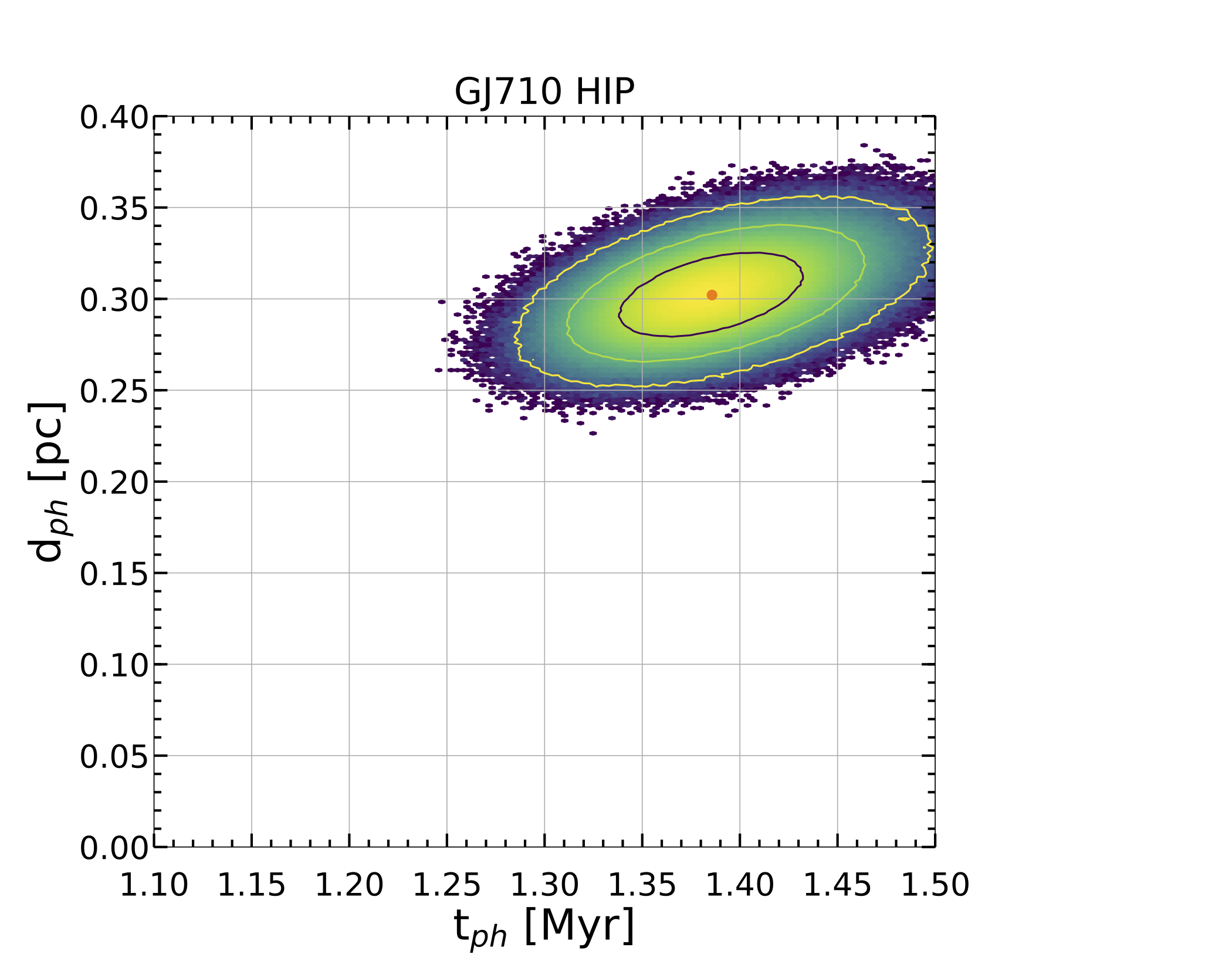}
  \includegraphics[width=0.31\textwidth,trim=0 0 5cm 0,clip]{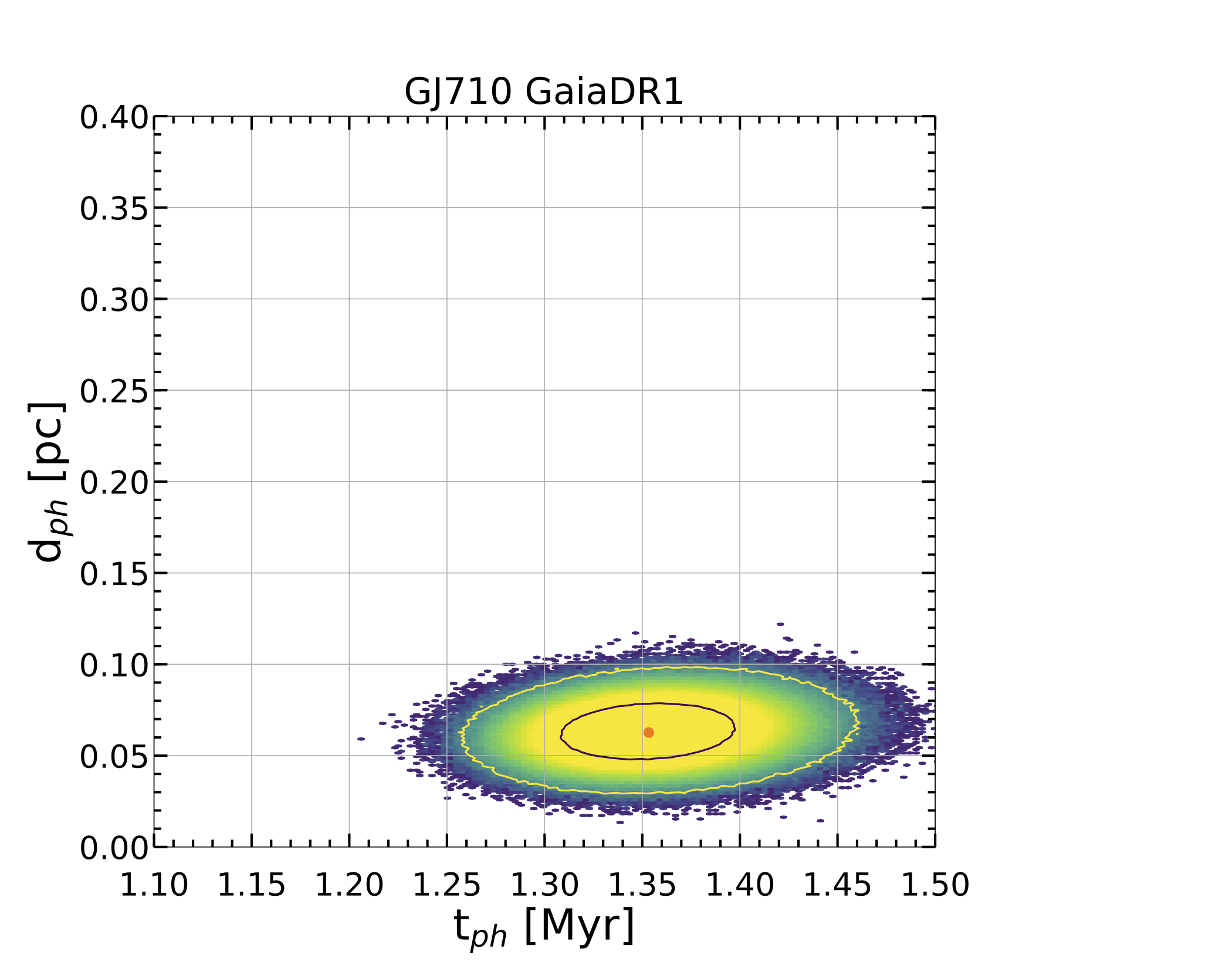}
  \includegraphics[width=0.31\textwidth,trim=0 0 5cm 0,clip]{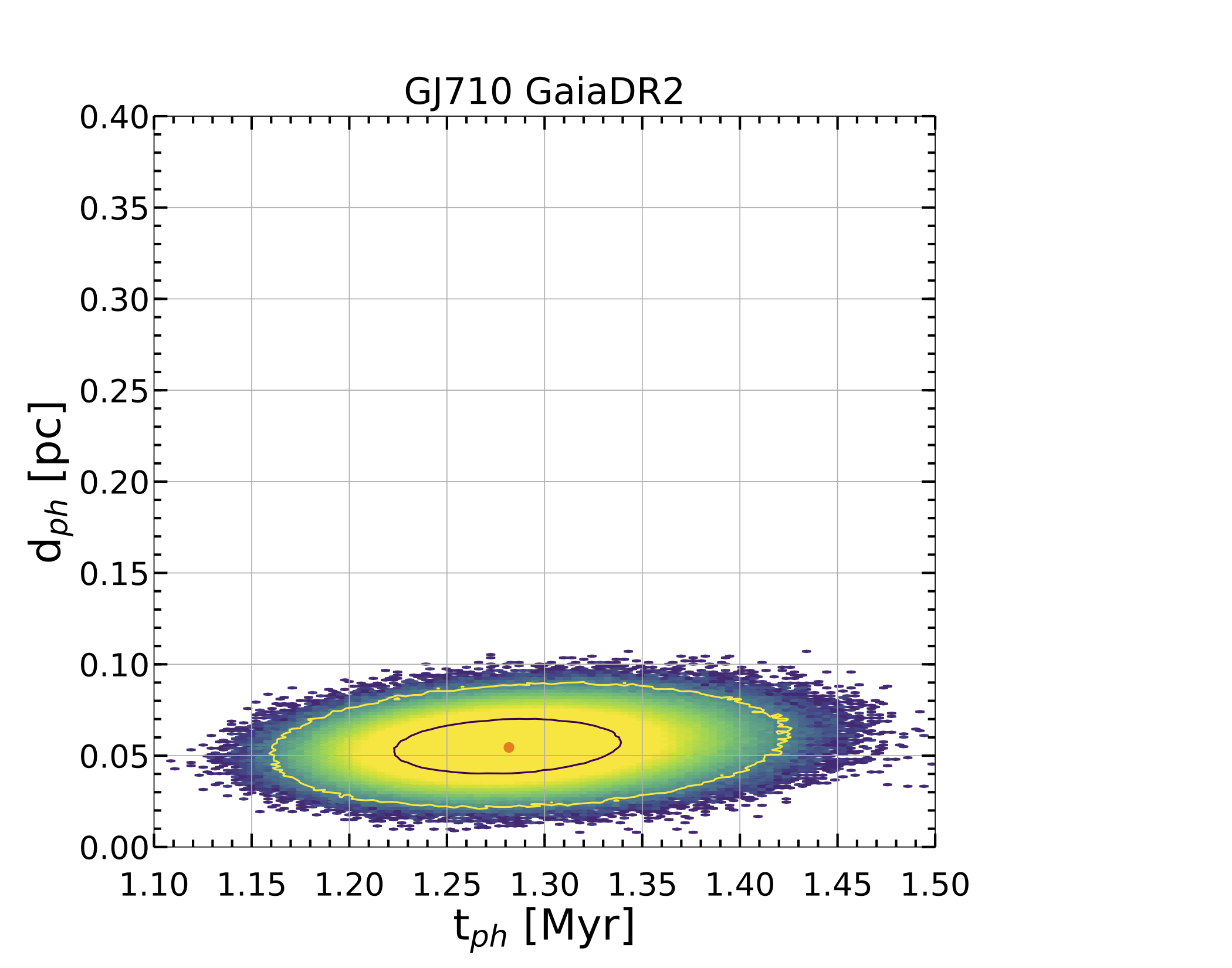}
   \includegraphics[width=0.05\textwidth,trim=0 -30cm 0 0,clip]{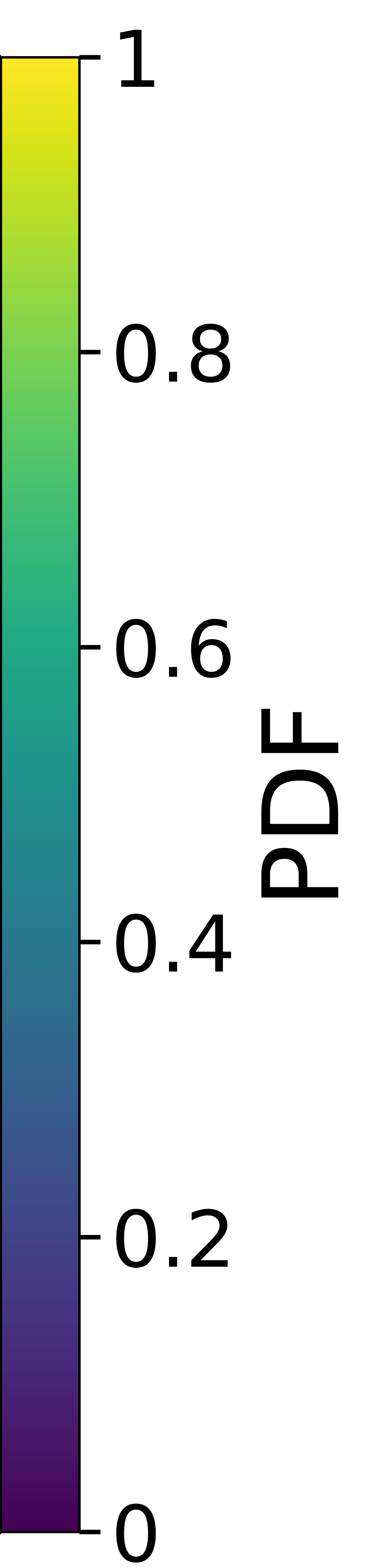}
   
  \caption{Joint distribution of the time and distance of closest approach for GJ~710. The leftmost
  panel shows results obtained with \hip\ data, while the middle and right panels show the results
for \gdr{1} and \gdr{2}, respectively (see \tabref{table_gj710}). The
    contour levels indicate regions enclosing 0.60, 0.90, and 0.99\% cumulative probability (color bar).}
  \label{gj710}
\end{figure*}

For decades GJ~710 has been pointed out as the major future perturber of the Oort cloud. The first
calculations using Hipparcos catalogue led to an encounter distance of $0.33$~pc, $1.38$~Myr from the
present time \citep[see e.g.][]{Garcia-Sanchez2001,Torres2018}. Using \gdr{1,} \citet{Torres2018}
pointed out that the encounter distance is even smaller, at $0.062$~pc, $1.35$~Myr from today
\citep[see also][]{Berski2016,Bobylev2017,Bailer-Jones_1}. With the data from \gdr{2} in hand, the
distance and time of closest approach have again slightly decreased to $0.054$~pc and $1.28$~Myr
(\tabref{catalogue},\tabref{table_gj710}).  \citet{Marcos2018} and \citet{Bailer-Jones2018} found
similar but slightly discrepant results. The small discrepancy in the various results is mainly due
to the orbit integration method and the Galactic potential used in their calculations, considering
that the input data are the same. 

\begin{figure}[ht]      
\includegraphics[width=\columnwidth]{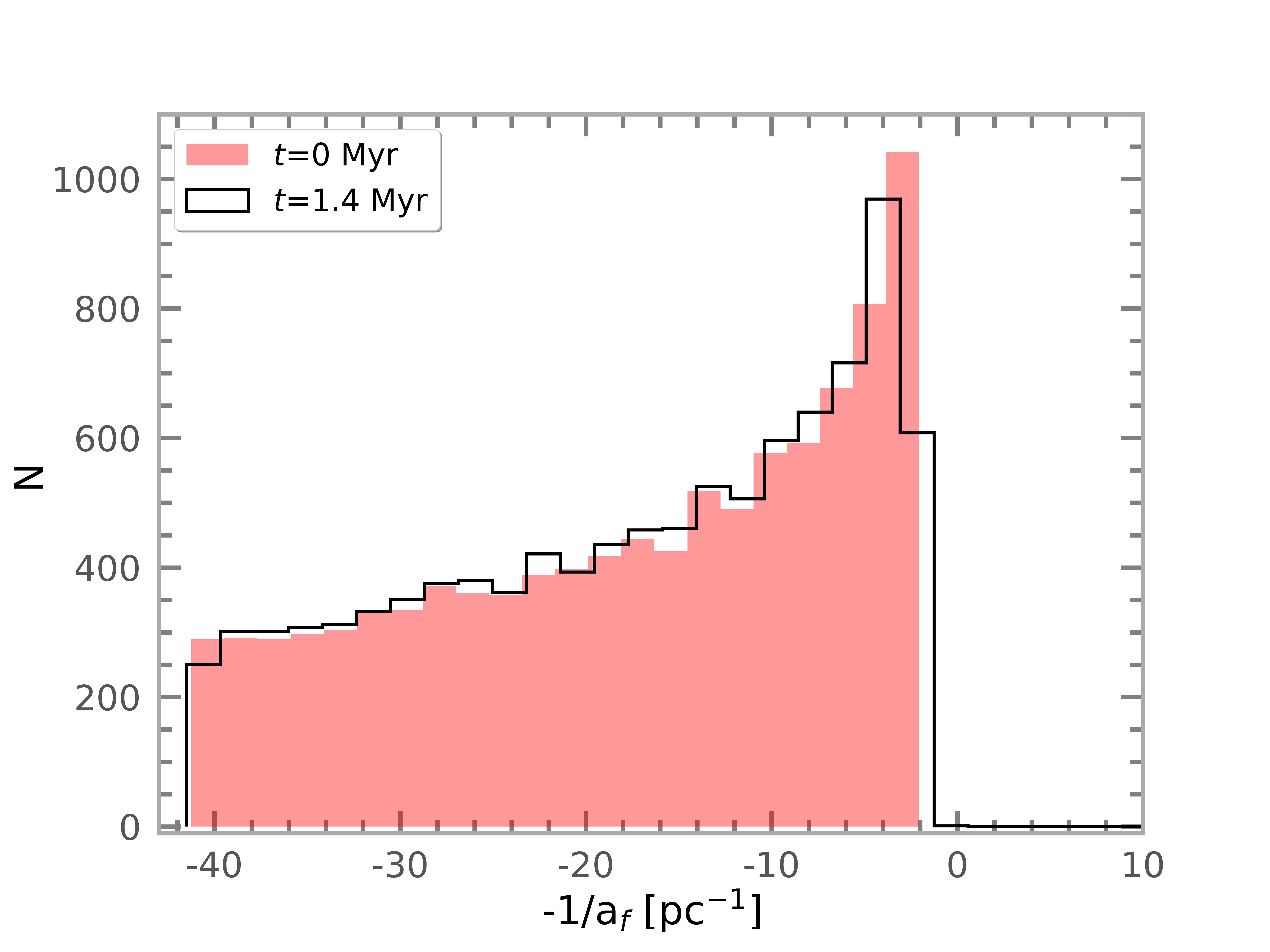}
\caption{ Histogram of the orbital energy distribution of the particles in the Oort cloud, 
before and after the encounter with the star GJ710. The red  histogram
corresponds to the  initial semi-major axis distribution, and the black curve to the final one.} 
\label{perilif_gj710}
\end{figure}

A comparison of the results obtained for GJ~710 from \hip, \gdr{1,} and \gdr{2} data is shown in
\figref{gj710}. Calculations were performed following the method described in
Sect.\ref{obs_model}, using the astrometric data described in Table \ref{table_gj710}. The discrepancy
between \hip\ and {\gaia} is due to the difference in the value of the astrometric parameters and
radial velocity. This results in a shift in the perihelion distance of GJ~710. Using different
parameters for the Galactic potential will also lead to slightly different values (see e.g.
\citealt{Bailer-Jones2018}). We note that the time of perihelion is more uncertain for the \gdr{2}
data, which is caused by the larger uncertainty in the radial velocity. 
 Following the method described in  Sect.\ref{pert_oc} and Sect.\ref{sec4}, we investigated the effect of  GJ~710 on
 a simulated Oort cloud (\figref{ef_af_qf}, first row). The perturbation because of GJ 710 lifts the semi-major axis 
 of the comets within the region between $\sim$ 10~000 and 100~000 AU (\figref{perilif_gj710}), creating  $\sim$ 0.01\% hyperbolic 
 objects, while $\sim$ 0.30\% of the comets gain a semi-major axis beyond the edge of the Oort cloud  ($a > 100\,000$ AU).

\begin{table*}
\centering
 \caption{Comparison of the different astrometric parameters and radial velocities obtained for GJ~710 from \hip, \gdr{1,} and \gdr{2} data. The last two rows represents the time and the closest approach distance to the Sun of GJ~710  with its respective confidence interval.}
 \label{table_gj710}
  {\normalsize

\begin{tabular}{lccc}
 \hline
 \hline
Parameters & Hipparcos & Gaia DR1 & Gaia DR2 \\
  \hline
$ \varpi$ [mas] & 51.12$\pm$1.63 & 52.35$\pm$0.27 & 52.51$\pm$0.04 \\
 $\mu_{\alpha*}$ [mas/yr] & 1.15$\pm$1.66 & -0.47$\pm$0.13 &-0.45$\pm$0.08  \\
 $\mu_{\delta}$ [mas/yr] & 1.99$\pm$1.22 & -0.18$\pm$0.09 & -0.02$\pm$0.07 \\
 $\mu_{tot}$ [mas/yr]& 2.30$\pm$2.06 & 0.50$\pm$0.16 &  0.46$\pm$0.11 \\
 $v_{rad}$ [km/s] & -13.80$\pm$0.30$^{*}$ & -13.80$\pm$0.30$^{*}$  & -14.52$\pm$0.43$^{**}$ \\
  $t_{ph}$ [Myr] & 1.385 [1.109,1.500] & 1.353 [1.219, 1.541] & 1.281 [1.109, 1.500] \\
 $d_{ph}$ [pc] & 0.302 [0.302,0.324] & 0.062 [0.014 , 0.116] & 0.054 [0.003, 0.106] \\
 \hline 
 \hline 
 \end{tabular}
 \tablefoot{$^{*}${\footnotesize Pulkovo catalogue (\citealt{pulkovo_2006})}, $^{**}${\footnotesize \gdr{2} radial velocity catalogue (\citealt{gdr2_rv}).}}
 }
\end{table*}

\section{Dynamical evolution of the Oort cloud}
 \label{sec4}

As pointed out by \cite{Heisler1986a} the Galactic tidal field is a major contributor to Oort cloud
perturbations at large distances, while as we showed in \secref{pert_oc} the cumulative effect of
passing stars can also lead to substantial perturbations of the Oort cloud comets. In this section
we study the cumulative effects of the known stellar encounters (\tabref{se_2pc}) and the
Galactic tidal field over the interval of 20~Myr centred on the present time. This provides a
lower limit to the combined effect of stellar encounters and the Galactic tidal field on the
dynamical evolution of the Oort cloud.

 \subsection{Numerical model  \label{gt+gs}}

\begin{figure*}
\includegraphics[width=0.5\textwidth,trim=-10 6cm 0 0,clip]{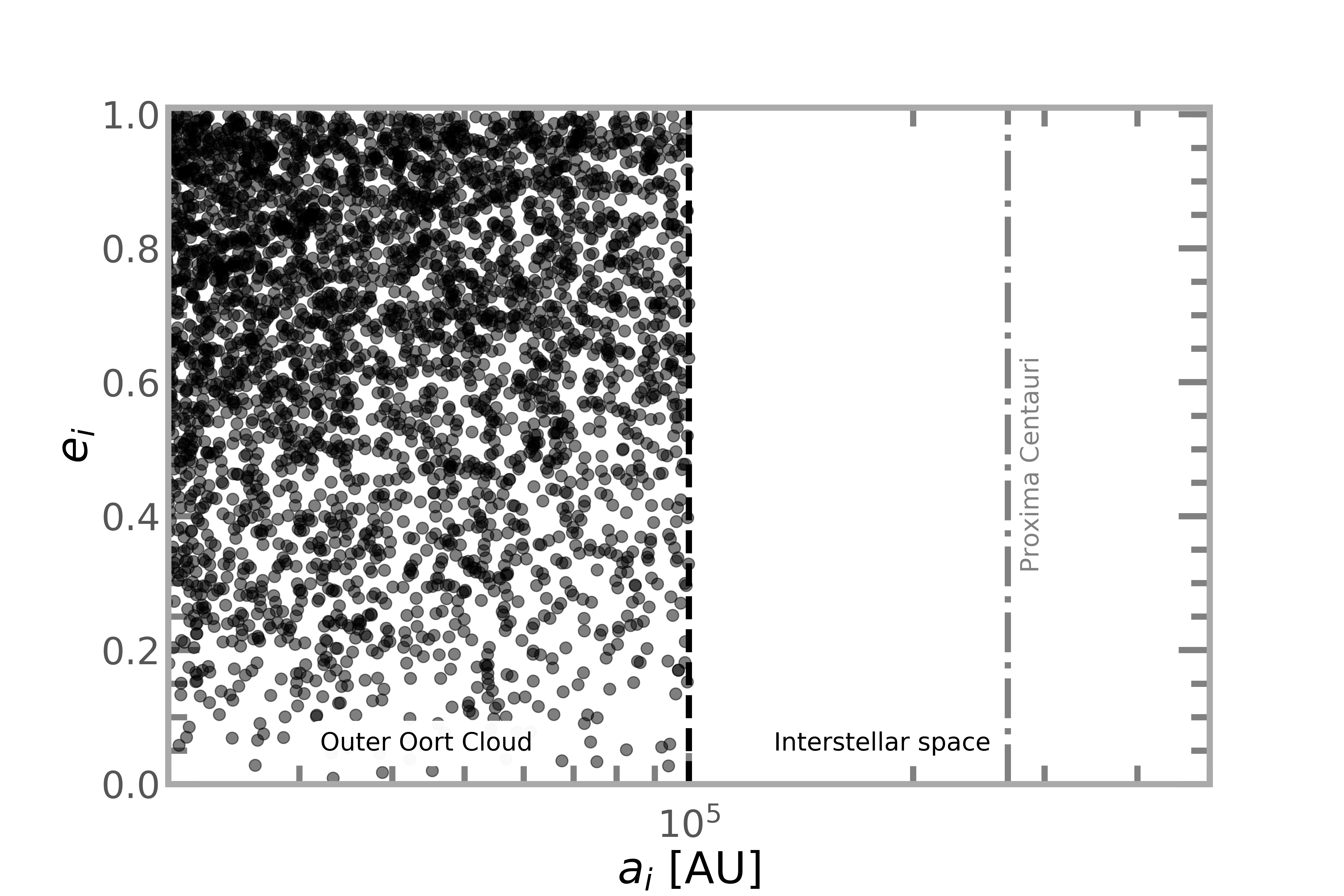}
\includegraphics[width=0.5\textwidth,trim=-10 6cm 0 0,clip]{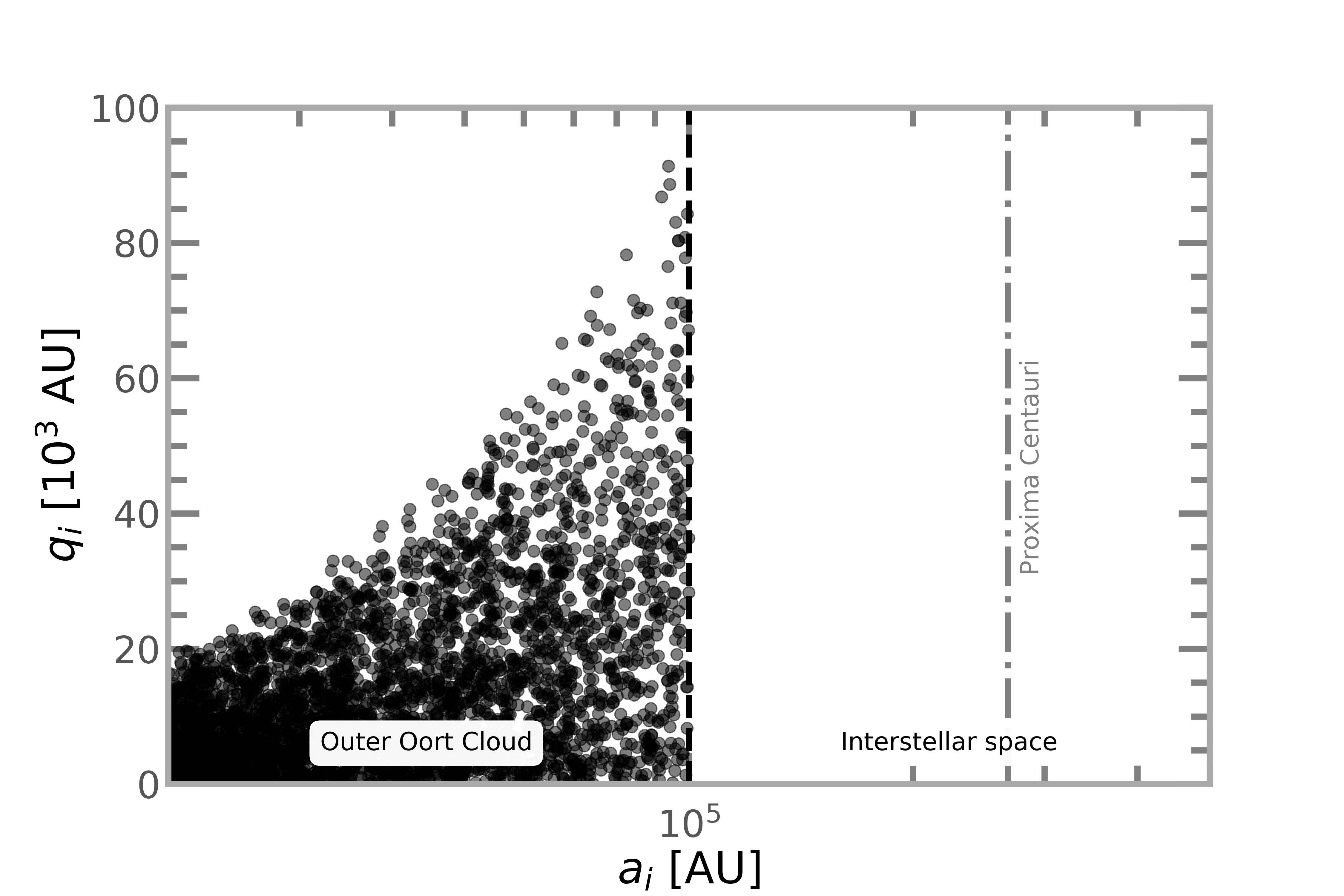}\\
\includegraphics[width=0.5\textwidth,trim=-10 6cm 0 0,clip]{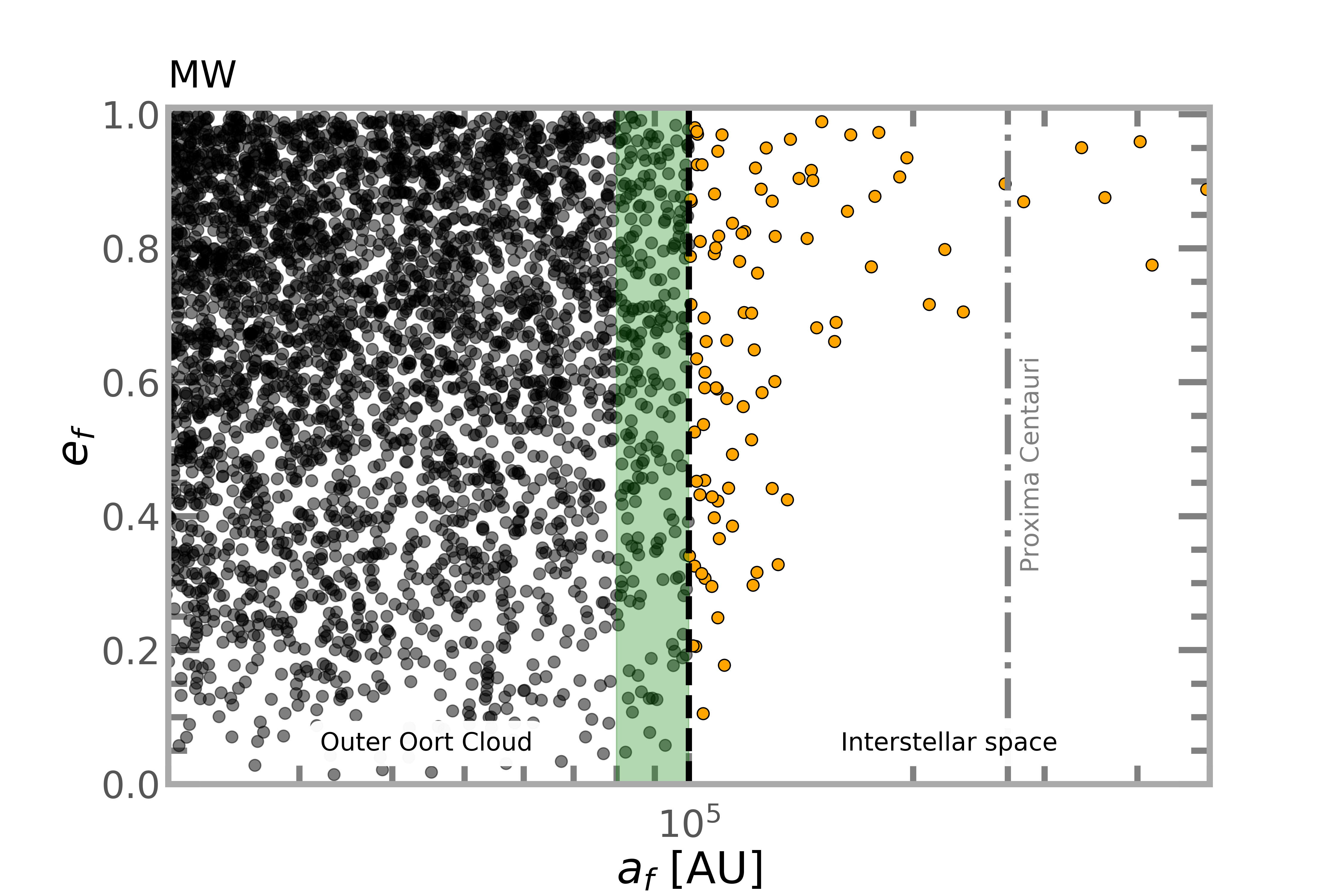}
\includegraphics[width=0.5\textwidth,trim=-10 6cm 0 0,clip]{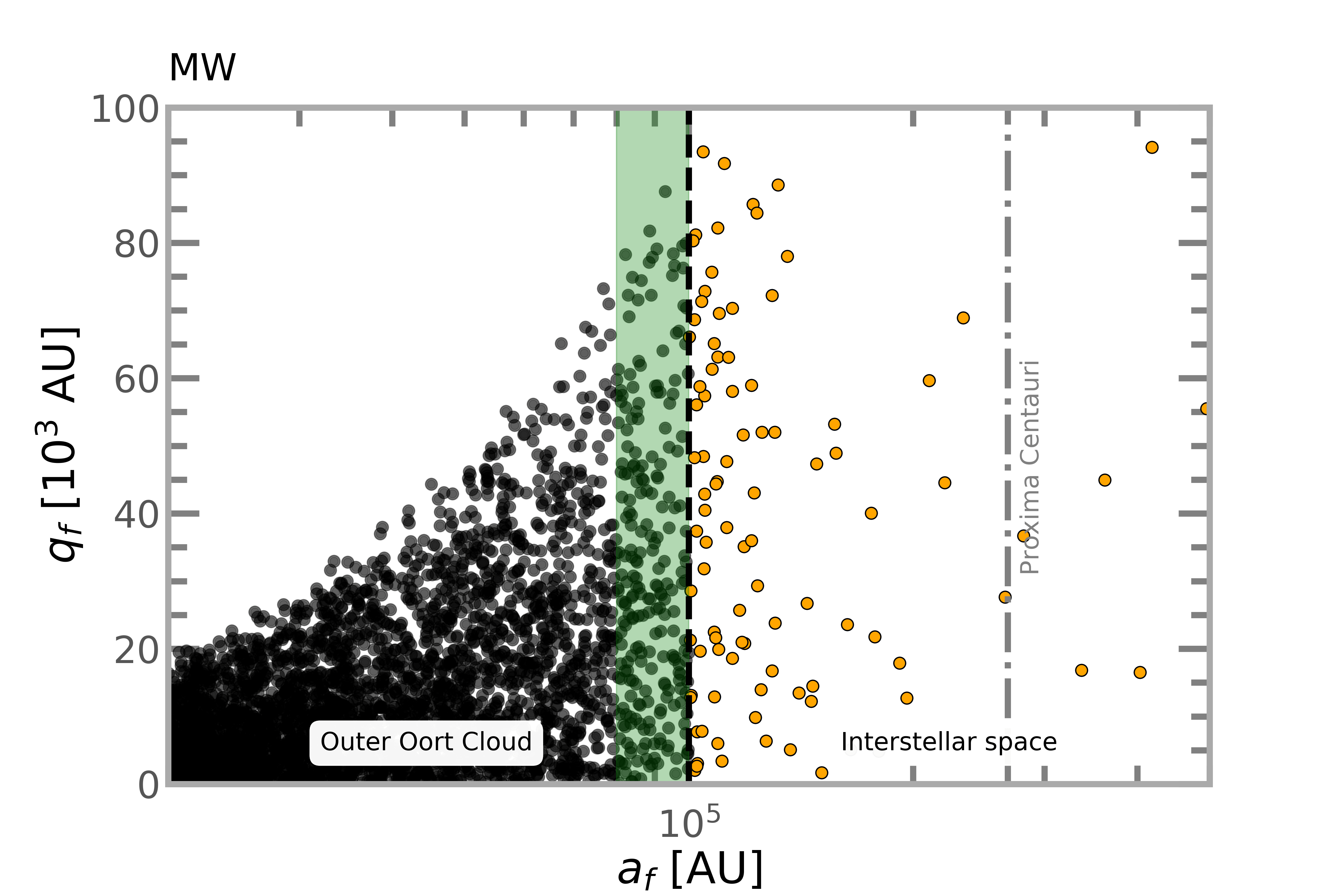}\\
\includegraphics[width=0.5\textwidth,trim=-10 6cm 0 0,clip]{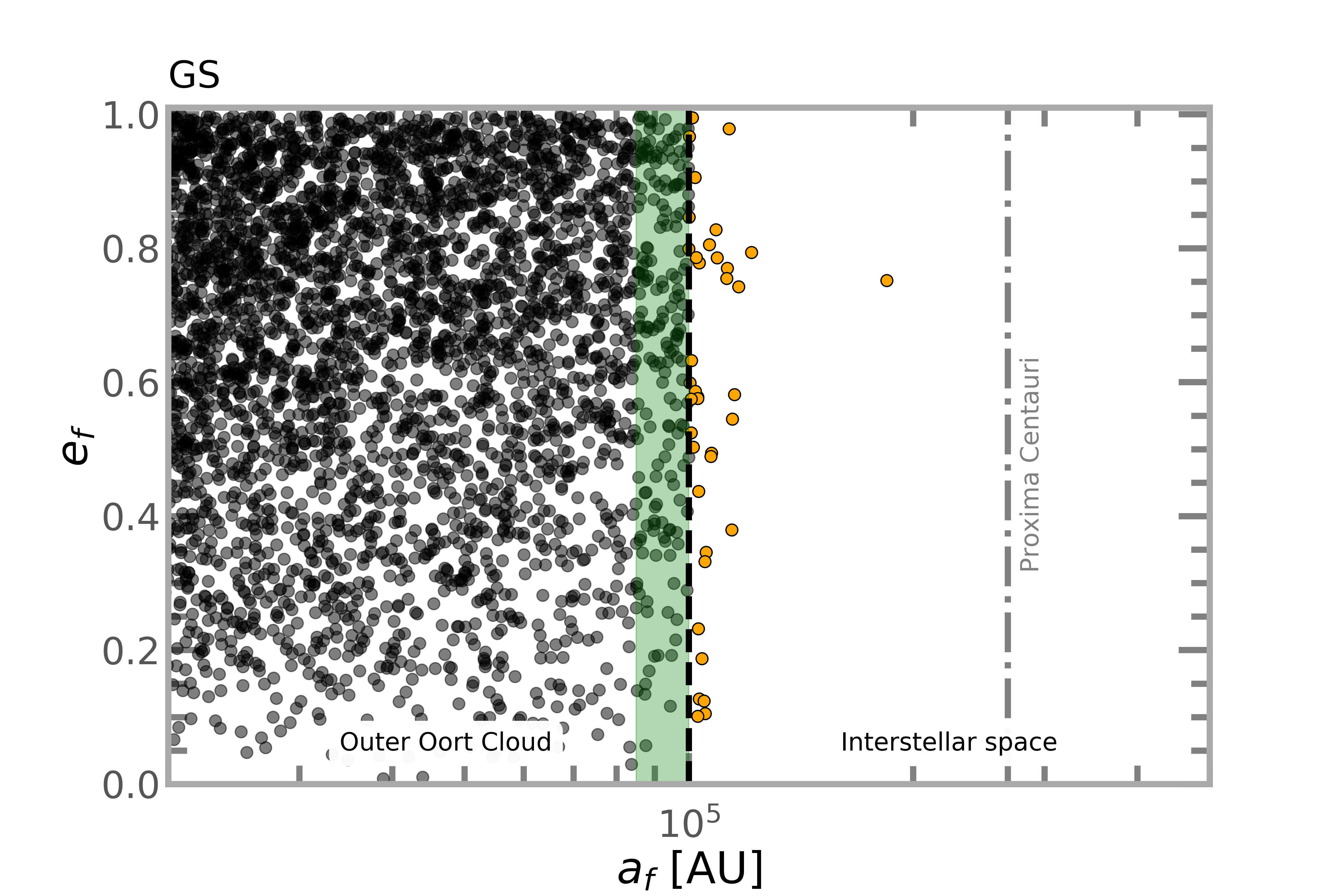}
\includegraphics[width=0.5\textwidth,trim=-10 6cm 0 0,clip]{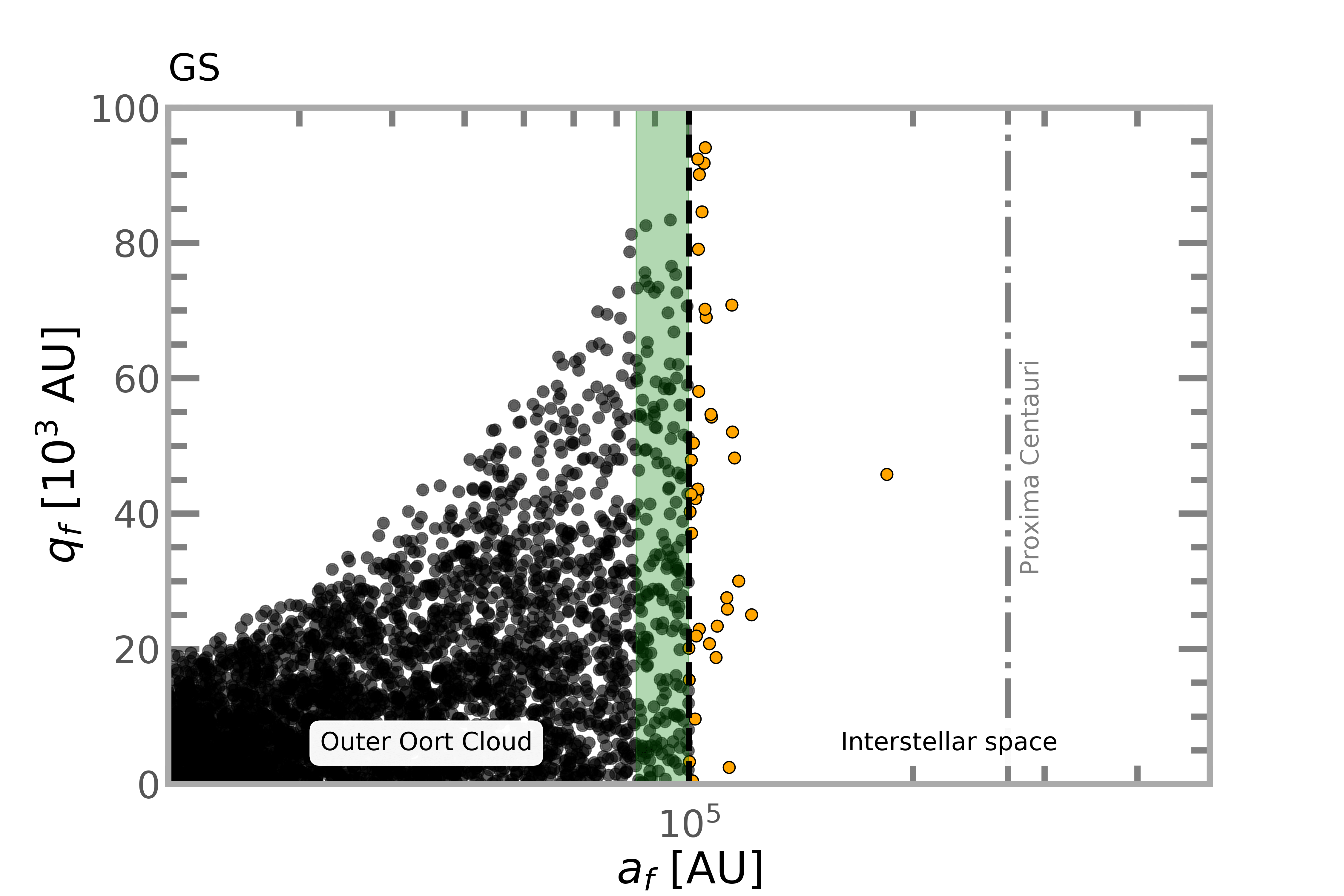}\\
\includegraphics[width=0.5\textwidth,trim=-10 0cm 0 0,clip]{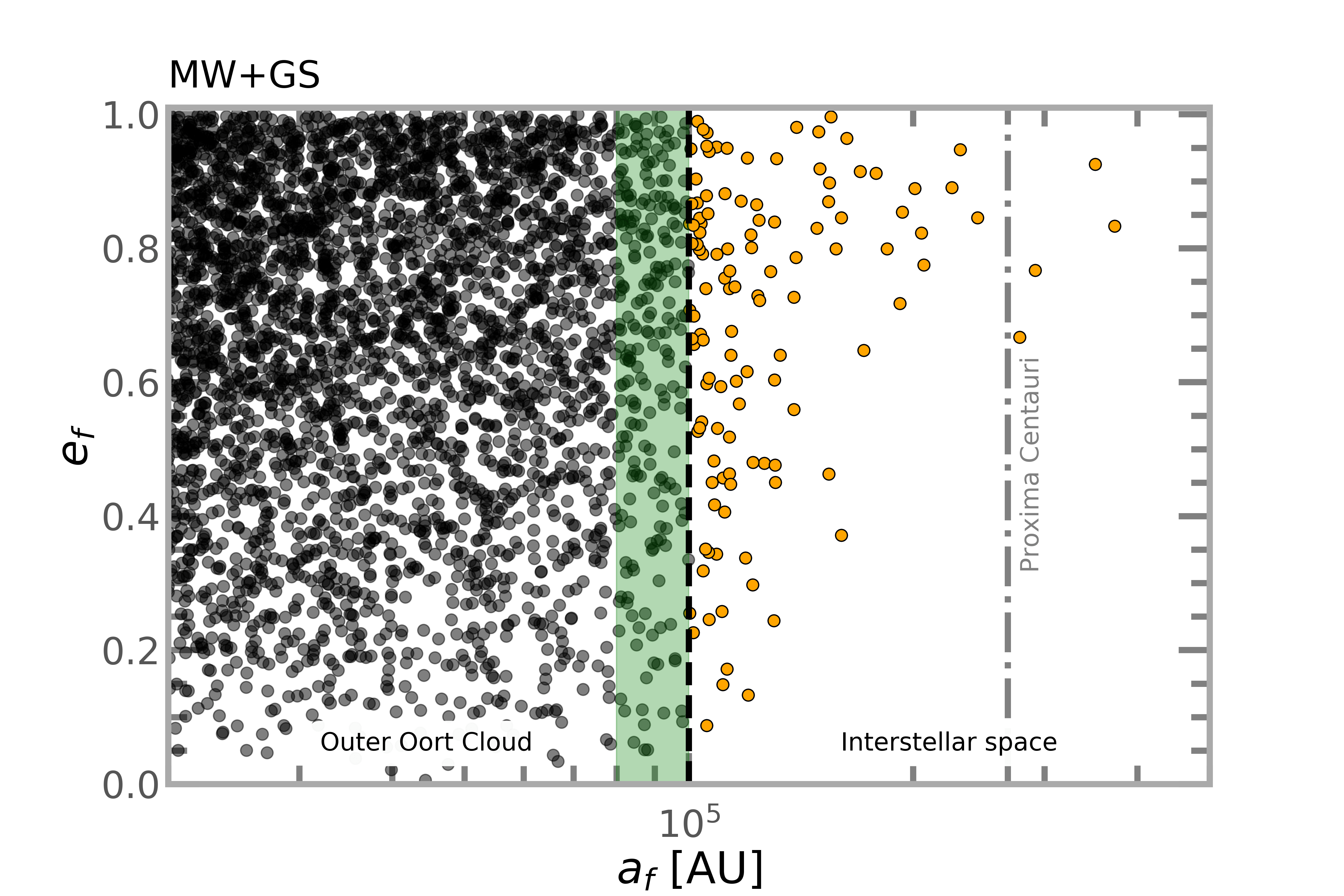}
\includegraphics[width=0.5\textwidth,trim=-10 0cm 0 0,clip]{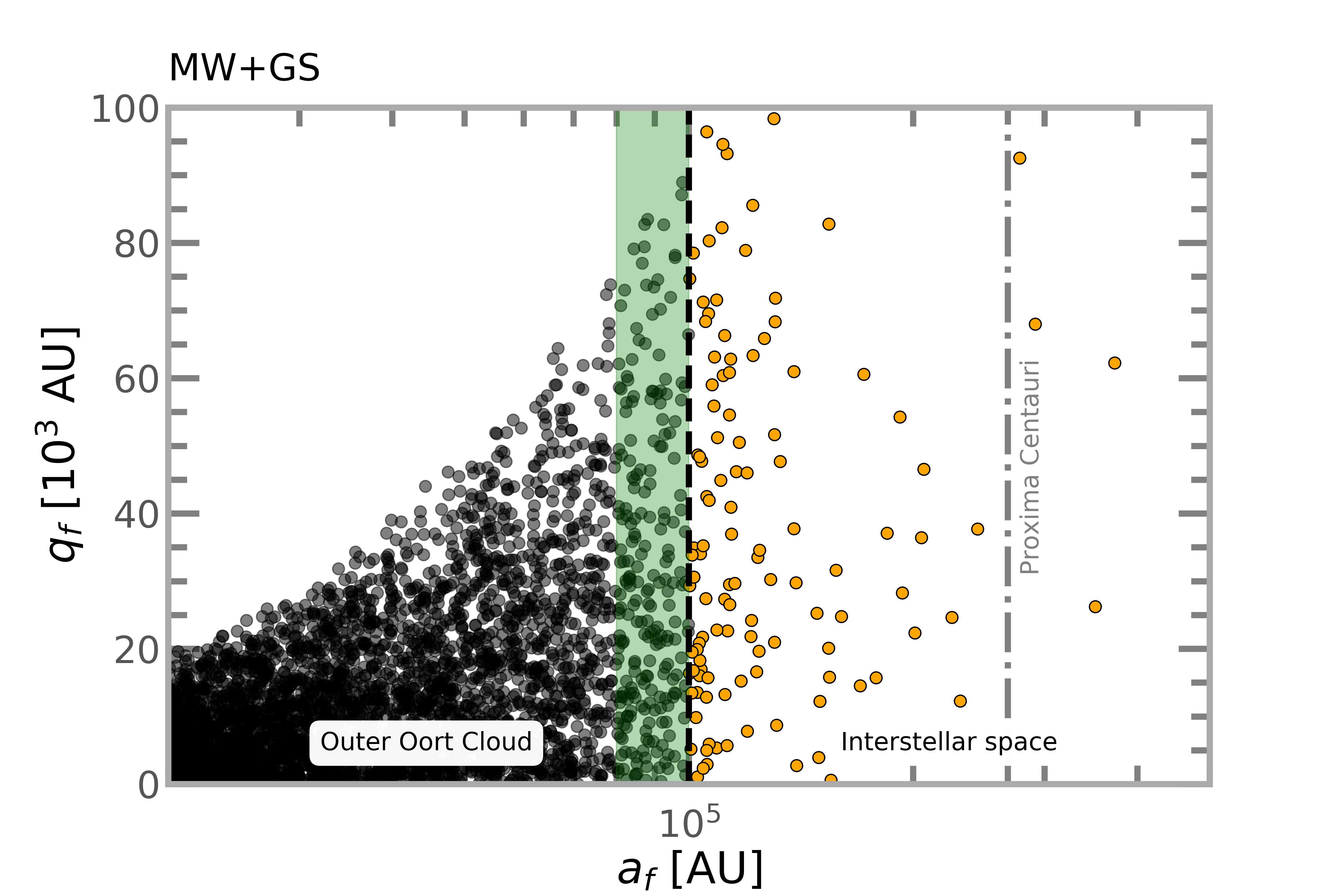}
\caption{ Final eccentricity and perihelion as a function of final semi-major-axis, for a
total integration of 20~Myr. The first row corresponds to the initial conditions. The second and third
rows show the effect of the Galactic tidal field and \gaia\  stars, respectively. The last row
corresponds to the combined effects of the Galactic tides and \gaia\ stars. The green area corresponds to
the initial position of the ejected particles, coloured yellow.} 
\label{ef_af_qf}
\end{figure*}

We use the Astrophysical Multi-purpose Software Environment -- AMUSE \citep{Portegies2009,
Pelupessy2013a, Portegies2013, amusebook} -- for our calculations. Following the works of \cite{Rickman2008} and
\cite{Hanse2016} we first construct an isotropic Oort cloud of $10,000$ test particles
(\figref{ef_af_qf}, first row). The distribution of Oort cloud particles is spherically symmetric and isotropic, and
they follow a uniform distribution in the orbital elements $\cos i$, $\omega$, $\Omega$, and $M$.
The initial eccentricities, $e$, are selected with a probability density distribution $p(e)\propto
e$ and the perihelia, $q$, are chosen outside of the planetary region ($q>32$ AU). The semi-major
axes, $a$, are distributed proportional to $a^{-1.5}$ over the range $3\times 10^{3}$--$10^{5}$ AU.
In order to ensure a thermalised Oort cloud (e.g. \cite{DuncanM.J.QuinnT.1987, Dybczynski2002, Rickman2008}) we used a
radial density profile of $r^{-3.5}$ (where $r$ is the distance between the comets and the Sun).

Subsequently we used the GPU-accelerated direct N-body code \texttt{ABIE} (Cai et al.\ in
preparation) with a fifteenth-order Gauss-Radau integrator \citep{Everhart1985} optimised for close
encounters. We couple \texttt{ABIE} and the \texttt{Gala} package in such a way that \texttt{ABIE}
advances the positions of the Oort cloud particles and \texttt{Gala} calculates the accelerations on
each particle due to the Galactic tidal field, based on the positions provided by \texttt{ABIE}.
The calculated accelerations are subsequently inserted into the Gauss-Radau integrator in
\texttt{ABIE} as additional forces. Using the catalogue of nearby stars (\tabref{catalogue}), we
selected all the stars (31) that are predicted to pass within $2.5$~pc of the Sun
(\tabref{se_2pc}) $\pm10$~Myr from today. These stars are included in the integrator with their
present-day positions and velocities with respect to the Sun. Hence we evolve a system for a period
of $20$~Myr which consists of one host star (the Sun) surrounded by $10,000$ test particles (Oort
cloud) under the influence of external perturbations due to passing stars and the Galactic tidal
field.

 \subsection{Galactic tide and Gaia star perturbation    \label{gt+gs}}

In order to disentangle the effects of the Galactic tidal field and the encounters with stars
identified in \gdr{2}  we considered three main cases for
external perturbations, \textit{Galactic tidal field, \gaia\ stars, Galactic tidal field + \gaia\
stars}.  We focus now on the effect of the external perturbations considering an  extended
  Oort cloud ($a\leq 100\,000$ AU). The first row of  \figref{ef_af_qf} shows the initial conditions followed by 
  the final perihelion distance as a function of the  final semi-major axis  for the three scenarios previously discussed. 
  Considering a short integration of 20 Myr (10 Myr in the past, and 10 Myr in the future). The green area represents
  the original location of  the ejected particles (yellow dots).

 The effect of the Galactic tidal field on the Oort cloud decreases from the outskirts to the
  inner regions of the cloud (second row \figref{ef_af_qf}). The particles at the edge of the
  cloud suffer a considerable change in their orbital elements. Specifically, for $\sim 0.91$\% (yellow
  dots, \figref{ef_af_qf}) of the objects, their semi-major axes increase up to interstellar distance (
    $a >$ 100\,000 AU).  The particles in the inner Oort cloud remain unaffected. A small fraction of the
    particles ($\sim0.02$\%) acquire hyperbolic orbits. When \gaia\ stars are the only perturber
    (\figref{ef_af_qf}, third row) their effect is much less pronounced than that of the Galactic tidal field in particular in the
    outskirts on the Oort cloud. The effect of \gaia\ stars is dominated by the star GJ~710 (\figref{perilif_gj710}). However,  the cumulative effect of relatively  distant encounters ($\sim$ 1pc) helps to change the semi-major axis of $\sim$ $0.38$\% of the comets  in the outer Oort cloud, whereas $0.01$\% of the outer Oort cloud objects acquire hyperbolic orbits.

 The combination of the  Galactic tidal field  and \gaia\ stars (\figref{ef_af_qf},
  last row) enhances the perturbations on the Oort cloud, causing  $0.03$\% of the initial objects to become 
  unbound from the solar system, while  $\sim1.12$\% acquire orbits with semi-major axis in 
  the interstellar regions, i.e., $a \geq 100\,000$ AU. In all three scenarios for
  external perturbations, a considerable population of objects with $a \geq 100\,000$ AU is created.
  Their orbits remain elliptic, but the effect of external perturbations lifts their semi-major axis
  beyond the Oort cloud (yellow dots, \figref{ef_af_qf}). This effect is only
relevant for the outermost regions of the Oort cloud ($\sim 80\,000$--$100\,000$ AU). The orbital elements of the particles in 
the inner parts of the cloud will not be affected as strongly.

\begin{figure*}
  \includegraphics[width=0.333\textwidth,trim=0 0 3cm 0,clip]{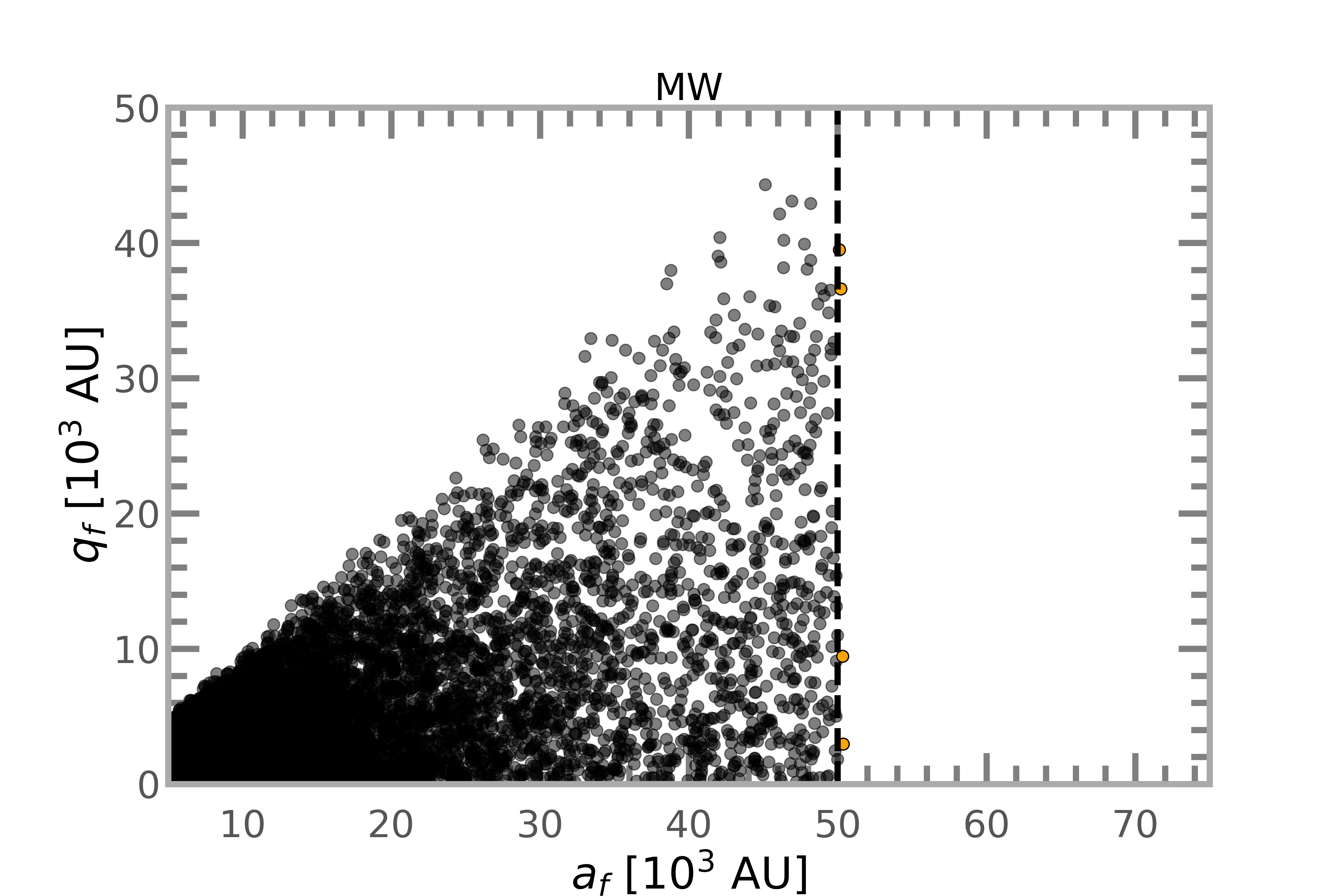}
  \includegraphics[width=0.333\textwidth,trim=0 0 3cm 0,clip]{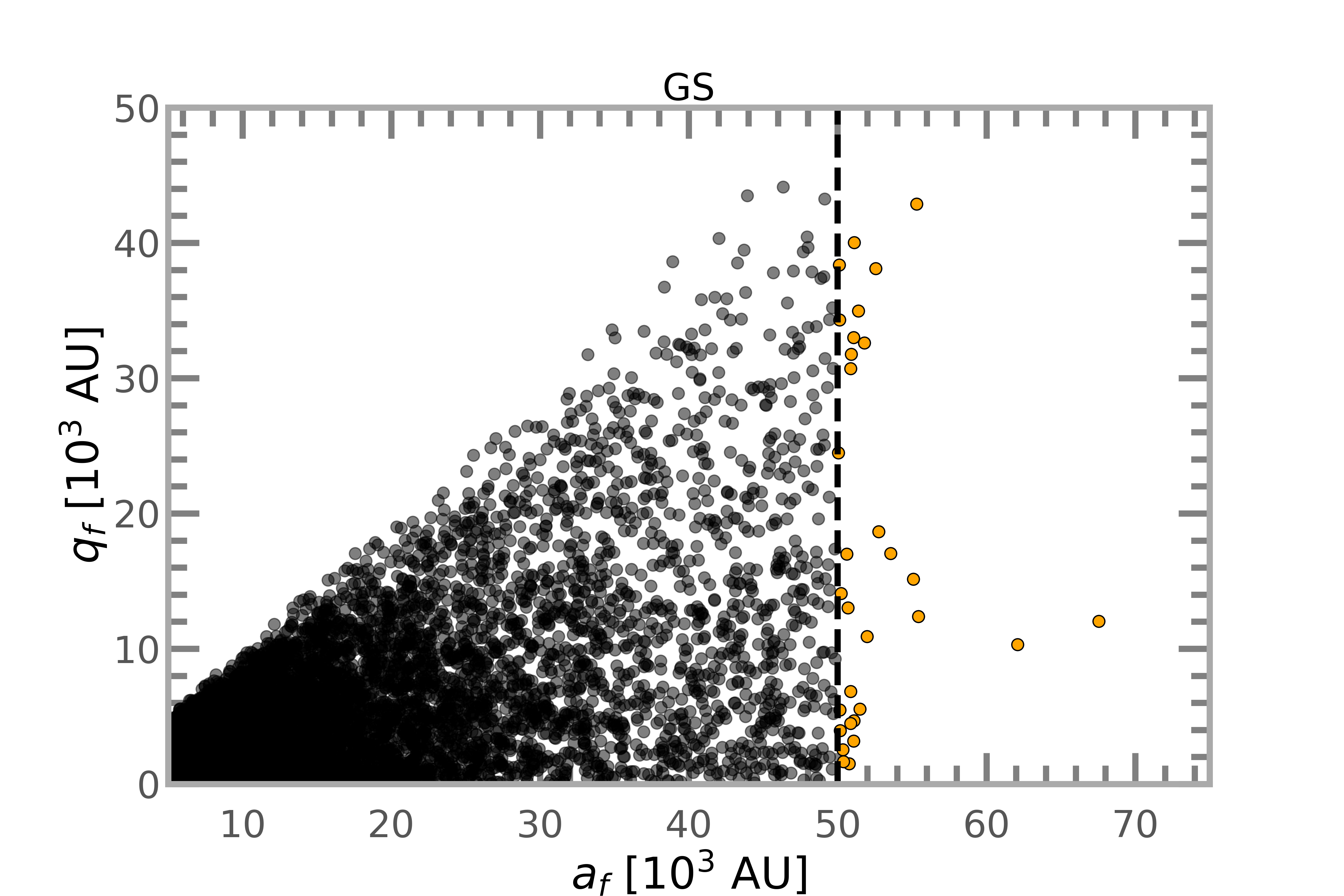}
  \includegraphics[width=0.333\textwidth,trim=0 0 3cm 0,clip]{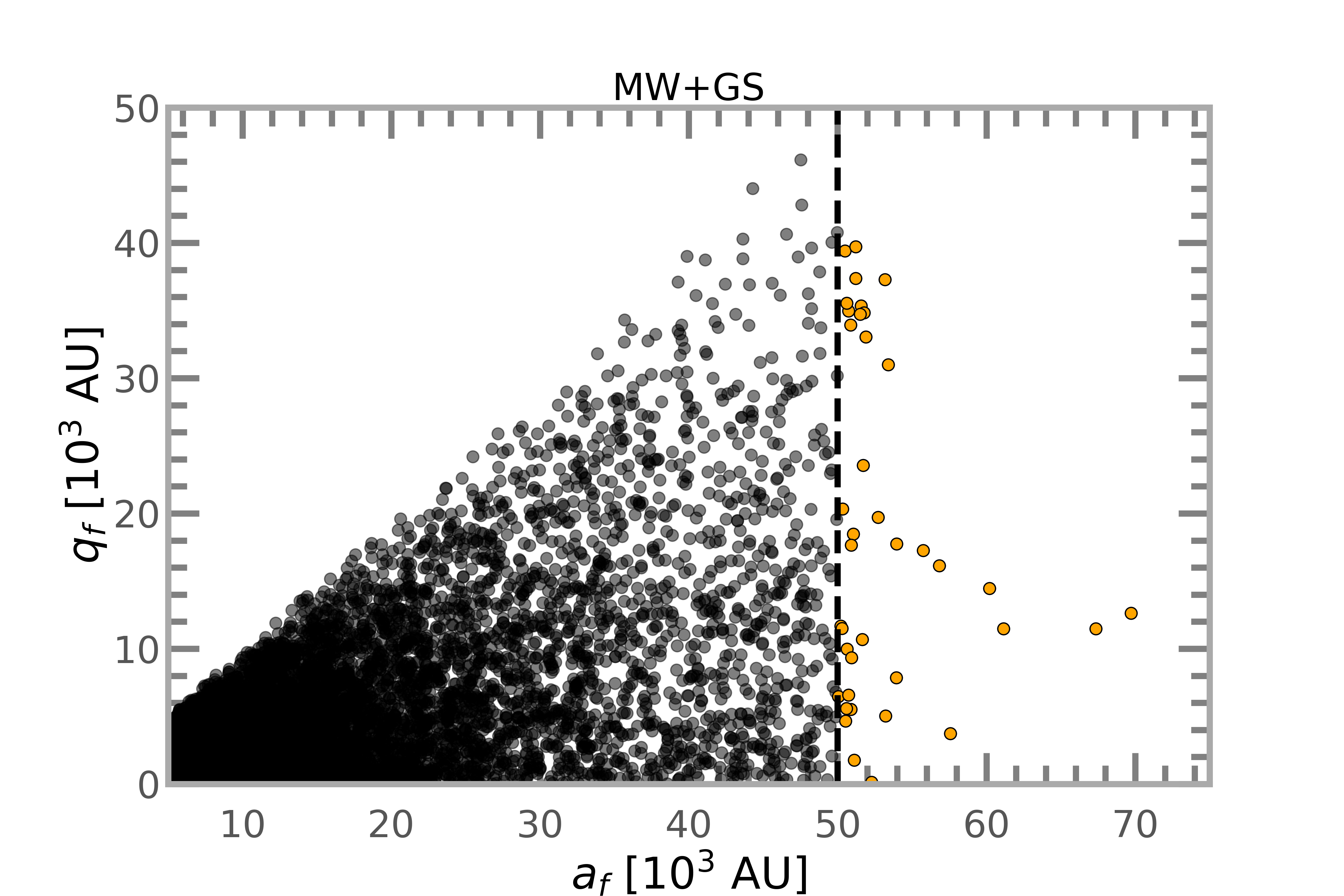}
  \caption{Final perihelion as a function of final semi-major axis for an Oort cloud size of
  $50,000$ AU. The first panel shows the effect of the Galactic tidal field (marked as a MW).
The second and third panels show the effect of Gaia stars (GS) and the combination of both the Galactic tide and Gaia
stars (MW+GS), respectively. Black dots represent the particles in the Oort cloud, while the yellow dots represent the
ejected ones.}
  \label{50k_oc}
\end{figure*}

 We now consider a compact Oort cloud with semi-major axes up to $50\,000$ AU. 
 The effect of the Galactic tidal field has a negligible effect (first panel,
\figref{50k_oc}) over the particles in the cloud. The second panel in \figref{50k_oc} shows the effect of the Gaia stars. The
  effect of GJ~710 is prominent,  and causes a major perturbation. The last panel in
\figref{50k_oc} shows that the effect of the \gaia\ stars dominates 
over the Galactic tidal field, however  it is the combined effect which efficiently increases the number of particles (by $\sim1.20$\%) with semi-major axis beyond the limits of the cloud ($a > 50\,000$ AU).
 
 We conclude that the cumulative effect of passing stars  and the Galactic  tidal field
  are efficient mechanisms in the creation of  comets for which the semi-major axis is larger than the extent  of the Oort cloud ($a$ > 100\,000 AU), but with bound and eccentric orbits. Hereafter we refer to such objects as transitional interstellar comets (TICs). If we consider an Oort cloud with $a\leq100\,000$ AU, the Galactic tidal
field is the major perturber, while for an Oort cloud with $a\leq 50\,000$ AU passing stars
provide the major effect, mainly due to the close encounter with GJ~710 (\figref{perilif_gj710}).

\begin{figure*}
\centering
\includegraphics[width=0.8\textwidth]{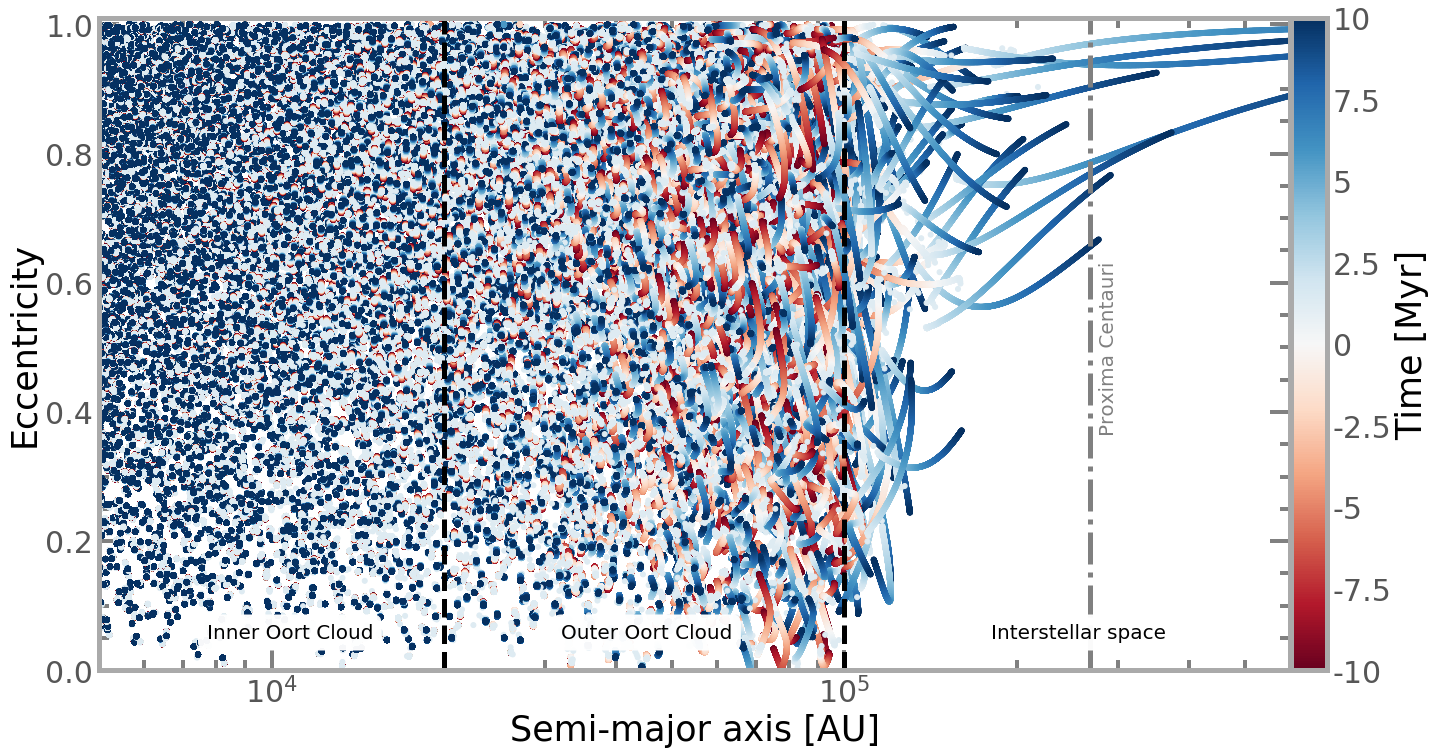}
\caption{ Orbital evolution of the particles in the simulated Oort cloud. The figure shows
    the evolution of the eccentricity as a function of the semi-major axis over the period of 20 Myr
    ($\pm10$~Myrs.). The colour bar represents the integration time. An animation can be found at:
\url{https://home.strw.leidenuniv.nl/~storres/\#Research}}. 
\label{peri_lift}
\end{figure*}

\begin{figure}[h]
\includegraphics[width=\columnwidth]{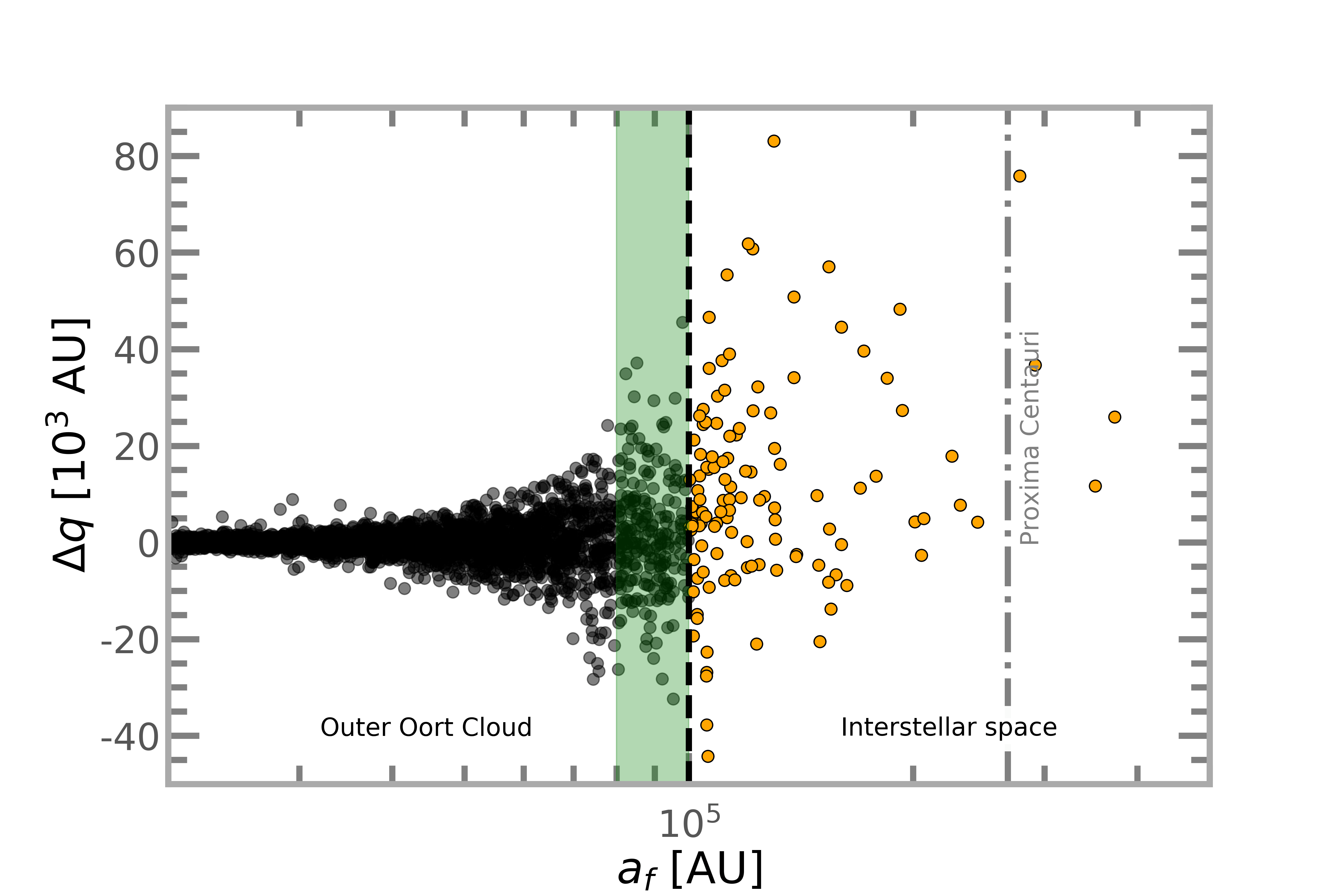}\\
\includegraphics[width=\columnwidth]{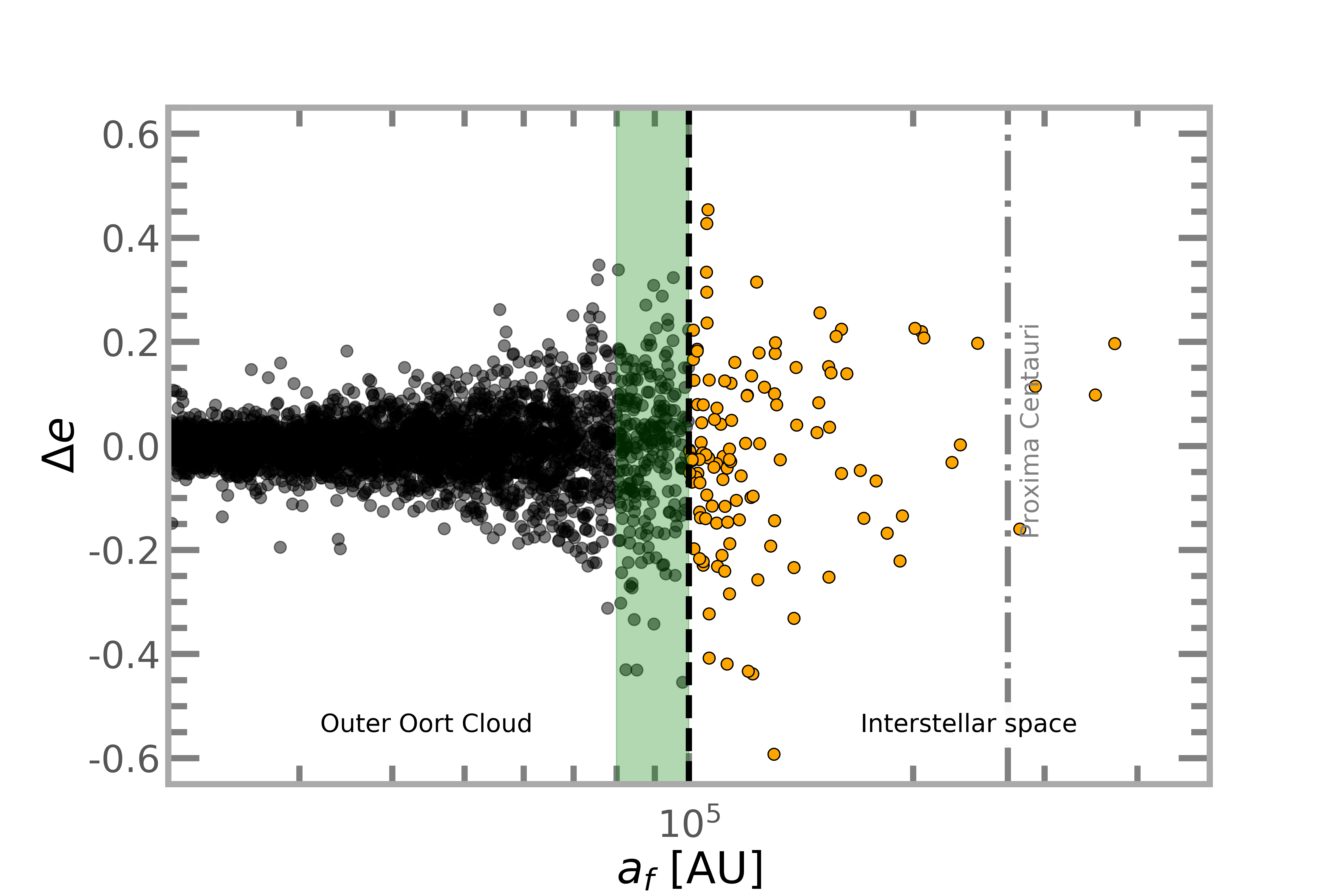}
\caption{ Mean perihelion and eccentricity changes as a function of the semi-major axis of
the comets. The green area corresponds to the region of the initial position of the particles
ejected, represented as a yellow dots.} 
\label{d_e_q_a}
\end{figure}

\begin{figure}[ht]      
\includegraphics[width=\columnwidth]{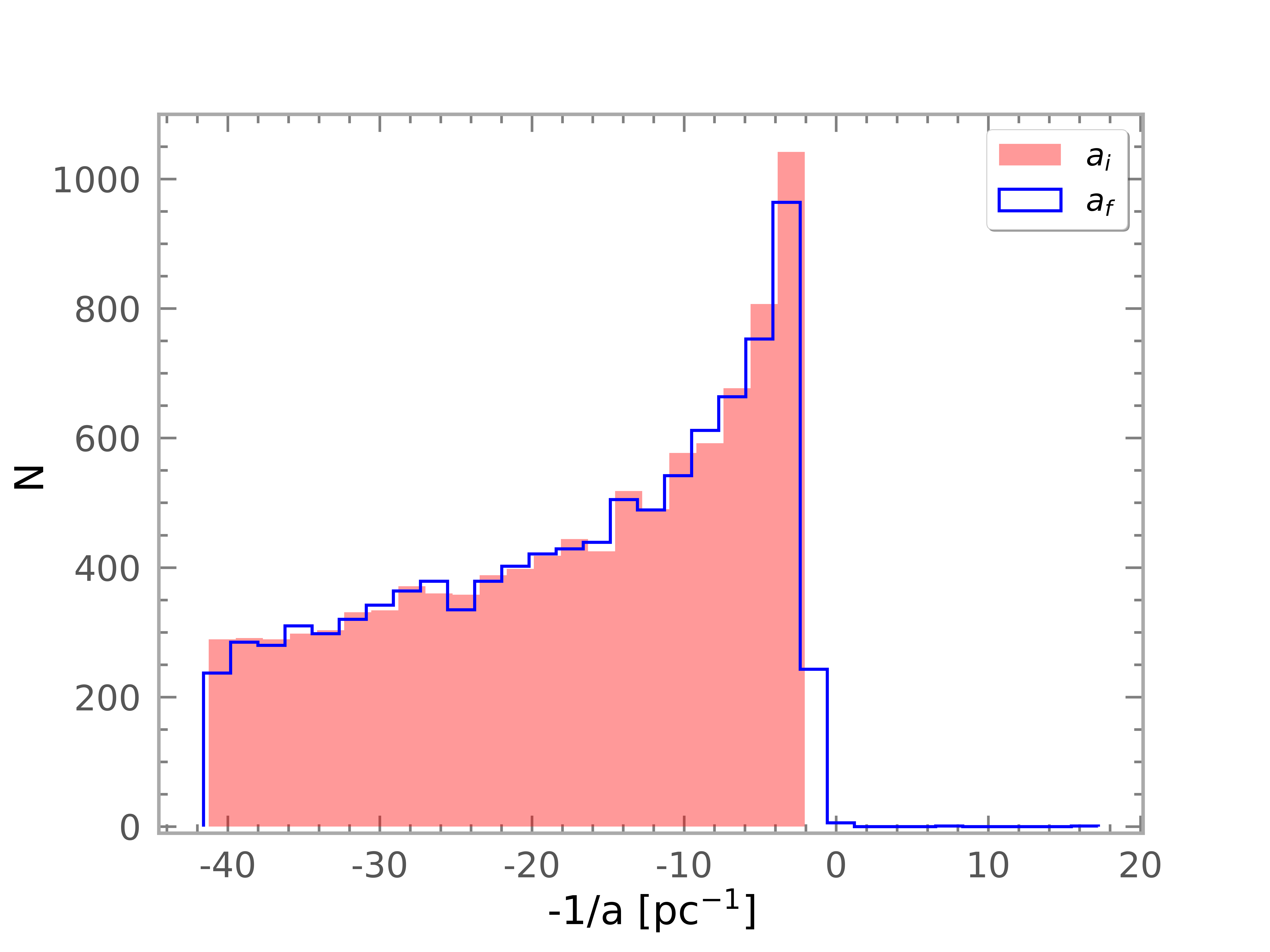}
\caption{ Histogram of the orbital energy distribution of the particles in the Oort cloud. The blue curve
corresponds to the  final semi-major axis, while the red curve shows the initial distribution.} 
\label{energy}
\end{figure}

 For long timescales (on the order of gigayears), the synergy between Galactic tides and
  stellar encounters to bring comets into the observable zone is now well understood
  (\citealt{Rickman2008, Fouchard2011}). Both perturbations strongly depend on the semi-major axis
  of the comets. In general the Galactic tidal field rapidly changes the perihelia of the outer regions, while passing stars are a good mechanism to eject or inject particles when a  close
  encounter happened (see e.g, \citealt{PortegiesZwart2015a}). For  short timescales
  ($\sim20$ Myr),  the Galactic tide and stellar encounters prove  to be an efficient mechanism for the creation of TICs. The outermost part of the
  cloud ($\sim 80\,000$--$100\,000$ AU, \figref{peri_lift}) is heavily perturbed, whereas the
  innermost part remains unchanged ($3000$--$50\,000$ AU, \figref{peri_lift}).  This implies that the
  edge of the Oort cloud is sensitive to external perturbations and is relatively easy to
  strip. The particles in the outermost part of the Oort cloud have a considerable
  change in their orbital elements. The change of the perihelion and eccentricity increases as a function 
  of the semi-major axes (\figref{d_e_q_a}), whereas the semi-major axes reach interstellar distances. 
  These objects previously referred to as {transitional interstellar comets} remain bound to the Sun with
eccentric orbits (\figref{energy}). The detailed effects of subsequent perturbations due to passing
stars and the Galactic tidal filed will determine if these objects will return to the solar system or
become unbound. 

 Considering the efficiency of external perturbations on circumstellar comet clouds in the
  creation of interstellar objects, and noting that \citet{Valtonen1982} pointed out that objects
  with a relative velocity above 0.5~km/s can probably enter and leave the solar system, we
  speculate that a `cloud' of objects exists in interstellar space which overlaps with our Oort
  cloud and constantly exchanges material with it.  An indication that this may be the case was
  provided by the first interstellar comet detected, 'Oumuamua \citep{Williams2017}, which
  opened a new era in the study of interstellar objects. Estimates of the local density of
  interstellar objects range from $10^{14}$ pc$^{-3}$ \citep{Zwart2017}, to $8\times10^{14}$
  pc$^{-3}$ \citep{Jewitt2017}, to $2\times10^{15}$ pc$^{-3}$ \citep{Do2018}. The existence of an
  interstellar comet cloud could partly explain the slightly hyperbolic comets and 
potential interstellar objects that might have been detected in the solar system but not yet classified as such
(see e.g. \citealt{Ashton2018, delaFuenteMarcos2018, Siraj2019}). A future detailed study of the evolution of
the TICs created by the  tides and stellar encounters is needed
to draw more solid conclusions.

  The results presented here are based on the assumption of a hypothetical present day
   spheroidal cloud of comets extending up to $100,000$~AU from the Sun. If we consider a smaller structure
   (\figref{50k_oc}),  passing stars are the main perturbers, while the Galactic tidal field barely
   influences the orbit of the comets. In addition we stress that our sample of stars considered as
   perturbers of the Oort cloud is incomplete due to the \gaia\ survey limits combined with our data
   quality filtering and the upper limit we imposed on the distance to the stars in our sample. A
 more complete inventory of Oort cloud perturbers would increase the effects of the stellar
 encounters.

\section{Summary and Conclusions}
\label{sec5}

In this work we present a study of the combined effect of the
Galactic tidal field and close stellar encounters predicted to occur over a time interval of 20~Myr
around the present on the Oort cloud of the solar system. Our focus is on the loss of comets to interstellar space. Following
\cite{Rickman2008}, we first presented a simple model of stellar encounters based on data compiled
for 13 spectral types of the stars in the solar neighbourhood. We confirm that individual
perturbations of randomly passing stars cannot  alter the orbits of the comets in the
Oort cloud unless a very close encounter occurs. However, from a consideration of the stellar
encounter statistics we show that the comets in the cloud may be lost to interstellar space over a
short period of time due to the cumulative effect of stellar encounters.

Motivated by this result we used \gdr{2} data to identify $14,659$ stars passing within $50$~pc
of the Sun over the time period of $\pm10$~Myr centred on the present. Out of this sample 31 stars
are predicted to be major perturbers of the Oort cloud, approaching the Sun to within $2.5$~pc. This
catalogue of perturbing stars (presented in \tabref{catalogue}) constitutes an astrometrically clean
sample, which is nevertheless incomplete due to the \gaia\ survey limitations, the upper limit imposed on
the distance to the stars in the sample (50~pc), and the strict data-quality filtering.  Our
estimates of the effect of { known} stellar encounters is therefore conservative  \citep[we note
that][find a larger number of stellar encounters from \gdr{2} due to their less stringent data-quality
filtering]{Bailer-Jones2018}.  Using the impulse approximation (\equref{impulse_aprox-comet}) we
then calculated the impulse that each star passing within $2.5$~pc of the Sun imparts to a comet
in the Oort cloud. We found that (as expected) the effect of individual encounters is relatively
small (on the order of $10^{-3}$ to $10^{-4}$ \kms). The cumulative effect of \gaia\ stars
was then investigated. We found that the collective effect of stars passing within $\sim$ 1pc can
lift the perihelion of members of the Oort cloud in a relatively short period of time.    

Finally, we focused our study on the combined effect of multiple stellar encounters and the Galactic
tidal field on a simulated Oort cloud. To achieve this we used a detailed $N$-body simulations,
evolving a system of one host star (the Sun) surrounded by 10,000 test particles (the Oort cloud)
and affected by three different sets of external perturbations (\gaia\ stars only, Galactic tidal
field only, and the combination of both), over a period of $20$~Myr ($\pm10$~Myr centred on today).
 When we consider an extended Oort cloud  ($a\leq$ 100\,000~AU), we find that the effect of the
  Galactic tidal field alone leads to the creation of {TICs} of
  around $0.91$\% of the initial comets, while the collective effect of the passing stars only leads to a smaller
  fraction of $0.38$\%. For the compact model of the Oort cloud ($a\leq$ 50\,000~AU), passing stars
  dominate the perturbations, mainly due to the star GJ~710, while the effect of the Galactic tidal
  field is almost negligible. Overall for an extended cloud, the Galactic tide dominates over the passing stars, for the case
   of a compact cloud the opposite is true. However, it is the combined effect of passing stars and the Galactic tidal field which significantly
    increases the perturbation on the Oort cloud. These combined effects raise the semi-major axis of around $1.12$\% of the 
    initial particles for the compact model, and $\sim1.20$\% for the extended one, up to the interstellar regions (i.e. a > 100\,000 AU).  
The estimates presented in this work are conservative and 
  based on a small sample of known stars  that pass near to the Sun during $\pm10$~Myr. The effects of a more complete sample will 
 increase the number of TICs. Overall the external perturbations are an efficient mechanisms in the
formation of interstellar comets over a short period of time (in the order of tens of megayears).

The further evolution of the transitional interstellar comets  depends on the perturbations introduced by 
passing stars and the Galactic tidal field.  These perturbations determine  
whether  the transitional interstellar comets will remain bound to the solar system or eventually become
interstellar comets. Under the hypothesis that other planetary systems also
possess Oort cloud-like structures, they most probably experience the same 
mechanism of erosion due to external perturbations. This leads us to speculate 
that there is a large population of cometary bodies that occupy interstellar space. 
Therefore, visits to the solar system by interstellar comets such as ’Oumuamua 
may well be a frequent  occurrence.

\begin{acknowledgements}

We thank the referee for the constructive reports and helpful suggestions to improve the present work. 
ST expresses his gratitude to the Mexican National Council for Science and Technology (CONACYT) for
the grant \#291004-410780; to Leiden Observatory for the unconditional support; and to Daniel Hestroffer, Eric Mamajek, Konstantin Batygin, and Ylva
G\"{o}tberg, for the discussions and comments on the manuscript. This project was shaped in part at the 2018 New York Gaia Sprint. 
This work was supported by the Netherlands Research School for Astronomy (NOVA) and by NWO (grant \#621.016.701 [LGM-II]).
This work has made use of data from the European Space Agency (ESA) mission  {\it Gaia} (\url{https://www.cosmos.esa.int/gaia}), 
processed by the {\it Gaia} Data Processing and Analysis Consortium  (DPAC, \url{https://www.cosmos.esa.int/web/gaia/dpac/consortium}). 
Funding for the DPAC has been provided by national institutions, in particular the institutions participating in the {\it Gaia} Multilateral Agreement. 
This work was carried out on the Dutch national  e-infrastructure with the use of {\it Cartesius} the Dutch national supercomputer and 
the support of SURF Cooperative.

\end{acknowledgements}

\bibliographystyle{aa}
\bibliography{Bibliography.bib}

\begin{thebibliography}{85}
\expandafter\ifx\csname natexlab\endcsname\relax\def\natexlab#1{#1}\fi

\bibitem[{Abolfathi {et~al.}(2017)Abolfathi, Aguado, Aguilar, Prieto, Almeida,
  Ananna, Anders, Anderson, Andrews, Anguiano, Aragon-Salamanca,
  Argudo-Fernandez, Armengaud, Ata, Aubourg, Avila-Reese, Badenes, Bailey,
  Balland, Barger, Barrera-Ballesteros, Bartosz, Bastien, Bates, Baumgarten,
  Bautista, Beaton, Beers, Belfiore, Bender, Bernardi, Bershady, Beutler, Bird,
  Bizyaev, Blanc, Blanton, Blomqvist, Bolton, Boquien, Borissova, Bovy, Diaz,
  Brandt, Brinkmann, Brownstein, Bundy, Burgasser, Burtin, Busca, Canas,
  Cano-Diaz, Cappellari, Carrera, Casey, Sodi, Chen, Cherinka, Chiappini, Choi,
  Chojnowski, Chuang, Chung, Clerc, Cohen, Comerford, Comparat, do~Nascimento,
  da~Costa, Cousinou, Covey, Crane, Cruz-Gonzalez, Cunha, Ilha, Damke, Darling,
  Davidson, Dawson, Lizaola, de~la Macorra, de~la Torre, {De Lee}, Agathe,
  Machado, Dell'Agli, Delubac, Diamond-Stanic, Donor, Downes, Drory, des
  Bourboux, Duckworth, Dwelly, Dyer, Ebelke, Eigenbrot, Eisenstein, Elsworth,
  Emsellem, Eracleous, Erfanianfar, Escoffier, Fan, Alvar, Fernandez-Trincado,
  Cirolini, Feuillet, Finoguenov, Fleming, Font-Ribera, Freischlad, Frinchaboy,
  Fu, Chew, Galbany, Perez, Garcia-Dias, Garcia-Hernandez, Oehmichen, Gaulme,
  Gelfand, Gil-Marin, Gillespie, Goddard, Hernandez, Gonzalez-Perez, Grabowski,
  Green, Grier, Gueguen, Guo, Guy, Hagen, Hall, Harding, Hasselquist, Hawley,
  Hayes, Hearty, Hekker, Hernandez, Toledo, Hogg, Holley-Bockelmann, Holtzman,
  Hou, Hsieh, Hunt, Hutchinson, Hwang, Angel, Johnson, Jones, Jonsson, Jullo,
  Khan, Kinemuchi, Kirkby, Kirkpatrick, Kitaura, Knapp, Kneib, Kollmeier,
  Lacerna, Lane, Lang, Law, Goff, Lee, Li, Li, Lian, Liang, Lima, Lin, Long,
  Lucatello, Lundgren, Mackereth, MacLeod, Mahadevan, Maia, Majewski, Manchado,
  Maraston, Mariappan, Marques-Chaves, Masseron, Masters, McDermid, McGreer,
  Melendez, Meneses-Goytia, Merloni, Merrifield, Meszaros, Meza, Minchev,
  Minniti, Mueller, Muller-Sanchez, Muna, Munoz, Myers, Nair, Nandra, Ness,
  Newman, Nichol, Nidever, Nitschelm, Noterdaeme, O'Connell, Oelkers, Oravetz,
  Oravetz, Ortiz, Osorio, Pace, Padilla, Palanque-Delabrouille, Palicio, Pan,
  Pan, Parikh, Paris, Park, Peirani, Pellejero-Ibanez, Penny, Percival,
  Perez-Fournon, Petitjean, Pieri, Pinsonneault, Pisani, Prada, Prakash,
  Queiroz, Raddick, Raichoor, Rembold, Richstein, Riffel, Riffel, Rix, Robin,
  Torres, Roman-Zuniga, Ross, Rossi, Ruan, Ruggeri, Ruiz, Salvato, Sanchez,
  Sanchez, Almeida, Sanchez-Gallego, Rojas, Santiago, Schiavon, Schimoia,
  Schlafly, Schlegel, Schneider, Schuster, Schwope, Seo, Serenelli, Shen, Shen,
  Shetrone, Shull, Aguirre, Simon, Skrutskie, Slosar, Smethurst, Smith, Sobeck,
  Somers, Souter, Souto, Spindler, Stark, Stassun, Steinmetz, Stello,
  Storchi-Bergmann, Streblyanska, Stringfellow, Suarez, Sun, Szigeti,
  Taghizadeh-Popp, Talbot, Tang, Tao, Tayar, Tembe, Teske, Thaker, Thomas,
  Tissera, Tojeiro, Tremonti, Troup, Urry, Valenzuela, van~den Bosch,
  Vargas-Gonzalez, Vargas-Magana, Vazquez, Villanova, Vogt, Wake, Wang, Weaver,
  Weijmans, Weinberg, Westfall, Whelan, Wilcots, Wild, Williams, Wilson,
  Wood-Vasey, Wylezalek, Xiao, Yan, Yang, Ybarra, Yeche, Zakamska, Zamora,
  Zarrouk, Zasowski, Zhang, Zhao, Zhao, Zheng, Zheng, Zhou, Zhu, Zinn, \&
  Zou}]{Abolfathi2017}
Abolfathi, B., Aguado, D.~S., Aguilar, G., {et~al.} 2017, Astron. J., 42
  [\eprint[arXiv]{1707.09322}]

\bibitem[{Anderson \& Francis(2012)}]{Anderson2012}
Anderson, E. \& Francis, C. 2012, Astron. Lett., 38, 331

\bibitem[{Andrae {et~al.}(2018)Andrae, Fouesneau, Creevey, Ordenovic, Mary,
  Burlacu, Chaoul, Jean-Antoine-Piccolo, Kordopatis, Korn, Lebreton, Panem,
  Pichon, Th{\'{e}}venin, Walmsley, \& Bailer-Jones}]{Andrae2018}
Andrae, R., Fouesneau, M., Creevey, O., {et~al.} 2018, Astron. Astrophys., 616,
  A8

\bibitem[{{Ashton} {et~al.}(2018){Ashton}, {Gladman}, {Kavelaars}, \&
  {Williams}}]{Ashton2018}
{Ashton}, E., {Gladman}, B., {Kavelaars}, J., \& {Williams}, G. 2018, in
  AAS/Division for Planetary Sciences Meeting Abstracts, Vol.~50, AAS/Division
  for Planetary Sciences Meeting Abstracts \#50, 201.02

\bibitem[{Bailer-Jones(2015)}]{Bailer-Jones2015}
Bailer-Jones, C. A.~L. 2015, Astron. Astrophys., 575, A35

\bibitem[{Bailer-Jones(2018)}]{Bailer-Jones_1}
Bailer-Jones, C. A.~L. 2018, Astron. Astrophys., 609, A8

\bibitem[{Bailer-Jones {et~al.}(2018)Bailer-Jones, Rybizki, Andrae, \&
  Fouesneau}]{Bailer-Jones2018}
Bailer-Jones, C. A.~L., Rybizki, J., Andrae, R., \& Fouesneau, M. 2018, Astron.
  Astrophys., 1

\bibitem[{Berski \& Dybczy{\'{n}}ski(2016)}]{Berski2016}
Berski, F. \& Dybczy{\'{n}}ski, P.~A. 2016, Astron. Astrophys., 15, 4

\bibitem[{Bobylev \& Bajkova(2017)}]{Bobylev2017}
Bobylev, V.~V. \& Bajkova, A.~T. 2017, Astron. Lett., 43, 559

\bibitem[{Bovy(2015)}]{Bovy2015}
Bovy, J. 2015, Astron. J., 29

\bibitem[{Bovy(2017)}]{Bovy2017}
Bovy, J. 2017, Mon. Not. R. Astron. Soc., 1387, 1360

\bibitem[{Brasser {et~al.}(2006)Brasser, Duncan, \& Levison}]{Brasser2006}
Brasser, R., Duncan, M.~J., \& Levison, H.~F. 2006, Icarus, 184, 59

\bibitem[{Brasser \& Morbidelli(2013)}]{Brasser2013}
Brasser, R. \& Morbidelli, A. 2013, Icarus, 225, 40

\bibitem[{Buder {et~al.}(2018)Buder, Asplund, Duong, Kos, Lind, Ness, Sharma,
  Bland-Hawthorn, Casey, {De Silva}, D'Orazi, Freeman, Lewis, Lin, Martell,
  Schlesinger, Simpson, Zucker, Zwitter, Amarsi, Anguiano, Carollo, Casagrande,
  {\v{C}}otar, Cottrell, Costa, Gao, Hayden, Horner, Ireland, Kafle, Munari,
  Nataf, Nordlander, Stello, Ting, Traven, Watson, Wittenmyer, Wyse, Yong,
  Zinn, \& {\v{Z}}erjal}]{Buder2018}
Buder, S., Asplund, M., Duong, L., {et~al.} 2018, Mon. Not. R. Astron. Soc.,
  478, 4513

\bibitem[{{de la Fuente Marcos} {et~al.}(2018){de la Fuente Marcos}, {de la
  Fuente Marcos}, \& {Aarseth}}]{delaFuenteMarcos2018}
{de la Fuente Marcos}, C., {de la Fuente Marcos}, R., \& {Aarseth}, S.~J. 2018,
  \mnras, 476, L1

\bibitem[{{de la Fuente Marcos} \& {de la Fuente Marcos}(2018)}]{Marcos2018}
{de la Fuente Marcos}, R. \& {de la Fuente Marcos}, C. 2018, Res. Notes AAS, 2,
  30

\bibitem[{Dehnen \& Binney(1998)}]{Dehnen1998}
Dehnen, W. \& Binney, J.~J. 1998, Mon. Not. R. Astron. Soc., 298, 387

\bibitem[{Do {et~al.}(2018)Do, Tucker, \& Tonry}]{Do2018}
Do, A., Tucker, M.~A., \& Tonry, J. 2018, Astrophys. J., 855, L10

\bibitem[{Dones {et~al.}(2015)Dones, Brasser, Kaib, \& Rickman}]{Dones2015}
Dones, L., Brasser, R., Kaib, N., \& Rickman, H. 2015, Space Sci. Rev., 197,
  191

\bibitem[{Dones {et~al.}(2004)Dones, Weissman, Levison, \& Duncan}]{Dones2004a}
Dones, L., Weissman, P.~R., Levison, H.~F., \& Duncan, M.~J. 2004, Comets II,
  323, 153

\bibitem[{Duncan(2008)}]{Duncan2008}
Duncan, M.~J. 2008, Space Sci. Rev., 109

\bibitem[{Duncan {et~al.}(1987)Duncan, T., \& S.}]{DuncanM.J.QuinnT.1987}
Duncan, M.~J., T., Q., \& S., T. 1987, Astron. J.
  [\eprint[arXiv]{2008SSRv..138..109D}]

\bibitem[{Dybczy{\'{n}}ski(2002)}]{Dybczynski2002}
Dybczy{\'{n}}ski, P.~A. 2002, Earth, Moon Planets, 90, 483

\bibitem[{Dybczy{\'{n}}ski \& Berski(2015)}]{Dybczynski2015}
Dybczy{\'{n}}ski, P.~A. \& Berski, F. 2015, Mon. Not. R. Astron. Soc., 2471, 13

\bibitem[{Everhart(1985)}]{Everhart1985}
Everhart, E. 1985, Dyn. Comets Their Orig. Evol. Proc. IAU Colloq. 83, 115

\bibitem[{Feng \& Bailer-Jones(2015)}]{Feng2015}
Feng, F. \& Bailer-Jones, C.~A. 2015, Mon. Not. R. Astron. Soc., 454, 3267

\bibitem[{Fouchard {et~al.}(2006)Fouchard, Froeschl{\'{e}}, Valsecchi, \&
  Rickman}]{Fouchard2006}
Fouchard, M., Froeschl{\'{e}}, C., Valsecchi, G., \& Rickman, H. 2006, 299

\bibitem[{Fouchard {et~al.}(2018)Fouchard, Higuchi, Ito, \&
  Maquet}]{Fouchard2018}
Fouchard, M., Higuchi, A., Ito, T., \& Maquet, L. 2018, Astron. Astrophys., 45,
  1

\bibitem[{Fouchard {et~al.}(2011)Fouchard, Rickman, Froeschle, \&
  Valsecchi}]{Fouchard2011}
Fouchard, M., Rickman, H., Froeschle, C., \& Valsecchi, G.~B. 2011, Astron.
  Astrophys., 535, 86

\bibitem[{{Gaia Collaboration} {et~al.}(2018){Gaia Collaboration}, Brown,
  Vallenari, Prusti, de~Bruijne, Babusiaux, Bailer-Jones, Biermann, Evans,
  Eyer, Jansen, Jordi, Klioner, Lammers, Lindegren, Luri, Mignard, Panem,
  Pourbaix, Randich, Sartoretti, Siddiqui, Soubiran, van Leeuwen, Walton,
  Arenou, Bastian, Cropper, Drimmel, Katz, Lattanzi, Bakker, Cacciari,
  Casta{\~{n}}eda, Chaoul, Cheek, {De Angeli}, Fabricius, Guerra, Holl, Masana,
  Messineo, Mowlavi, Nienartowicz, Panuzzo, Portell, Riello, Seabroke, Tanga,
  Th{\'{e}}venin, Gracia-Abril, Comoretto, Garcia-Reinaldos, Teyssier, Altmann,
  Andrae, Audard, Bellas-Velidis, Benson, Berthier, Blomme, Burgess, Busso,
  Carry, Cellino, Clementini, Clotet, Creevey, Davidson, {De Ridder},
  Delchambre, Dell'Oro, Ducourant, Fern{\'{a}}ndez-Hern{\'{a}}ndez, Fouesneau,
  Fr{\'{e}}mat, Galluccio, Garc{\'{i}}a-Torres,
  Gonz{\'{a}}lez-N{\'{u}}{\~{n}}ez, Gonz{\'{a}}lez-Vidal, Gosset, Guy,
  Halbwachs, Hambly, Harrison, Hern{\'{a}}ndez, Hestroffer, Hodgkin, Hutton,
  Jasniewicz, Jean-Antoine-Piccolo, Jordan, Korn, Krone-Martins, Lanzafame,
  Lebzelter, L{\"{o}}ffler, Manteiga, Marrese, Mart{\'{i}}n-Fleitas, Moitinho,
  Mora, Muinonen, Osinde, Pancino, Pauwels, Petit, Recio-Blanco, Richards,
  Rimoldini, Robin, Sarro, Siopis, Smith, Sozzetti, S{\"{u}}veges, Torra, van
  Reeven, Abbas, {Abreu Aramburu}, Accart, Aerts, Altavilla, {\'{A}}lvarez,
  Alvarez, Alves, Anderson, Andrei, {Anglada Varela}, Antiche, Antoja, Arcay,
  Astraatmadja, Bach, Baker, Balaguer-N{\'{u}}{\~{n}}ez, Balm, Barache, Barata,
  Barbato, Barblan, Barklem, Barrado, Barros, Barstow, {Bartholom{\'{e}}
  Mu{\~{n}}oz}, Bassilana, Becciani, Bellazzini, Berihuete, Bertone, Bianchi,
  Bienaym{\'{e}}, Blanco-Cuaresma, Boch, Boeche, Bombrun, Borrachero, Bossini,
  Bouquillon, Bourda, Bragaglia, Bramante, Breddels, Bressan, Brouillet,
  Br{\"{u}}semeister, Brugaletta, Bucciarelli, Burlacu, Busonero, Butkevich,
  Buzzi, Caffau, Cancelliere, Cannizzaro, Cantat-Gaudin, Carballo, Carlucci,
  Carrasco, Casamiquela, Castellani, Castro-Ginard, Charlot, Chemin, Chiavassa,
  Cocozza, Costigan, Cowell, Crifo, Crosta, Crowley, Cuypers†, Dafonte,
  Damerdji, Dapergolas, David, David, de~Laverny, {De Luise}, {De March},
  de~Martino, de~Souza, de~Torres, Debosscher, del Pozo, Delbo, Delgado,
  Delgado, {Di Matteo}, Diakite, Diener, Distefano, Dolding, Drazinos,
  Dur{\'{a}}n, Edvardsson, Enke, Eriksson, Esquej, {Eynard Bontemps}, Fabre,
  Fabrizio, Faigler, Falc{\~{a}}o, {Farr{\`{a}}s Casas}, Federici, Fedorets,
  Fernique, Figueras, Filippi, Findeisen, Fonti, Fraile, Fraser,
  Fr{\'{e}}zouls, Gai, Galleti, Garabato, Garc{\'{i}}a-Sedano, Garofalo,
  Garralda, Gavel, Gavras, Gerssen, Geyer, Giacobbe, Gilmore, Girona,
  Giuffrida, Glass, Gomes, Granvik, Gueguen, Guerrier, Guiraud,
  Guti{\'{e}}rrez-S{\'{a}}nchez, Haigron, Hatzidimitriou, Hauser, Haywood,
  Heiter, Helmi, Heu, Hilger, Hobbs, Hofmann, Holland, Huckle, Hypki, Icardi,
  Jan{\ss}en, {Jevardat de Fombelle}, Jonker, Juh{\'{a}}sz, Julbe, Karampelas,
  Kewley, Klar, Kochoska, Kohley, Kolenberg, Kontizas, Kontizas, Koposov,
  Kordopatis, Kostrzewa-Rutkowska, Koubsky, Lambert, Lanza, Lasne, Lavigne, {Le
  Fustec}, {Le Poncin-Lafitte}, Lebreton, Leccia, Leclerc, Lecoeur-Taibi,
  Lenhardt, Leroux, Liao, Licata, Lindstr{\o}m, Lister, Livanou, Lobel,
  L{\'{o}}pez, Managau, Mann, Mantelet, Marchal, Marchant, Marconi, Marinoni,
  Marschalk{\'{o}}, Marshall, Martino, Marton, Mary, Massari, Matijevi{\v{c}},
  Mazeh, McMillan, Messina, Michalik, Millar, Molina, Molinaro, Moln{\'{a}}r,
  Montegriffo, Mor, Morbidelli, Morel, Morris, Mulone, Muraveva, Musella,
  Nelemans, Nicastro, Noval, O'Mullane, Ord{\'{e}}novic,
  Ord{\'{o}}{\~{n}}ez-Blanco, Osborne, Pagani, Pagano, Pailler, Palacin,
  Palaversa, Panahi, Pawlak, Piersimoni, Pineau, Plachy, Plum, Poggio,
  Poujoulet, Pr{\v{s}}a, Pulone, Racero, Ragaini, Rambaux, Ramos-Lerate,
  Regibo, Reyl{\'{e}}, Riclet, Ripepi, Riva, Rivard, Rixon, Roegiers, Roelens,
  Romero-G{\'{o}}mez, Rowell, Royer, Ruiz-Dern, Sadowski, {Sagrist{\`{a}}
  Sell{\'{e}}s}, Sahlmann, Salgado, Salguero, Sanna, Santana-Ros, Sarasso,
  Savietto, Schultheis, Sciacca, Segol, Segovia, S{\'{e}}gransan, Shih,
  Siltala, Silva, Smart, Smith, Solano, Solitro, Sordo, {Soria Nieto}, Souchay,
  Spagna, Spoto, Stampa, Steele, Steidelm{\"{u}}ller, Stephenson, Stoev, Suess,
  Surdej, Szabados, Szegedi-Elek, Tapiador, Taris, Tauran, Taylor, Teixeira,
  Terrett, Teyssandier, Thuillot, Titarenko, {Torra Clotet}, Turon, Ulla,
  Utrilla, Uzzi, Vaillant, Valentini, Valette, van Elteren, {Van Hemelryck},
  van Leeuwen, Vaschetto, Vecchiato, Veljanoski, Viala, Vicente, Vogt, von
  Essen, Voss, Votruba, Voutsinas, Walmsley, Weiler, Wertz, Wevers,
  Wyrzykowski, Yoldas, {\v{Z}}erjal, Ziaeepour, Zorec, Zschocke, Zucker,
  Zurbach, \& Zwitter}]{Gaiacollaboration2018}
{Gaia Collaboration}, Brown, A. G.~A., Vallenari, A., {et~al.} 2018, Astron.
  Astrophys., 616, A1

\bibitem[{{Gaia Collaboration, Brown} {et~al.}(2016){Gaia Collaboration,
  Brown}, Vallenari, Prusti, de~Bruijne, Mignard, Drimmel, Babusiaux,
  Bailer-Jones, Bastian, Biermann, Evans, Eyer, Jansen, Jordi, Katz, Klioner,
  Lammers, Lindegren, Luri, O'Mullane, Panem, Pourbaix, Randich, Sartoretti,
  Siddiqui, Soubiran, Valette, van Leeuwen, Walton, Aerts, Arenou, Cropper,
  H{\o}g, Lattanzi, Grebel, Holland, Huc, Passot, Perryman, Bramante, Cacciari,
  Casta{\~{n}}eda, Chaoul, Cheek, {De Angeli}, Fabricius, Guerra,
  Hern{\'{a}}ndez, Jean-Antoine-Piccolo, Masana, Messineo, Mowlavi,
  Nienartowicz, Ord{\'{o}}{\~{n}}ez-Blanco, Panuzzo, Portell, Richards, Riello,
  Seabroke, Tanga, Th{\'{e}}venin, Torra, Els, Gracia-Abril, Comoretto,
  Garcia-Reinaldos, Lock, Mercier, Altmann, Andrae, Astraatmadja,
  Bellas-Velidis, Benson, Berthier, Blomme, Busso, Carry, Cellino, Clementini,
  Cowell, Creevey, Cuypers, Davidson, {De Ridder}, de~Torres, Delchambre,
  Dell'Oro, Ducourant, Fr{\'{e}}mat, Garc{\'{i}}a-Torres, Gosset, Halbwachs,
  Hambly, Harrison, Hauser, Hestroffer, Hodgkin, Huckle, Hutton, Jasniewicz,
  Jordan, Kontizas, Korn, Lanzafame, Manteiga, Moitinho, Muinonen, Osinde,
  Pancino, Pauwels, Petit, Recio-Blanco, Robin, Sarro, Siopis, Smith, Smith,
  Sozzetti, Thuillot, van Reeven, Viala, Abbas, {Abreu Aramburu}, Accart,
  Aguado, Allan, Allasia, Altavilla, {\'{A}}lvarez, Alves, Anderson, Andrei,
  {Anglada Varela}, Antiche, Antoja, Ant{\'{o}}n, Arcay, Bach, Baker,
  Balaguer-N{\'{u}}{\~{n}}ez, Barache, Barata, Barbier, Barblan, {Barrado y
  Navascu{\'{e}}s}, Barros, Barstow, Becciani, Bellazzini, {Bello
  Garc{\'{i}}a}, Belokurov, Bendjoya, Berihuete, Bianchi, Bienaym{\'{e}},
  Billebaud, Blagorodnova, Blanco-Cuaresma, Boch, Bombrun, Borrachero,
  Bouquillon, Bourda, Bouy, Bragaglia, Breddels, Brouillet, Br{\"{u}}semeister,
  Bucciarelli, Burgess, Burgon, Burlacu, Busonero, Buzzi, Caffau, Cambras,
  Campbell, Cancelliere, Cantat-Gaudin, Carlucci, Carrasco, Castellani,
  Charlot, Charnas, Chiavassa, Clotet, Cocozza, Collins, Costigan, Crifo,
  Cross, Crosta, Crowley, Dafonte, Damerdji, Dapergolas, David, David, {De
  Cat}, de~Felice, de~Laverny, {De Luise}, {De March}, de~Martino, de~Souza,
  Debosscher, del Pozo, Delbo, Delgado, Delgado, {Di Matteo}, Diakite,
  Distefano, Dolding, {Dos Anjos}, Drazinos, Duran, Dzigan, Edvardsson, Enke,
  Evans, {Eynard Bontemps}, Fabre, Fabrizio, Faigler, Falc{\~{a}}o,
  {Farr{\`{a}}s Casas}, Federici, Fedorets, Fern{\'{a}}ndez-Hern{\'{a}}ndez,
  Fernique, Fienga, Figueras, Filippi, Findeisen, Fonti, Fouesneau, Fraile,
  Fraser, Fuchs, Gai, Galleti, Galluccio, Garabato, Garc{\'{i}}a-Sedano,
  Garofalo, Garralda, Gavras, Gerssen, Geyer, Gilmore, Girona, Giuffrida,
  Gomes, Gonz{\'{a}}lez-Marcos, Gonz{\'{a}}lez-N{\'{u}}{\~{n}}ez,
  Gonz{\'{a}}lez-Vidal, Granvik, Guerrier, Guillout, Guiraud, G{\'{u}}rpide,
  Guti{\'{e}}rrez-S{\'{a}}nchez, Guy, Haigron, Hatzidimitriou, Haywood, Heiter,
  Helmi, Hobbs, Hofmann, Holl, Holland, Hunt, Hypki, Icardi, Irwin, {Jevardat
  de Fombelle}, Jofr{\'{e}}, Jonker, Jorissen, Julbe, Karampelas, Kochoska,
  Kohley, Kolenberg, Kontizas, Koposov, Kordopatis, Koubsky, Krone-Martins,
  Kudryashova, Kull, Bachchan, Lacoste-Seris, Lanza, Lavigne, {Le
  Poncin-Lafitte}, Lebreton, Lebzelter, Leccia, Leclerc, Lecoeur-Taibi,
  Lemaitre, Lenhardt, Leroux, Liao, Licata, Lindstr{\o}m, Lister, Livanou,
  Lobel, L{\"{o}}ffler, L{\'{o}}pez, Lorenz, MacDonald, {Magalh{\~{a}}es
  Fernandes}, Managau, Mann, Mantelet, Marchal, Marchant, Marconi, Marinoni,
  Marrese, Marschalk{\'{o}}, Marshall, Mart{\'{i}}n-Fleitas, Martino, Mary,
  Matijevi{\v{c}}, Mazeh, McMillan, Messina, Michalik, Millar, Miranda, Molina,
  Molinaro, Molinaro, Moln{\'{a}}r, Moniez, Montegriffo, Mor, Mora, Morbidelli,
  Morel, Morgenthaler, Morris, Mulone, Muraveva, Musella, Narbonne, Nelemans,
  Nicastro, Noval, Ord{\'{e}}novic, Ordieres-Mer{\'{e}}, Osborne, Pagani,
  Pagano, Pailler, Palacin, Palaversa, Parsons, Pecoraro, Pedrosa,
  Pentik{\"{a}}inen, Pichon, Piersimoni, Pineau, Plachy, Plum, Poujoulet,
  Pr{\v{s}}a, Pulone, Ragaini, Rago, Rambaux, Ramos-Lerate, Ranalli, Rauw,
  Read, Regibo, Reyl{\'{e}}, Ribeiro, Rimoldini, Ripepi, Riva, Rixon, Roelens,
  Romero-G{\'{o}}mez, Rowell, Royer, Ruiz-Dern, Sadowski, {Sagrist{\`{a}}
  Sell{\'{e}}s}, Sahlmann, Salgado, Salguero, Sarasso, Savietto, Schultheis,
  Sciacca, Segol, Segovia, Segransan, Shih, Smareglia, Smart, Solano, Solitro,
  Sordo, {Soria Nieto}, Souchay, Spagna, Spoto, Stampa, Steele,
  Steidelm{\"{u}}ller, Stephenson, Stoev, Suess, S{\"{u}}veges, Surdej,
  Szabados, Szegedi-Elek, Tapiador, Taris, Tauran, Taylor, Teixeira, Terrett,
  Tingley, Trager, Turon, Ulla, Utrilla, Valentini, van Elteren, {Van
  Hemelryck}, van Leeuwen, Varadi, Vecchiato, Veljanoski, Via, Vicente, Vogt,
  Voss, Votruba, Voutsinas, Walmsley, Weiler, Weingrill, Wevers, Wyrzykowski,
  Yoldas, {\v{Z}}erjal, Zucker, Zurbach, Zwitter, Alecu, Allen, {Allende
  Prieto}, Amorim, Anglada-Escud{\'{e}}, Arsenijevic, Azaz, Balm, Beck,
  Bernstein, Bigot, Bijaoui, Blasco, Bonfigli, Bono, Boudreault, Bressan,
  Brown, Brunet, Bunclark, Buonanno, Butkevich, Carret, Carrion, Chemin,
  Ch{\'{e}}reau, Corcione, Darmigny, de~Boer, de~Teodoro, de~Zeeuw, {Delle
  Luche}, Domingues, Dubath, Fodor, Fr{\'{e}}zouls, Fries, Fustes, Fyfe,
  Gallardo, Gallegos, Gardiol, Gebran, Gomboc, G{\'{o}}mez, Grux, Gueguen,
  Heyrovsky, Hoar, Iannicola, {Isasi Parache}, Janotto, Joliet, Jonckheere,
  Keil, Kim, Klagyivik, Klar, Knude, Kochukhov, Kolka, Kos, Kutka, Lainey,
  LeBouquin, Liu, Loreggia, Makarov, Marseille, Martayan, Martinez-Rubi,
  Massart, Meynadier, Mignot, Munari, Nguyen, Nordlander, Ocvirk, O'Flaherty,
  {Olias Sanz}, Ortiz, Osorio, Oszkiewicz, Ouzounis, Palmer, Park, Pasquato,
  Peltzer, Peralta, P{\'{e}}turaud, Pieniluoma, Pigozzi, Poels, Prat,
  Prod'homme, Raison, Rebordao, Risquez, Rocca-Volmerange, Rosen, Ruiz-Fuertes,
  Russo, Sembay, {Serraller Vizcaino}, Short, Siebert, Silva, Sinachopoulos,
  Slezak, Soffel, Sosnowska, Strai{\v{z}}ys, ter Linden, Terrell, Theil, Tiede,
  Troisi, Tsalmantza, Tur, Vaccari, Vachier, Valles, {Van Hamme}, Veltz,
  Virtanen, Wallut, Wichmann, Wilkinson, Ziaeepour, \&
  Zschocke}]{Gaiacollaboration2017}
{Gaia Collaboration, Brown}, Vallenari, A., Prusti, T., {et~al.} 2016, Astron.
  Astrophys., 595, A2

\bibitem[{{Gaia Collaboration, Prusti} {et~al.}(2016){Gaia Collaboration,
  Prusti}, de~Bruijne, Brown, Vallenari, Babusiaux, Bailer-Jones, Bastian,
  Biermann, Evans, Eyer, Jansen, Jordi, Klioner, Lammers, Lindegren, Luri,
  Mignard, Milligan, Panem, Poinsignon, Pourbaix, Randich, Sarri, Sartoretti,
  Siddiqui, Soubiran, Valette, van Leeuwen, Walton, Aerts, Arenou, Cropper,
  Drimmel, H{\o}g, Katz, Lattanzi, O'Mullane, Grebel, Holland, Huc, Passot,
  Bramante, Cacciari, Casta{\~{n}}eda, Chaoul, Cheek, {De Angeli}, Fabricius,
  Guerra, Hern{\'{a}}ndez, Jean-Antoine-Piccolo, Masana, Messineo, Mowlavi,
  Nienartowicz, Ord{\'{o}}{\~{n}}ez-Blanco, Panuzzo, Portell, Richards, Riello,
  Seabroke, Tanga, Th{\'{e}}venin, Torra, Els, Gracia-Abril, Comoretto,
  Garcia-Reinaldos, Lock, Mercier, Altmann, Andrae, Astraatmadja,
  Bellas-Velidis, Benson, Berthier, Blomme, Busso, Carry, Cellino, Clementini,
  Cowell, Creevey, Cuypers, Davidson, {De Ridder}, de~Torres, Delchambre,
  Dell'Oro, Ducourant, Fr{\'{e}}mat, Garc{\'{i}}a-Torres, Gosset, Halbwachs,
  Hambly, Harrison, Hauser, Hestroffer, Hodgkin, Huckle, Hutton, Jasniewicz,
  Jordan, Kontizas, Korn, Lanzafame, Manteiga, Moitinho, Muinonen, Osinde,
  Pancino, Pauwels, Petit, Recio-Blanco, Robin, Sarro, Siopis, Smith, Smith,
  Sozzetti, Thuillot, van Reeven, Viala, Abbas, {Abreu Aramburu}, Accart,
  Aguado, Allan, Allasia, Altavilla, {\'{A}}lvarez, Alves, Anderson, Andrei,
  {Anglada Varela}, Antiche, Antoja, Ant{\'{o}}n, Arcay, Atzei, Ayache, Bach,
  Baker, Balaguer-N{\'{u}}{\~{n}}ez, Barache, Barata, Barbier, Barblan, Baroni,
  {Barrado y Navascu{\'{e}}s}, Barros, Barstow, Becciani, Bellazzini, Bellei,
  {Bello Garc{\'{i}}a}, Belokurov, Bendjoya, Berihuete, Bianchi,
  Bienaym{\'{e}}, Billebaud, Blagorodnova, Blanco-Cuaresma, Boch, Bombrun,
  Borrachero, Bouquillon, Bourda, Bouy, Bragaglia, Breddels, Brouillet,
  Br{\"{u}}semeister, Bucciarelli, Budnik, Burgess, Burgon, Burlacu, Busonero,
  Buzzi, Caffau, Cambras, Campbell, Cancelliere, Cantat-Gaudin, Carlucci,
  Carrasco, Castellani, Charlot, Charnas, Charvet, Chassat, Chiavassa, Clotet,
  Cocozza, Collins, Collins, Costigan, Crifo, Cross, Crosta, Crowley, Dafonte,
  Damerdji, Dapergolas, David, David, {De Cat}, de~Felice, de~Laverny, {De
  Luise}, {De March}, de~Martino, de~Souza, Debosscher, del Pozo, Delbo,
  Delgado, Delgado, di~Marco, {Di Matteo}, Diakite, Distefano, Dolding, {Dos
  Anjos}, Drazinos, Dur{\'{a}}n, Dzigan, Ecale, Edvardsson, Enke, Erdmann,
  Escolar, Espina, Evans, {Eynard Bontemps}, Fabre, Fabrizio, Faigler,
  Falc{\~{a}}o, {Farr{\`{a}}s Casas}, Faye, Federici, Fedorets,
  Fern{\'{a}}ndez-Hern{\'{a}}ndez, Fernique, Fienga, Figueras, Filippi,
  Findeisen, Fonti, Fouesneau, Fraile, Fraser, Fuchs, Furnell, Gai, Galleti,
  Galluccio, Garabato, Garc{\'{i}}a-Sedano, Gar{\'{e}}, Garofalo, Garralda,
  Gavras, Gerssen, Geyer, Gilmore, Girona, Giuffrida, Gomes,
  Gonz{\'{a}}lez-Marcos, Gonz{\'{a}}lez-N{\'{u}}{\~{n}}ez,
  Gonz{\'{a}}lez-Vidal, Granvik, Guerrier, Guillout, Guiraud, G{\'{u}}rpide,
  Guti{\'{e}}rrez-S{\'{a}}nchez, Guy, Haigron, Hatzidimitriou, Haywood, Heiter,
  Helmi, Hobbs, Hofmann, Holl, Holland, Hunt, Hypki, Icardi, Irwin, {Jevardat
  de Fombelle}, Jofr{\'{e}}, Jonker, Jorissen, Julbe, Karampelas, Kochoska,
  Kohley, Kolenberg, Kontizas, Koposov, Kordopatis, Koubsky, Kowalczyk,
  Krone-Martins, Kudryashova, Kull, Bachchan, Lacoste-Seris, Lanza, Lavigne,
  {Le Poncin-Lafitte}, Lebreton, Lebzelter, Leccia, Leclerc, Lecoeur-Taibi,
  Lemaitre, Lenhardt, Leroux, Liao, Licata, Lindstr{\o}m, Lister, Livanou,
  Lobel, L{\"{o}}ffler, L{\'{o}}pez, Lopez-Lozano, Lorenz, Loureiro, MacDonald,
  {Magalh{\~{a}}es Fernandes}, Managau, Mann, Mantelet, Marchal, Marchant,
  Marconi, Marie, Marinoni, Marrese, Marschalk{\'{o}}, Marshall,
  Mart{\'{i}}n-Fleitas, Martino, Mary, Matijevi{\v{c}}, Mazeh, McMillan,
  Messina, Mestre, Michalik, Millar, Miranda, Molina, Molinaro, Molinaro,
  Moln{\'{a}}r, Moniez, Montegriffo, Monteiro, Mor, Mora, Morbidelli, Morel,
  Morgenthaler, Morley, Morris, Mulone, Muraveva, Musella, Narbonne, Nelemans,
  Nicastro, Noval, Ord{\'{e}}novic, Ordieres-Mer{\'{e}}, Osborne, Pagani,
  Pagano, Pailler, Palacin, Palaversa, Parsons, Paulsen, Pecoraro, Pedrosa,
  Pentik{\"{a}}inen, Pereira, Pichon, Piersimoni, Pineau, Plachy, Plum,
  Poujoulet, Pr{\v{s}}a, Pulone, Ragaini, Rago, Rambaux, Ramos-Lerate, Ranalli,
  Rauw, Read, Regibo, Renk, Reyl{\'{e}}, Ribeiro, Rimoldini, Ripepi, Riva,
  Rixon, Roelens, Romero-G{\'{o}}mez, Rowell, Royer, Rudolph, Ruiz-Dern,
  Sadowski, {Sagrist{\`{a}} Sell{\'{e}}s}, Sahlmann, Salgado, Salguero,
  Sarasso, Savietto, Schnorhk, Schultheis, Sciacca, Segol, Segovia, Segransan,
  Serpell, Shih, Smareglia, Smart, Smith, Solano, Solitro, Sordo, {Soria
  Nieto}, Souchay, Spagna, Spoto, Stampa, Steele, Steidelm{\"{u}}ller,
  Stephenson, Stoev, Suess, S{\"{u}}veges, Surdej, Szabados, Szegedi-Elek,
  Tapiador, Taris, Tauran, Taylor, Teixeira, Terrett, Tingley, Trager, Turon,
  Ulla, Utrilla, Valentini, van Elteren, {Van Hemelryck}, van Leeuwen, Varadi,
  Vecchiato, Veljanoski, Via, Vicente, Vogt, Voss, Votruba, Voutsinas,
  Walmsley, Weiler, Weingrill, Werner, Wevers, Whitehead, Wyrzykowski, Yoldas,
  {\v{Z}}erjal, Zucker, Zurbach, Zwitter, Alecu, Allen, {Allende Prieto},
  Amorim, Anglada-Escud{\'{e}}, Arsenijevic, Azaz, Balm, Beck, Bernstein,
  Bigot, Bijaoui, Blasco, Bonfigli, Bono, Boudreault, Bressan, Brown, Brunet,
  Bunclark, Buonanno, Butkevich, Carret, Carrion, Chemin, Ch{\'{e}}reau,
  Corcione, Darmigny, de~Boer, de~Teodoro, de~Zeeuw, {Delle Luche}, Domingues,
  Dubath, Fodor, Fr{\'{e}}zouls, Fries, Fustes, Fyfe, Gallardo, Gallegos,
  Gardiol, Gebran, Gomboc, G{\'{o}}mez, Grux, Gueguen, Heyrovsky, Hoar,
  Iannicola, {Isasi Parache}, Janotto, Joliet, Jonckheere, Keil, Kim,
  Klagyivik, Klar, Knude, Kochukhov, Kolka, Kos, Kutka, Lainey, LeBouquin, Liu,
  Loreggia, Makarov, Marseille, Martayan, Martinez-Rubi, Massart, Meynadier,
  Mignot, Munari, Nguyen, Nordlander, Ocvirk, O'Flaherty, {Olias Sanz}, Ortiz,
  Osorio, Oszkiewicz, Ouzounis, Palmer, Park, Pasquato, Peltzer, Peralta,
  P{\'{e}}turaud, Pieniluoma, Pigozzi, Poels, Prat, Prod'homme, Raison,
  Rebordao, Risquez, Rocca-Volmerange, Rosen, Ruiz-Fuertes, Russo, Sembay,
  {Serraller Vizcaino}, Short, Siebert, Silva, Sinachopoulos, Slezak, Soffel,
  Sosnowska, Strai{\v{z}}ys, ter Linden, Terrell, Theil, Tiede, Troisi,
  Tsalmantza, Tur, Vaccari, Vachier, Valles, {Van Hamme}, Veltz, Virtanen,
  Wallut, Wichmann, Wilkinson, Ziaeepour, \& Zschocke}]{GC_Prusti2016}
{Gaia Collaboration, Prusti}, de~Bruijne, J. H.~J., Brown, A. G.~A., {et~al.}
  2016, Astron. Astrophys., 595, A1

\bibitem[{Garc{\'{i}}a-S{\'{a}}nchez {et~al.}(1999)Garc{\'{i}}a-S{\'{a}}nchez,
  Preston, Jones, Weissman, Lestrade, Latham, \& Stefanik}]{Garcia-Sanchez1999}
Garc{\'{i}}a-S{\'{a}}nchez, J., Preston, R.~A., Jones, D.~L., {et~al.} 1999,
  Astron. J., 117, 1042

\bibitem[{Garc{\'{i}}a-S{\'{a}}nchez {et~al.}(2001)Garc{\'{i}}a-S{\'{a}}nchez,
  Weissman, Preston, Jones, Lestrade, Latham, Stefanik, \&
  Paredes}]{Garcia-Sanchez2001}
Garc{\'{i}}a-S{\'{a}}nchez, J., Weissman, P.~R., Preston, R.~A., {et~al.} 2001,
  Astron. Astrophys., 379, 634

\bibitem[{{Gontcharov}(2006)}]{pulkovo_2006}
{Gontcharov}, G.~A. 2006, Astronomy Letters, 32, 759

\bibitem[{Hanse {et~al.}(2016)Hanse, J{\'{i}}lkov{\'{a}}, {Portegies Zwart}, \&
  Pelupessy}]{Hanse2016}
Hanse, J., J{\'{i}}lkov{\'{a}}, L., {Portegies Zwart}, S.~F., \& Pelupessy,
  F.~I. 2016, Mon. Not. R. Astron. Soc., 473, 5432

\bibitem[{Heisler \& Tremaine(1986)}]{Heisler1986a}
Heisler, J. \& Tremaine, S. 1986, Icarus, 65, 13

\bibitem[{Hernquist(1990)}]{Hernquist1990}
Hernquist, L. 1990, Astron. J., 359

\bibitem[{Higuchi \& Kokubo(2015)}]{Higuchi2015}
Higuchi, A. \& Kokubo, E. 2015, Astron. J., 150, 26

\bibitem[{Hills(1981)}]{Hills1981}
Hills, J. 1981, Astron. J.

\bibitem[{Jewitt {et~al.}(2017)Jewitt, Luu, Rajagopal, Kotulla, Ridgway, Liu,
  \& Augusteijn}]{Jewitt2017}
Jewitt, D., Luu, J., Rajagopal, J., {et~al.} 2017, Astrophys. J., 850, L36

\bibitem[{J{\'{i}}lkov{\'{a}} {et~al.}(2016)J{\'{i}}lkov{\'{a}}, Hamers,
  Hammer, \& Zwart}]{Jilkova2016a}
J{\'{i}}lkov{\'{a}}, L., Hamers, A.~S., Hammer, M., \& Zwart, S.~P. 2016, Mon.
  Not. R. Astron. Soc., 457, 4218

\bibitem[{J{\'{i}}lkov{\'{a}} {et~al.}(2015)J{\'{i}}lkov{\'{a}}, {Portegies
  Zwart}, Pijloo, \& Hammer}]{Jilkova2015}
J{\'{i}}lkov{\'{a}}, L., {Portegies Zwart}, S., Pijloo, T., \& Hammer, M. 2015,
  Mon. Not. R. Astron. Soc., 453, 3158

\bibitem[{Jim{\'{e}}nez-Esteban {et~al.}(2018)Jim{\'{e}}nez-Esteban, Torres,
  Rebassa-Mansergas, Skorobogatov, Solano, Cantero, \&
  Rodrigo}]{Jimenez-Esteban2018}
Jim{\'{e}}nez-Esteban, F.~M., Torres, S., Rebassa-Mansergas, A., {et~al.} 2018,
  Mon. Not. R. Astron. Soc., 4518, 4505

\bibitem[{Jimenez-Torres {et~al.}(2011)Jimenez-Torres, Pichardo, Lake, \&
  Throop}]{Jimenez-Torres2011a}
Jimenez-Torres, J.~J., Pichardo, B., Lake, G., \& Throop, H. 2011, Mon. Not. R.
  Astron. Soc., 418, 1272

\bibitem[{Kaib \& Quinn(2008)}]{Kaib2008}
Kaib, N.~A. \& Quinn, T. 2008, Icarus, 197, 221

\bibitem[{Kunder {et~al.}(2017)Kunder, Kordopatis, Steinmetz, Zwitter,
  McMillan, Casagrande, Enke, Wojno, Valentini, Chiappini, Matijevi{\v{c}},
  Siviero, de~Laverny, Recio-Blanco, Bijaoui, Wyse, Binney, Grebel, Helmi,
  Jofre, Antoja, Gilmore, Siebert, Famaey, Bienaym{\'{e}}, Gibson, Freeman,
  Navarro, Munari, Seabroke, Anguiano, {\v{Z}}erjal, Minchev, Reid,
  Bland-Hawthorn, Kos, Sharma, Watson, Parker, Scholz, Burton, Cass, Hartley,
  Fiegert, Stupar, Ritter, Hawkins, Gerhard, Chaplin, Davies, Elsworth, Lund,
  Miglio, \& Mosser}]{Kunder2016}
Kunder, A., Kordopatis, G., Steinmetz, M., {et~al.} 2017, Astron. J., 153, 75

\bibitem[{Levison {et~al.}(2004)Levison, Morbidelli, \& Dones}]{Levison2004}
Levison, H.~F., Morbidelli, A., \& Dones, L. 2004, Astron. J., 128, 2553

\bibitem[{Lindegren {et~al.}(2018)Lindegren, Hern{\'{a}}ndez, Bombrun, Klioner,
  Bastian, \& Torres}]{Lindegren2018a}
Lindegren, L., Hern{\'{a}}ndez, J., Bombrun, A., {et~al.} 2018, Astron.
  Astrophys., 2 [\eprint[arXiv]{arXiv:1804.09366v1}]

\bibitem[{{M. Price-Whelan}(2017)}]{Gala2017}
{M. Price-Whelan}, A. 2017, J. Open Source Softw., 2, 388

\bibitem[{Mamajek(2018)}]{Mamajeck18}
Mamajek, E.~E. 2018, Priv. Commun.

\bibitem[{Mamajek {et~al.}(2015)Mamajek, Barenfeld, Ivanov, Kniazev,
  V{\"{a}}is{\"{a}}nen, Beletsky, \& Boffin}]{Mamajek2015}
Mamajek, E.~E., Barenfeld, S.~A., Ivanov, V.~D., {et~al.} 2015, Astrophys. J.,
  800, L17

\bibitem[{Mart{\'{i}}nez-Barbosa {et~al.}(2016)Mart{\'{i}}nez-Barbosa, Brown,
  Boekholt, {Portegies Zwart}, Antiche, \& Antoja}]{Martinez-Barbosa2016}
Mart{\'{i}}nez-Barbosa, C.~A., Brown, A. G.~A., Boekholt, T., {et~al.} 2016,
  Mon. Not. R. Astron. Soc., 457, 1062

\bibitem[{Mart{\'{i}}nez-Barbosa {et~al.}(2017)Mart{\'{i}}nez-Barbosa,
  J{\'{i}}lkov{\'{a}}, {Portegies Zwart}, \& Brown}]{Martinez-Barbosa2017}
Mart{\'{i}}nez-Barbosa, C.~A., J{\'{i}}lkov{\'{a}}, L., {Portegies Zwart}, S.,
  \& Brown, A.~G. 2017, Mon. Not. R. Astron. Soc., 464, 2290

\bibitem[{{Matthews}(1994)}]{Matthews1994}
{Matthews}, R.~A.~J. 1994, \qjras, 35, 1

\bibitem[{Miyamoto \& Nagai(1975)}]{Miyamoto-Nagai1975}
Miyamoto, M. \& Nagai, R. 1975, Astron. Soc. Japan, 27, 533

\bibitem[{Morbidelli(2008)}]{Morbidelli2008}
Morbidelli, A. 2008, 1

\bibitem[{Navarro {et~al.}(1996)Navarro, Frenk, \& White}]{NFW1995}
Navarro, J.~F., Frenk, C.~S., \& White, S. D.~M. 1996, Astron. J., 462, 563

\bibitem[{Oort(1950)}]{J.H.Oort1950}
Oort, J.~H. 1950, Commun. from Obs. Leiden, 408

\bibitem[{Pecaut \& Mamajek(2013)}]{Pecaut2013}
Pecaut, M.~J. \& Mamajek, E.~E. 2013, Astrophys. Journal, Suppl. Ser., 208
  [\eprint[arXiv]{1307.2657}]

\bibitem[{Pelupessy {et~al.}(2013)Pelupessy, van Elteren, de~Vries, McMillan,
  Drost, \& {Portegies Zwart}}]{Pelupessy2013a}
Pelupessy, F.~I., van Elteren, A., de~Vries, N., {et~al.} 2013, Astron.
  Astrophys., 557, A84

\bibitem[{Perryman {et~al.}(1997)Perryman, Lindegren, Kovalevsky, Hoeg,
  Bastian, Bernacca, Cr{\'{e}}z{\'{e}}, Donati, Grenon, van Leeuwen, van~der
  Marel, Mignard, Murray, {Le Poole}, Schrijver, Turon, Arenou,
  Froeschl{\'{e}}, \& Petersen}]{Perryman1997}
Perryman, M. A.~C., Lindegren, L., Kovalevsky, J., {et~al.} 1997, Astron.
  Astrophys., 323, L49

\bibitem[{{Portegies Zwart} \& {McMillan}(2018)}]{amusebook}
{Portegies Zwart}, S. \& {McMillan}, S. 2018, {Astrophysical Recipes; The art
  of AMUSE}

\bibitem[{{Portegies Zwart} {et~al.}(2009){Portegies Zwart}, McMillan, Harfst,
  Groen, Fujii, Nuall{\'{a}}in, Glebbeek, Heggie, Lombardi, Hut, Angelou,
  Banerjee, Belkus, Fragos, Fregeau, Gaburov, Izzard, Juri{\'{c}}, Justham,
  Sottoriva, Teuben, van Bever, Yaron, \& Zemp}]{Portegies2009}
{Portegies Zwart}, S., McMillan, S., Harfst, S., {et~al.} 2009, New Astron.,
  14, 369

\bibitem[{{Portegies Zwart} {et~al.}(2018){Portegies Zwart}, Torres, Pelupessy,
  B{\'{e}}dorf, \& Cai}]{Zwart2017}
{Portegies Zwart}, S., Torres, S., Pelupessy, I., B{\'{e}}dorf, J., \& Cai,
  M.~X. 2018, Mon. Not. R. Astron. Soc. Lett., 479, L17

\bibitem[{{Portegies Zwart} \& J{\'{i}}lkov{\'{a}}(2015)}]{PortegiesZwart2015a}
{Portegies Zwart}, S.~F. \& J{\'{i}}lkov{\'{a}}, L. 2015, Mon. Not. R. Astron.
  Soc., 451, 144

\bibitem[{Portegies~Zwart {et~al.}(2013)Portegies~Zwart, Mcmillan, Elteren,
  Pelupessy, \& Vries}]{Portegies2013}
Portegies~Zwart, S.~F., Mcmillan, S. L.~W., Elteren, A.~V., Pelupessy, F.~I.,
  \& Vries, N.~D. 2013, Comput. Phys. Commun., 184, 456

\bibitem[{Reid {et~al.}(2014)Reid, Menten, Brunthaler, Zheng, Dame, Xu, Wu,
  Zhang, Sanna, Sato, Hachisuka, Choi, Immer, Moscadelli, Rygl, \&
  Bartkiewicz}]{Reid2014}
Reid, M.~J., Menten, K.~M., Brunthaler, A., {et~al.} 2014, Astrophys. J., 783
  [\eprint[arXiv]{1401.5377}]

\bibitem[{Rickman(1976)}]{Rickman1976}
Rickman, H. 1976, Bull. Astron. Institutes Czechoslov., 27, 92

\bibitem[{Rickman {et~al.}(2008)Rickman, Fouchard, Froeschl{\'{e}}, \&
  Valsecchi}]{Rickman2008}
Rickman, H., Fouchard, M., Froeschl{\'{e}}, C., \& Valsecchi, G.~B. 2008,
  Celest. Mech. Dyn. Astron., 102, 111

\bibitem[{Rickman {et~al.}(2004)Rickman, Froeschl{\'{e}}, Froeschl{\'{e}}, \&
  Valsecchi}]{Rickman2004}
Rickman, H., Froeschl{\'{e}}, C., Froeschl{\'{e}}, C., \& Valsecchi, G.~B.
  2004, Astron. Astrophys., 428, 673

\bibitem[{Scholz(2014)}]{Scholz2014}
Scholz, R.-D. 2014, Astron. Astrophys., 561, A113

\bibitem[{Sch{\"{o}}nrich {et~al.}(2010)Sch{\"{o}}nrich, Binney, \&
  Dehnen}]{Schonrich2010}
Sch{\"{o}}nrich, R., Binney, J., \& Dehnen, W. 2010, Mon. Not. R. Astron. Soc.,
  403, 1829

\bibitem[{Shannon {et~al.}(2014)Shannon, Jackson, Veras, \&
  Wyatt}]{Shannon2014}
Shannon, A., Jackson, A.~P., Veras, D., \& Wyatt, M. 2014, Mon. Not. R. Astron.
  Soc., 446, 2059

\bibitem[{{Siraj} \& {Loeb}(2019)}]{Siraj2019}
{Siraj}, A. \& {Loeb}, A. 2019, \apjl, 872, L10

\bibitem[{{Soubiran} {et~al.}(2018){Soubiran}, {Jasniewicz}, {Chemin},
  {Zurbach}, {Brouillet}, {Panuzzo}, {Sartoretti}, {Katz}, {Le Campion},
  {Marchal}, {Hestroffer}, {Th{\'e}venin}, {Crifo}, {Udry}, {Cropper},
  {Seabroke}, {Viala}, {Benson}, {Blomme}, {Jean-Antoine}, {Huckle}, {Smith},
  {Baker}, {Damerdji}, {Dolding}, {Fr{\'e}mat}, {Gosset}, {Guerrier}, {Guy},
  {Haigron}, {Jan{\ss}en}, {Plum}, {Fabre}, {Lasne}, {Pailler}, {Panem},
  {Riclet}, {Royer}, {Tauran}, {Zwitter}, {Gueguen}, \& {Turon}}]{gdr2_rv}
{Soubiran}, C., {Jasniewicz}, G., {Chemin}, L., {et~al.} 2018, \aap, 616, A7

\bibitem[{{The Astropy Collaboration} {et~al.}(2018){The Astropy
  Collaboration}, Price-Whelan, Sipőcz, G{\"{u}}nther, Lim, Crawford, Conseil,
  Shupe, Craig, Dencheva, Ginsburg, VanderPlas, Bradley,
  P{\'{e}}rez-Su{\'{a}}rez, de~Val-Borro, Aldcroft, Cruz, Robitaille, Tollerud,
  Ardelean, Babej, Bach, Bachetti, Bakanov, Bamford, Barentsen, Barmby,
  Baumbach, Berry, Biscani, Boquien, Bostroem, Bouma, Brammer, Bray,
  Breytenbach, Buddelmeijer, Burke, Calderone, Rodr{\'{i}}guez, Cara, Cardoso,
  Cheedella, Copin, Corrales, Crichton, D'Avella, Deil, Depagne, Dietrich,
  Donath, Droettboom, Earl, Erben, Fabbro, Ferreira, Finethy, Fox, Garrison,
  Gibbons, Goldstein, Gommers, Greco, Greenfield, Groener, Grollier, Hagen,
  Hirst, Homeier, Horton, Hosseinzadeh, Hu, Hunkeler, Ivezi{\'{c}}, Jain,
  Jenness, Kanarek, Kendrew, Kern, Kerzendorf, Khvalko, King, Kirkby, Kulkarni,
  Kumar, Lee, Lenz, Littlefair, Ma, Macleod, Mastropietro, McCully, Montagnac,
  Morris, Mueller, Mumford, Muna, Murphy, Nelson, Nguyen, Ninan, N{\"{o}}the,
  Ogaz, Oh, Parejko, Parley, Pascual, Patil, Patil, Plunkett, Prochaska,
  Rastogi, Janga, Sabater, Sakurikar, Seifert, Sherbert, Sherwood-Taylor, Shih,
  Sick, Silbiger, Singanamalla, Singer, Sladen, Sooley, Sornarajah, Streicher,
  Teuben, Thomas, Tremblay, Turner, Terr{\'{o}}n, van Kerkwijk, de~la Vega,
  Watkins, Weaver, Whitmore, Woillez, \& Zabalza}]{TheAstropyCollaboration2018}
{The Astropy Collaboration}, Price-Whelan, A.~M., Sipőcz, B.~M., {et~al.}
  2018, Astron. J., 156, 123

\bibitem[{Torres {et~al.}(2018)Torres, Zwart, \& Brown}]{Torres2018}
Torres, S., Zwart, S.~P., \& Brown, A. G.~A. 2018, Proc. Int. Astron. Union,
  12, 269

\bibitem[{Valtonen \& Innanen(1982)}]{Valtonen1982}
Valtonen, M.~J. \& Innanen, K.~A. 1982, Astron. J., 307

\bibitem[{Veras {et~al.}(2013)Veras, Evans, Wyatt, \& Tout}]{Veras2013}
Veras, D., Evans, N.~W., Wyatt, M.~C., \& Tout, C.~A. 2013, Mon. Not. R.
  Astron. Soc., 437, 1127

\bibitem[{Veras {et~al.}(2014)Veras, Shannon, \& Boris}]{Veras2014}
Veras, D., Shannon, A., \& Boris, T.~G. 2014, 4185, 4175

\bibitem[{Weissman(1996)}]{Weissman1996}
Weissman, P.~R. 1996, Earth, Moon Planets, 72, 25

\bibitem[{Wiegert \& Tremaine(1999)}]{Wiegert1999}
Wiegert, P. \& Tremaine, S. 1999, Icarus, 137, 84

\bibitem[{Williams(2017)}]{Williams2017}
Williams, G. 2017, Minor Planet Electron. Circ., 2017-U181

\bibitem[{Zhao {et~al.}(2012)Zhao, Zhao, Chu, Jing, \& Deng}]{Zhao2012}
Zhao, G., Zhao, Y.~H., Chu, Y.~Q., Jing, Y.~P., \& Deng, L.~C. 2012, Res.
  Astron. Astrophys., 12, 723

\end{thebibliography}

\end{document}